\documentclass{jfm}
\usepackage{graphicx}
\usepackage{epstopdf, epsfig}
\usepackage{amsmath,amssymb}
\usepackage{bm}				
\usepackage[backref]{hyperref}
\usepackage{caption}
\usepackage{subcaption}
\usepackage{color} 
\usepackage[svgnames]{xcolor} 
\usepackage[normalem]{ulem}
\usepackage{cancel}

\newcommand{\beq}{\begin{equation}}
\newcommand{\eeq}{\end{equation}}
\newcommand\Red{\mbox{\textit{Re}}_\epsilon}   
\newcommand\Reu{\mbox{\textit{Re}}_u} 
\newcommand\Ref{\mbox{\textit{Re}}_f}
\newcommand\Rod{\mbox{\textit{Ro}}_\epsilon} 
\newcommand\Rou{\mbox{\textit{Ro}}_u}
\newcommand\Rof{\mbox{\textit{Ro}}_f}
\newcommand\Rol{\mbox{\textit{Ro}}_\lambda}
\newcommand\einv{\epsilon_{inv}}
\newcommand\UTD{U_{_{2D}}^2}

\shorttitle{Condensates in rotating turbulent flows}
\shortauthor{K. Seshasayanan, A. Alexakis}

\title{Condensates in rotating turbulent flows}

\author{Kannabiran Seshasayanan\aff{1}
  \corresp{\email{skannabiran@lps.ens.fr}},
  \and Alexandros Alexakis\aff{1}}

\affiliation{\aff{1} Laboratoire de  physique  statistique,  D{\'e}partement  de  physique  de  l'
{\'E}cole  normale  sup{\'e}rieure,  PSL Research University, Universit{\'e} Paris Diderot, Sorbonne Paris Cité{\'e}, 
Sorbonne Universit{\'e}s, UPMC Univ. Paris 06, CNRS, 75005 Paris, France
}

\begin{document}

\maketitle

\begin{abstract}

Using a large number of numerical simulations we examine 
the steady state of rotating turbulent flows in triple periodic domains,
varying the Rossby number $Ro$ (that measures the inverse rotation rate) 
and the Reynolds number $Re$ (that measures the strength of turbulence).
The examined flows are sustained by either a helical or a non-helical Roberts force, 
that is invariant along the axis of rotation. 
The forcing acts at a wavenumber $k_f$ such that $k_fL=4$, where $2\pi L$ is the size of the domain. 
Different flow behaviours were obtained as the parameters are varied. 
Above a critical rotation rate the flow becomes quasi two dimensional and 
transfers energy to the largest scales of the system forming large coherent 
structures known as condensates. 
We examine the behaviour of these condensates and their scaling properties close and away from this critical rotation rate. 
Close to the the critical rotation rate the system transitions super-critically to the condensate state
displaying a bimodal behaviour oscillating randomly between an incoherent-turbulent state and a condensate state.
Away from the critical rotation rate, it is shown that two distinct mechanisms can saturate the growth of the large scale energy.
The first mechanism is due to viscous forces and is similar to the saturation mechanism observed for the inverse cascade in two-dimensional flows.
The second mechanism is independent of viscosity and relies on the breaking of the two-dimensionalization condition of the rotating flow. 
The two mechanisms predict different scaling with respect to the control parameters of the system (Rossby and Reynolds), 
which are tested with the present results of the numerical simulations.
A phase space diagram in the $Re,Ro$ parameter plane is sketched.

\end{abstract}


\section{Introduction}


Turbulent rotating flows are met in a variety of contexts in nature. From 
the interior of stars, to planet atmospheres and industrial applications, rotation plays a 
dominant role in determining the properties of the underlying turbulence \citep{greenspan1968theory,hopfinger1993vortices,pedlosky2013geophysical}.
In its simplest form an incompressible turbulent flow in the presence of rotation is 
controlled by the incompressible Navier-Stokes equation, that in a rotating frame of reference reads:
\beq
\partial_t {\bf u} + {\bf u \cdot \nabla u} + 2{\bf \Omega \hat{e}_z \times u  } =-\nabla P + \nu \Delta {\bf u} + {\bf F}
\label{NSR}
\eeq
where ${\bf u}$ is the incompressible  velocity field, $\Omega$ is the rotation rate (assumed here to be in the $z$ direction
with $\hat{\bf e}_z$ its unit vector),
$P$ is the pressure that enforces the incompressibility condition ${\bf \nabla \cdot u}=0$,  $\nu$ is the viscosity and
$\bf F$ is a mechanical body force that acts at some length-scale $\ell_f$. Traditionally the strength of turbulence 
compared to viscous forces is measured by the Reynolds number $Re=U\ell_f/\nu$, while compared to the Coriolis force it is measured by the Rossby 
number $Ro=U/ \left( 2 \Omega\ell_f \right)$, where $U$ stands for the velocity amplitude. Precise definitions of these numbers will be given when we describe in 
detail the model under study.

It has been known for some time that 
when rotation is very strong, flows 
tend to become quasi-two-dimensional (quasi-2D) 
varying very weakly along the direction of rotation \citep{Hough1897,Proudman1916,Taylor1917}. 
The reason for this behaviour is that the incompressible projection of the Coriolis force 
$2{ \bf  \Omega  \hat{e}_z \times u -\nabla} P'= 2\Omega \Delta^{-1}\partial_z \nabla \times {\bf u} $ 
does not act on the part of the flow that is invariant along the rotation axis $\partial_z\bf u=0$.
At the same time velocity fluctuations that vary along this axis become inertial waves that satisfy the dispersion relation,
\begin{align}
\omega_{\bf k} =  \pm 2 \Omega \frac{k_z}{k},
\label{eqn:dispersion}
\end{align}    
where $\omega_{\bf k}$ is the wave frequency, $\bf k$ the wavenumber and the sign depends on the helicity of the mode.
Fast rotation leads to a de-correlation of inertial waves weakening their interactions. Thus, in the presence of strong rotation, 
fluid motions that are invariant along the direction of rotation  (often referred as the slow manifold)
become isolated from the remaining flow and if forced they dominate leading to the quasi-2D behaviour \citep{Chen2005,scott2014wave}.  
This quasi-2D behaviour has been realized 
in experiments \citep{ibbetson1975experiments,hopfinger1982turbulence,dickinson1983oscillating,
baroud2002anomalous,baroud2003scaling,Sugihara2005,ruppert2005extraction,
morize2006energy,Staplehurst2008,Bokhoven2009experiments,Yoshimatsu2011,Machicoane2016Two} 
and numerical simulations 
\citep{yeung1998numerical,Smith1999,godeferd1999direct,Chen2005,Thiele2009,Mininni2009,Mininni2010,favier2010space,
Mininni2012,marino2013inverse,alexakis2015rotating,biferale2016coherent,valente2017spectral}.

These arguments however have various limitations.
For large Reynolds numbers $Re$, the quasi-2D behaviour breaks down at scales $\ell$ smaller than 
the Zeman scale $\ell_{_Z}$ defined as the scale for which the vorticity $w_\ell \propto u_\ell/\ell$ 
is comparable to the rotation rate $\Omega$ \citep{zeman1994note}. 
Here $u_\ell$ stands for the typical velocity at scale $\ell$.
Thus for large $Re$ and low Rossby flows, such that $1 \ll 1/Ro \ll Re$, 
the large scales $\ell>\ell_{_Z}$ show a quasi-2D behaviour while 
smaller scales $\ell<\ell_{_Z}$ display three-dimensional (3D) behaviour.
Furthermore, the quasi-2D behaviour is also expected to break down 
even at large scales for sufficiently elongated boxes $H \gg \ell_f$, 
(where $H$ stands for the domain size in the direction of rotation). 
If $H$ is sufficiently large, the slowest inertial mode has a frequency 
$\omega \sim \Omega \ell_f/H$ comparable or smaller to the inverse eddy turnover time $\ell/u_\ell$.
This last limiting procedure, $1 \ll 1/Ro \ll H/\ell_f $ provided also that $Re \gg1$
corresponds to the weak wave turbulence limit, in which the nonlinear interactions can be treated in
a perturbative manner \cite{galtier2003weak,nazarenko2011wave}.
Finally,
for finite (fixed) heights $H$ and finite (fixed) Reynolds numbers, fast rotating flows become exactly 2D
above a critical rotation rate \citep{Gallet2015exact}. This corresponds to the limiting 
procedure $Re \ll 1/Ro$ and $H/\ell_f \ll 1/Ro$.
Thus, in general, the quasi-2D behaviour at low $Ro$ depends, on the scales under investigation, 
the  geometry of the system, and the relative amplitude of the Rossby and Reynolds number,
with different limits leading to different results.

The distinctive difference between 3D and 2D or quasi-2D flows is that the former one cascades energy to small scales 
while the later one cascades energy to large scales. 
Thus a significant change in the energy balance occurs when the rotation rate is increased and the flow becomes quasi-2D:
while in a forward cascade the energy that arrives at small scales gets dissipated, in an inverse cascade 
energy piles up at scales of the size of the domain size $L$. 
Indeed it has been shown both 
in numerical simulations \citep{Smith1999,smith1996crossover,pouquet2013inverse,Mininni2012,deusebio2014dimensional,biferale2016coherent} 
and experiments  \citep{Yarom2013experimental,Moisy2014direct,yarom2014experimental,campagne2015disentangling,campagne2016turbulent} 
that while for weak rotation the flow is close to isotropic state and cascades all energy to the small scales, 
for fast rotation the flow is in a quasi-2D state  that cascades at least part of the energy to the large scales.
This change in the direction of the cascade as a parameter is varied has been 
the subject of study of various investigations in different systems 
\citep{Smith1999,Celani2010turbulence,Alexakis2011two,deusebio2014dimensional,Sozza2015dimensional,
pouquet2013geophysical,marino2013inverse,seshasayanan2014edge,Marino2015resolving,
seshasayanan2016critical,benavides2017critical}. 
In particular for rotating flows it has been shown that the transition from a forward to an inverse cascade happens 
at critical rotation $\Omega_c$ above which the flow starts to cascade part of the injected energy $\epsilon$ inversely at a rate $\epsilon_{inv}$.
The fraction of the rate that cascades inversely $\epsilon_{inv}/\epsilon $  depends on the difference $\Omega-\Omega_c$ and the height of the domain $H$ \citep{deusebio2014dimensional} . 
This description holds at early times before the inverse cascading energy reaches scales the size of the domain.
At late times when energy starts to pile up and form a condensate the dynamics might change \citep{kraichnan1967inertial,smithr1994finite,xia2008turbulence}.

In this work, we try to determine the behaviour of a forced rotating flow at late times when the flow has
reached a steady state, in the absence of any large scale dissipative mechanism. 
Due to the long computational time required to reach a steady state, very few investigations have focused on this regime
like the early low resolution studies in \cite{Bartello1994} and more recently
the studies in \citep{alexakis2015rotating,dallas2016forcing,2017arXiv170108497Y}, 
where turbulent rotating flows at steady state were investigated. 
Experiments on the other hand for which long times are realizable have investigated this steady state limit 
\cite{Moisy2014direct,yarom2014experimental,campagne2016turbulent,Machicoane2016Two}. 

The rest of this paper is structured as follows.
In the next section \ref{Nsetup} we present our numerical setup and introduce our control parameters and observables.
In section \ref{theory} we discuss possible mechanisms for the saturation of the initial energy growth.
In section \ref{data} we present the results on global quantities from the numerical simulations and 
in section \ref{Dyn} we describe the spatial and spectral structures as well the dynamics involved.
In the final section \ref{cons} we summarize and draw our conclusions.

\section{Numerical set-up, and control parameters } 
\label{Nsetup} 

We consider the flow of a unit density liquid in a cubic triple periodic domain of size $2\pi L$ 
that is in a rotating frame with $z$ being the axis of rotation. The governing equation for the flow velocity $\bf u$ 
is given by eq. \eqref{NSR}.
The flow is driven by the body force $\bf F$, here we consider two cases given by,
\beq 
a) \quad
{\bf F} = f_0 \left[ 
\begin{array}{c}
-  \sin \left( k_f y \right), \\ 
+  \sin \left( k_f x \right), \\
\cos \left( k_f x \right) + \cos \left( k_f y \right) 
\end{array}
\right]
\quad \mathrm{and} \quad
b) \quad
{\bf F} = f_0 \left[ 
\begin{array}{c}
-  \sin \left( k_f y \right), \\ 
+  \sin \left( k_f x \right), \\
\sin \left( k_f x \right) + \sin \left( k_f y \right) 
\end{array}
\right].
\eeq
The first one is maximally helical $\langle \bf F \cdot \nabla \times F\rangle_S = k_f \langle F \cdot F \rangle_S $
and will be referred to as the helical forcing and the second one has zero helicity $\langle \bf F \cdot \nabla \times F\rangle_S =0$
and will be referred as the non-helical forcing. Here $\left\langle \right\rangle_S$ denotes spatial average. These forcing functions have been proposed by \citep{roberts1972dynamo} for dynamo studies and commonly
are referred to as Roberts flow. Helicity is known to play an important role in fast rotating turbulence since it has been shown that its forward cascade 
can control the dynamics at the small scales \cite{Mininni2010,Mininni2012}. In this work we will examine both 
cases with and without helicity in parallel.
It is also important to note that our forcing is invariant along the axis of rotation
and thus the forcing acts only on the slow manifold (that consists of all the Fourier velocity modes for which $k_z=0$). 
This in contrast with the case examined in \citep{alexakis2015rotating,2017arXiv170108497Y} where a Taylor-Green forcing was used that has zero 
average along the vertical direction. Thus, while the Taylor-Green forcing does not inject energy directly to the slow manifold, 
the Roberts forcing used here injects energy only to the slow manifold. 
The two cases can thus be considered as two extremes. 

This system was investigated using numerical simulations.
All runs were performed using the pseudo-spectral code {\sc Ghost} \citep{mininni2011hybrid}, where each component of ${\bf u}$ is represented as truncated Galerkin expansion in terms of the Fourier basis.
The non-linear terms are initially computed in physical space and then transformed to spectral space using fast-Fourier transforms. Aliasing errors are removed using the
2/3 de-aliasing rule. The temporal integration was performed using a fourth-order Runge-Kutta method. Further details on the code can be found in \cite{mininni2011hybrid}. 
The grid size varied depending on the value of $Re_F$ and $Ro_F$ from $64^3$ to $512^3$. A run was considered well resolved if the value of enstrophy spectrum at the
cut-off wavenumber was sufficiently smaller than its value at its peak. Each run started from a random multi-mode initial condition and was continued
for sufficiently long time so that long time averages in the steady state were obtained. 

The parameter $f_0$ gives the amplitude of the forcing, and $k_f$ is the wavenumber at which energy is injected into the flow. 
These two parameters define the length-scale $\ell_f=k_f^{-1}$, the time scale $\tau_f= {(k_ff_0)}^{-1/2}$  and velocity amplitude $U_f=\sqrt{f_0/k_f}$
which will be used to non-dimensionalize the control parameters in our system. The product $k_fL$ gives the scale separation between the forcing scale
and the box size. Throughout this work we have fixed the scale separation to $k_f \, L = 4$. We thus do not investigate the dependence on the box size.
%
The Reynolds number $\Ref$ and the Rossby number $\Rof$ based on $U_f$ are defined as, 
\beq
\Ref =\frac{\sqrt{f_0/k_f^3}}{\nu} \quad \mathrm{and } \quad \Rof = \frac{ \sqrt{f_0 k_f}}{2 \Omega }\,.
\eeq
The more classical definition of the Reynolds and Rossby number can be obtained using the root mean square amplitude of the velocity $U = \left\langle {\bf u} \cdot {\bf u} \right\rangle_{ST}^{1/2}$,
where $\left\langle \cdot \right\rangle_{ST}$ denotes spatial and temporal average.
This leads to the velocity based  Reynolds number $\Reu$ and the velocity based  Rossby number $\Rou$,
\beq
\Reu =\frac{U}{k_f \nu} \quad \mathrm{and } \quad \Rou = \frac{U k_f}{2 \Omega }\,.
\eeq
In many experiments as well as in many theoretical arguments it is the energy injection rate
\beq \epsilon=\left\langle {\bf u \cdot F} \right\rangle_{ST} = \nu \left\langle |{\bf \nabla \times u}|^2  \right\rangle_{ST} \eeq
per unit of volume that is controlled.
It is thus worth considering expressing the control parameters also in terms of $\epsilon$. This leads to the definition of the Reynolds number based on $\epsilon$,
\beq
\Red =\frac{\epsilon^{1/3}}{\nu k_f^{4/3}} \quad \mathrm{and } \quad \Rod = \frac{  \epsilon^{1/3} k_f^{2/3}}{2 \Omega }\,.
\eeq
Finally, the ratio of the square root of enstrophy to twice the rotation 
rate  is referred to as the micro-Rossby number $\Rol$ that in terms of $\Red$ and $\Rod$ can be expressed as,
\beq
\Rol = \frac{\langle|{\bf \nabla \times u}|^2\rangle_{ST} ^{1/2} }{2 \Omega} = \frac{\epsilon^{1/2}}{2\nu^{1/2} \Omega } = \Red^{1/2}\Rod.
\eeq
In the examined system only  $(\Ref,\Rof)$ are true control parameters, while 
$(\Reu,\Rou)$ and $(\Red,\Rod)$ can only be measured a posteriori. 

\begin{figure}
\begin{subfigure}[b]{0.5\textwidth}
\includegraphics[scale=0.335]{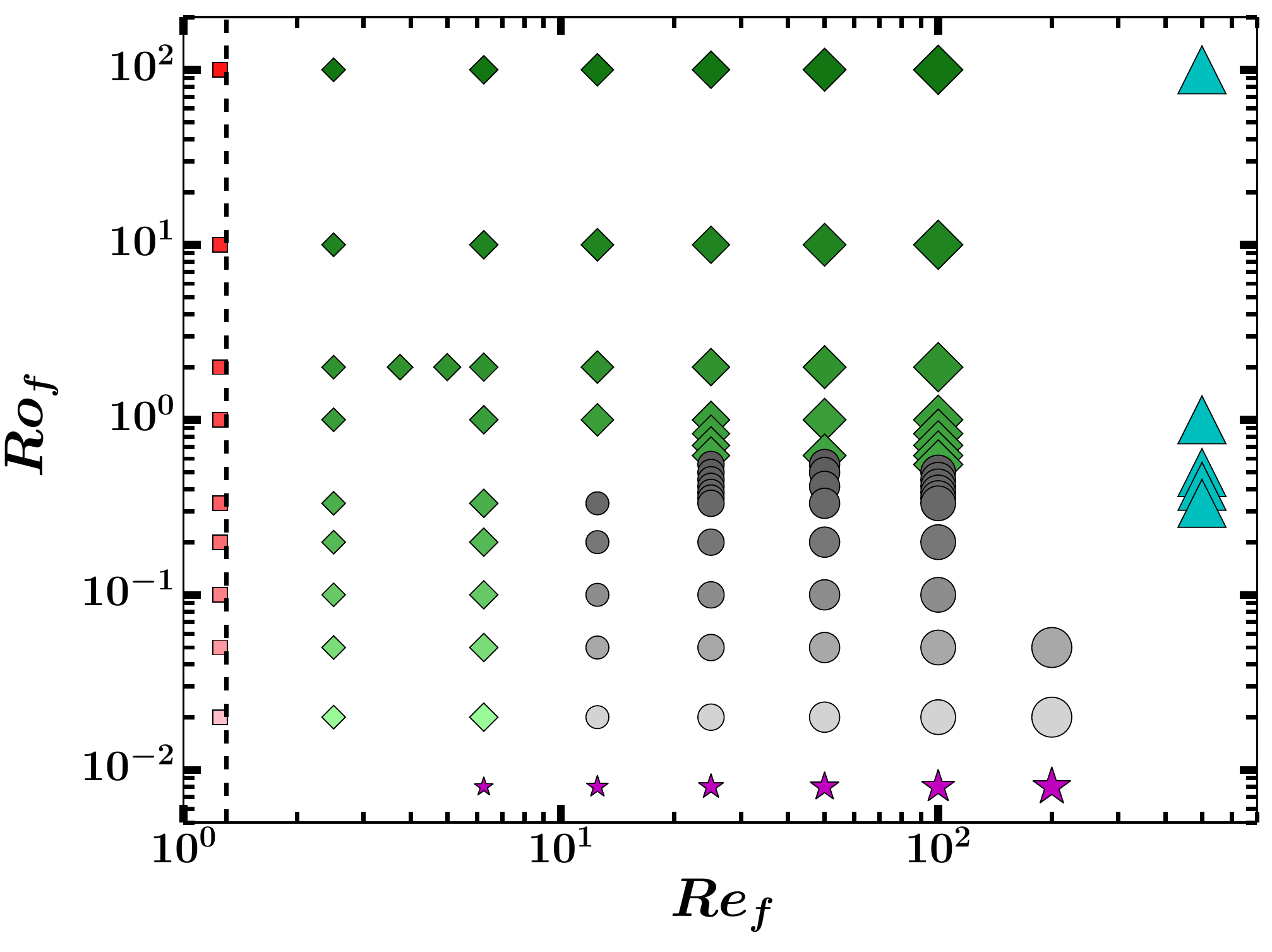}
\caption{}
\end{subfigure} 
\begin{subfigure}[b]{0.5\textwidth}
\includegraphics[scale=0.335]{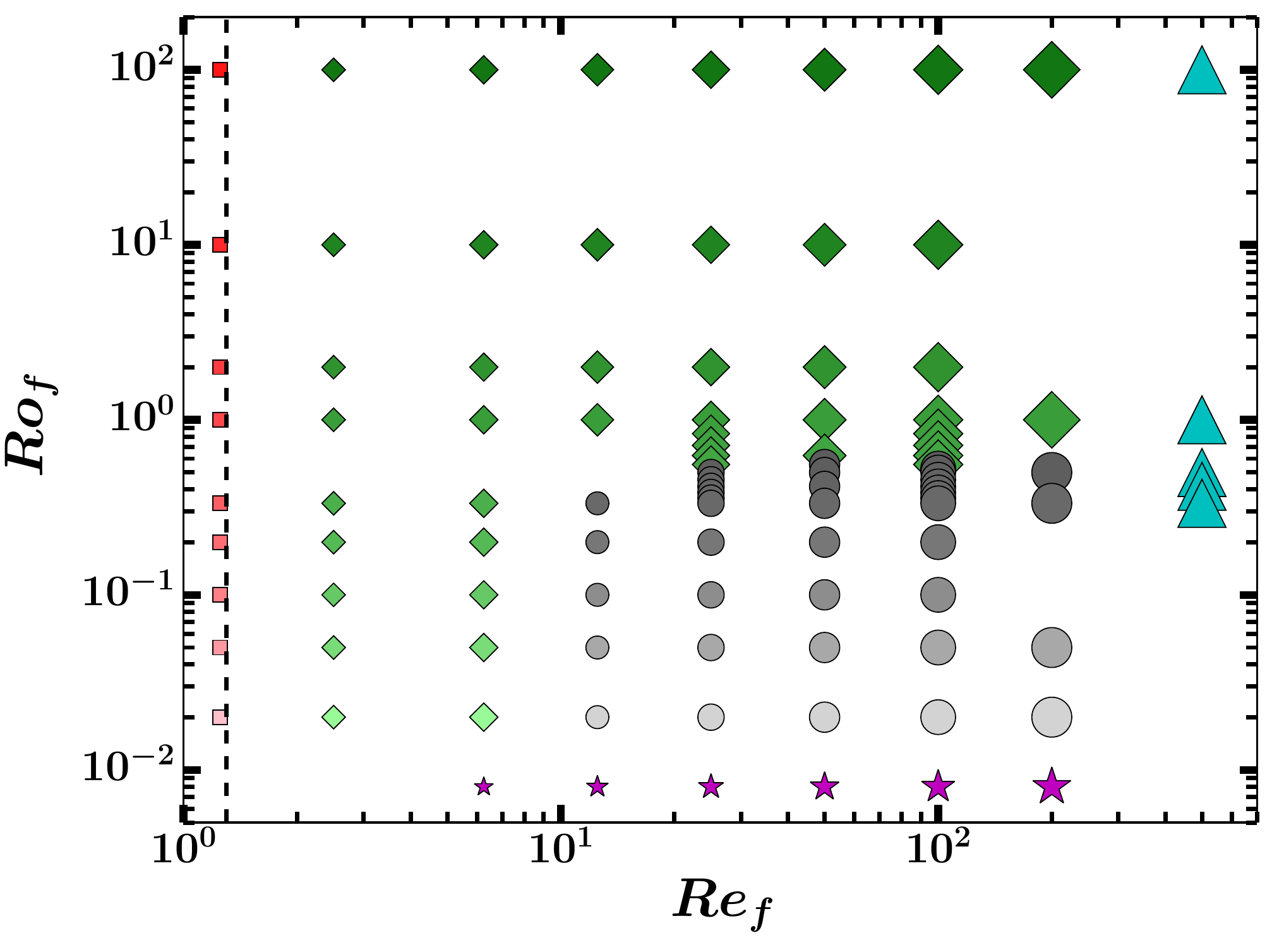}
\caption{}
\end{subfigure}
\caption{The figures show the location of all the examined numerical runs
in the $(\Rof,\Ref)$ plane for a) the case of helical flow and b) the case of the non-helical flow. Larger symbols denote larger values of $Re_f$ and lighter symbols correspond to smaller values of $Ro_f$. Different symbols correspond to different behaviour of the flow.  }
\label{map1}
\end{figure}
\begin{figure}
\begin{subfigure}[b]{0.5\textwidth}
\includegraphics[scale=0.335]{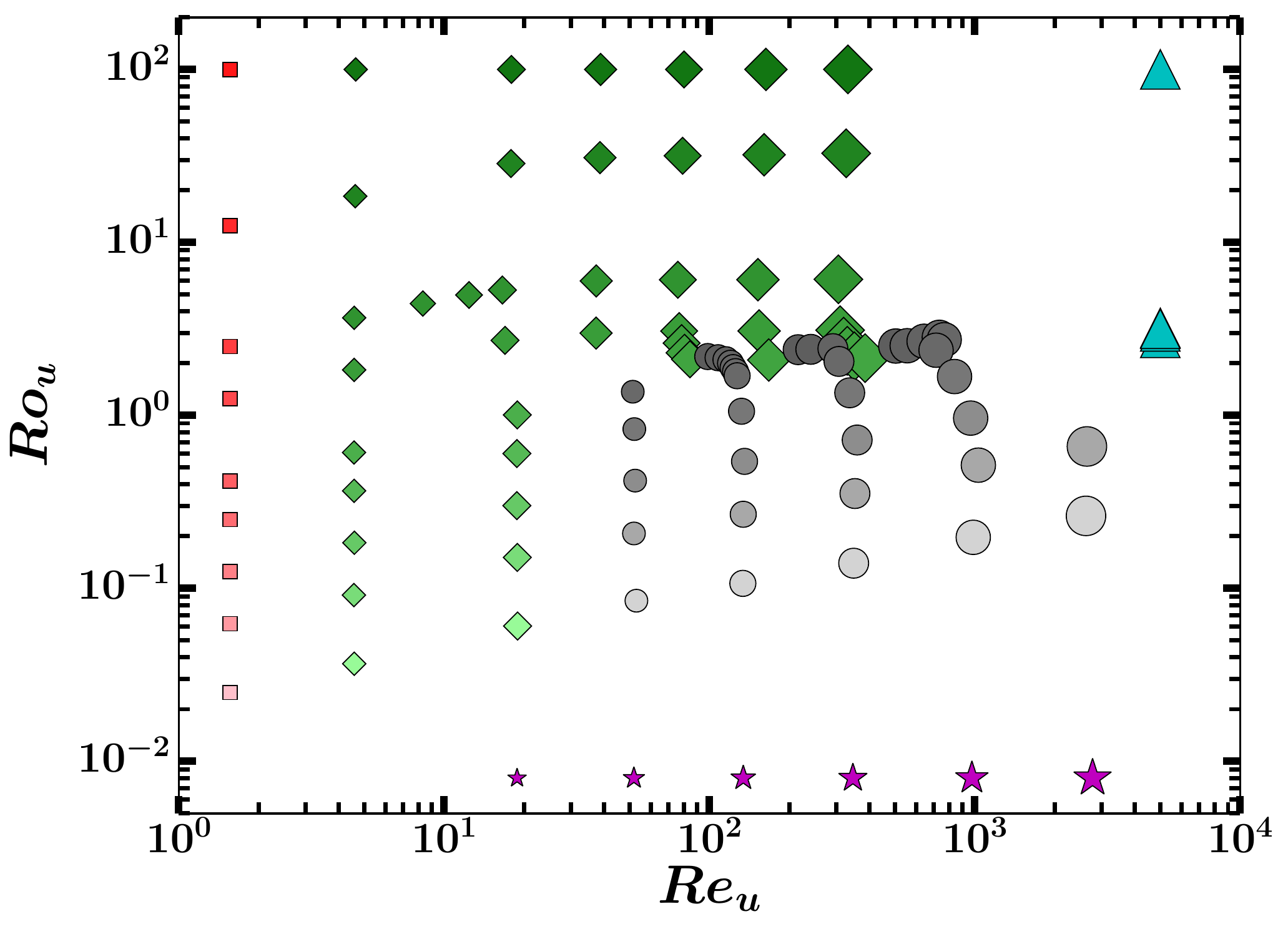}
\caption{}
\label{map2:1}
\end{subfigure} 
\begin{subfigure}[b]{0.5\textwidth}
\includegraphics[scale=0.335]{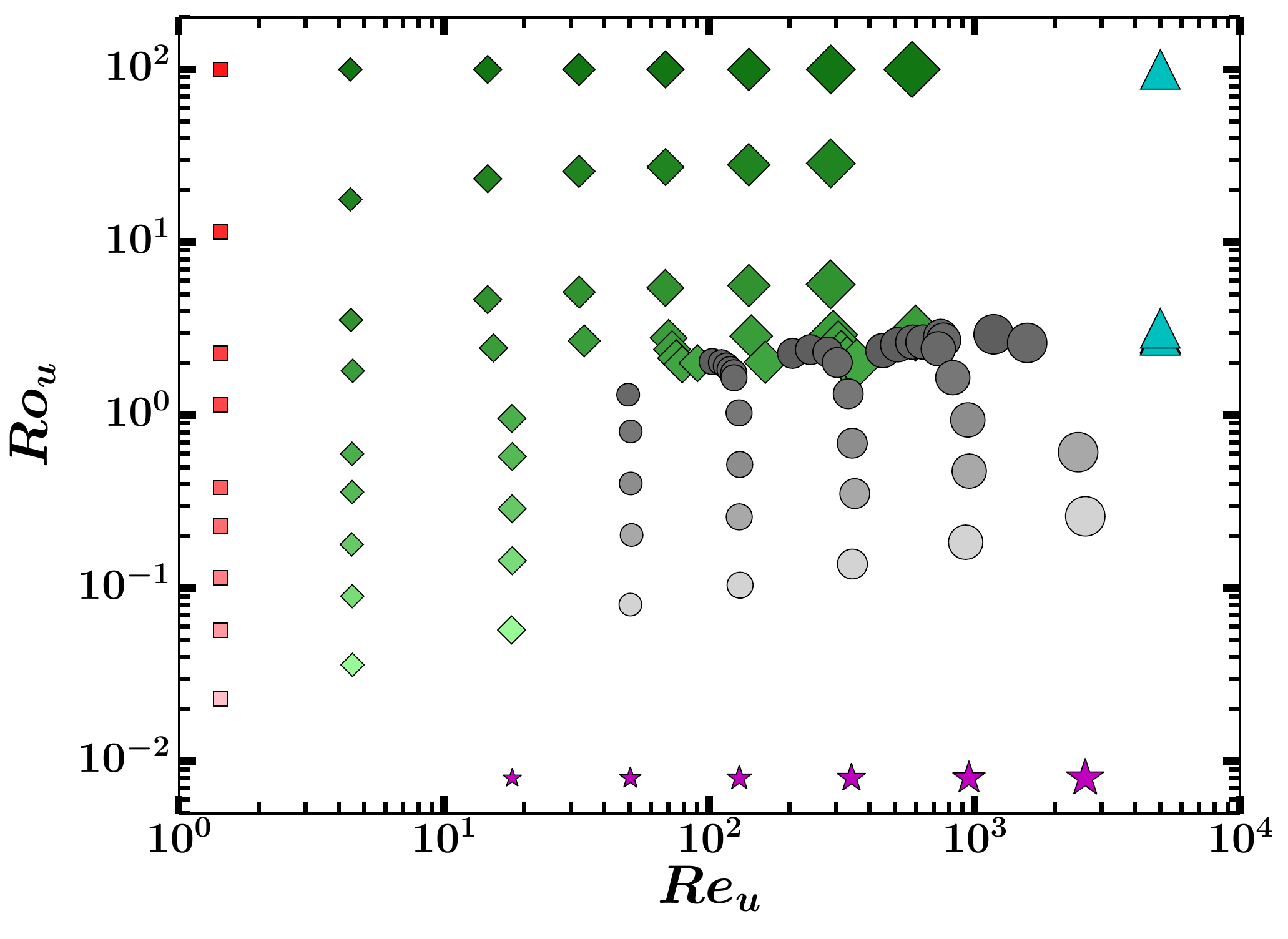}
\caption{}
\label{map2:2}
\end{subfigure}
\begin{subfigure}[b]{0.5\textwidth}
\includegraphics[scale=0.335]{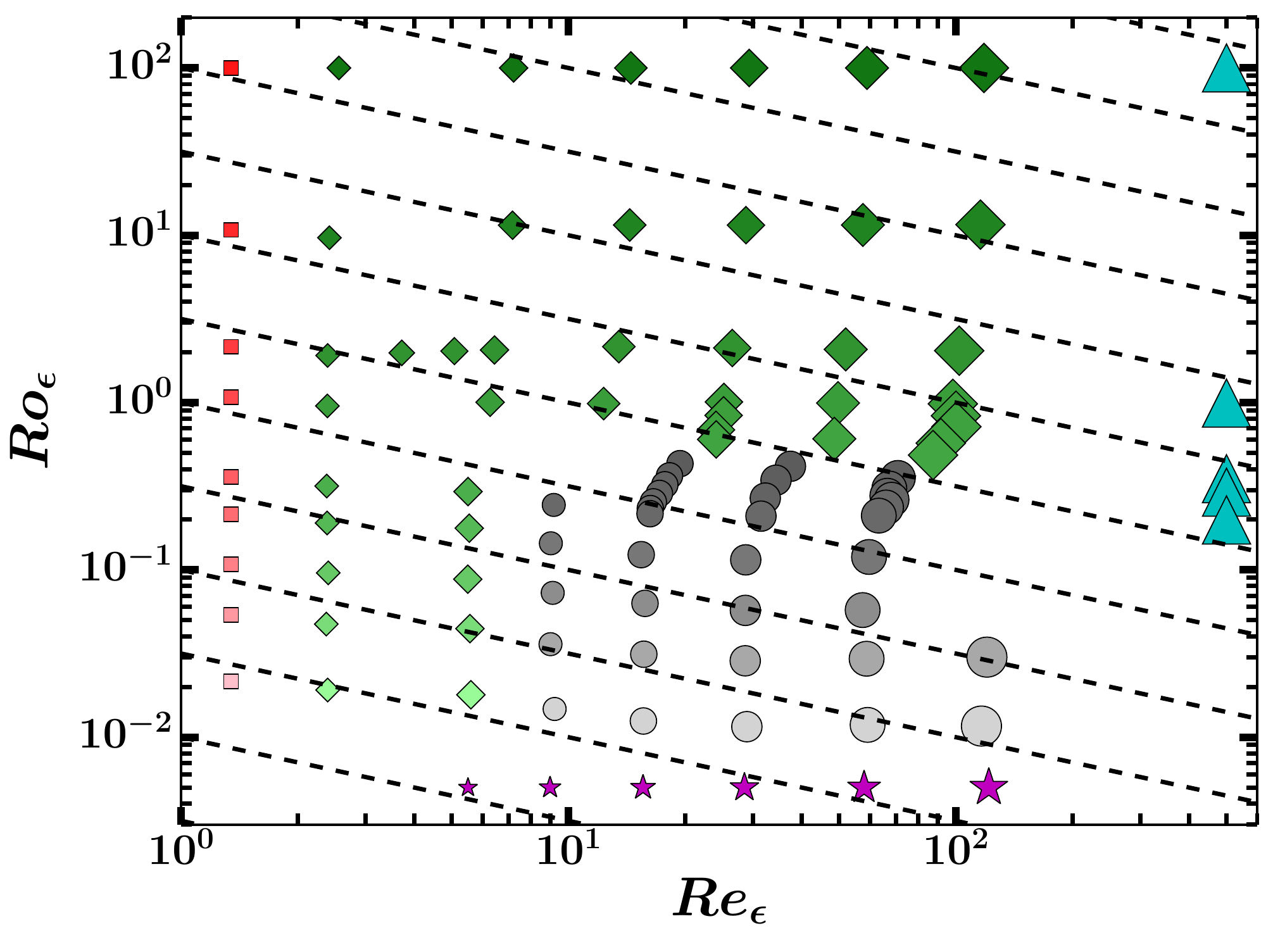}
\caption{}
\label{map2:3}
\end{subfigure} 
\begin{subfigure}[b]{0.5\textwidth}
\includegraphics[scale=0.335]{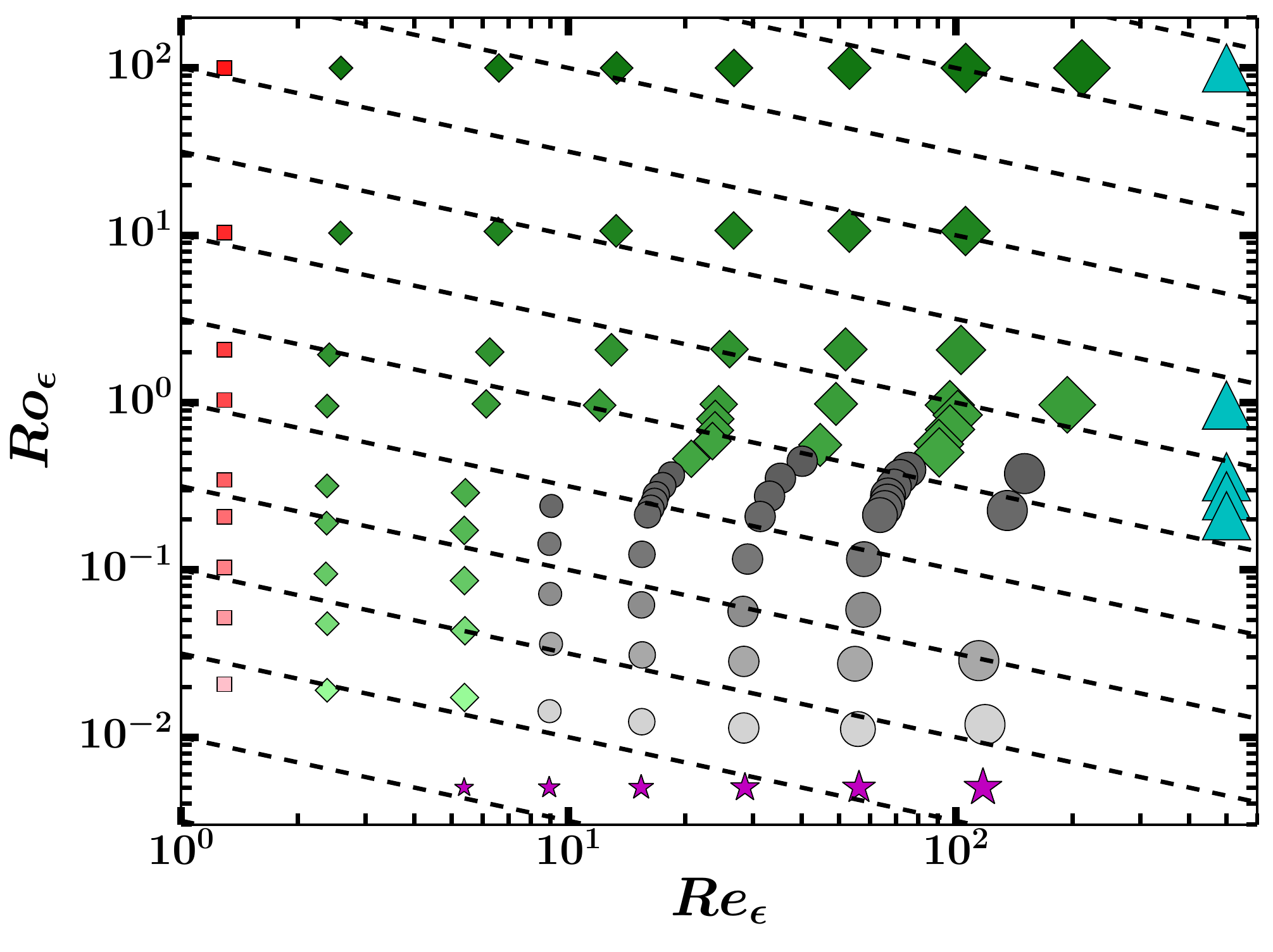}
\caption{}
\label{map2:4}
\end{subfigure}
\caption{The figures show the location of the numerical runs in the $(\Rou,\Reu)$ parameter space (panels a,b) and in the $(\Rod,\Red)$ plane (panel c,d) for the helical flows (panels a,c) and the non-helical flow (panels b,d). Larger symbols denote larger values of $Re_f$ and lighter symbols correspond to smaller values of $Ro_f$. Different symbols correspond to different behaviour of the flow.  }
\label{map2}
\end{figure}
The location of all of the performed runs in the ($\Ref, \Rof$) parameter space are shown in figure \ref{map1} 
for a) the helical flow, b) the non-helical flow in a log-log scale. 
The figure shows symbols that correspond to simulations that lead to different hydrodynamic steady states. 
Darker symbols correspond to larger values of $Ro_f$ while larger symbols correspond to larger values of $Re_f$. 
The largest symbols correspond to simulation runs of size $512^3$ points. 
The same symbols, sizes and shades (colours online) are used in some of the subsequent figures and thus the reader can refer to figure \ref{map1} to
estimate the value of $\Ref$ and $\Rof$. 
Each symbol corresponds to different behaviour of the flow: 
squares  \textcolor{red}{$\blacksquare$}    correspond to flows that are laminar, 
diamonds \textcolor{green}{$\blacklozenge$} correspond to unstable or turbulent flows that do not form a condensate,
circles  \textcolor{black}{$\bullet$}       correspond to turbulent flows that form a condensate.
%
We have shifted the points corresponding to $\Omega = 0, Ro_f = \infty$ to the values $Ro_f = 100$ in order for them to appear along with other points that correspond to finite rotation.

The star symbols \textcolor{purple}{$\star$} denote the simulations of the reduced two dimensional equations valid for $Ro_f \rightarrow 0$ given by,
\begin{align}
\partial_t {\bf u}_{_{2D}} + {\bf u}_{_{2D}} \cdot {\bm \nabla} {\bf u}_{_{2D}} &= -{\bm \nabla} p_{_{2D}} + \nu \Delta {\bf u}_{_{2D}} + {\bf f}_{_{2D}}, \label{R2D}  \\
\partial_t {\bf u}_z       + {\bf u}_{_{2D}} \cdot {\bm \nabla} {    u}_z       &=                         + \nu \Delta {    u}_z       + {\bf f}_z, \nonumber
\end{align}
where ${\bf u}_{_{2D}}$ stands for the horizontal components of the velocity field and $u_z$ for the vertical. 
All fields are independent of the vertical coordinate $z$.
The points for the 2D simulations ($Ro_f = 0$ limit) are placed at the position $Ro_f = 10^{-2}$. 
Finally the triangles \textcolor{cyan}{$\blacktriangle$}  denote hyper-viscous runs 
obtained when we replace the Laplacian in the equation \ref{map1} with the $\Delta^4$. 
Hyper-viscous runs model the limit $\Ref \rightarrow \infty$ and are place in figure \ref{map1} at the value  $Re_f = 1000$.

The vertical dashed line stands for the linear stability boundary of the laminar flow. For the chosen forcing the first unstable mode is $z$-independent 
and follows the linearized version of eq. \eqref{R2D}. Accordingly the unstable mode is independent of rotation and the vertical component of the laminar flow. 
As a result the laminar stability boundary is independent of $\Rof$ and is the same for the helical and the non-helical flow, that share the same
laminar ${\bf u}_{_{2D}}$ at $\Ref \simeq 1.278 $.

Figure \ref{map2} shows the same points in the parameter plane $(\Reu,\Rou)$ (top panels) and  $(\Red,\Rod)$ (bottom panels).
The dashed lines in figures \ref{map2:3},\ref{map2:4} indicate values of constant $\Rol$. 
For the range of examined parameters,
compared to the points in the $(\Ref,\Rof)$ plane there is a clear shift of the points to larger values of $\Reu$ as $\Rou$ is decreased in figures \ref{map2:1} and \ref{map2:3} while there is a 
decrease of $\Red$ as $\Rod$ is decreased in figures \ref{map2:2} and \ref{map2:4}.

Our principle goal in this work is using this large number of numerical simulations to determine 
the dependence of the large scale quantities of rotating turbulence like the saturation amplitude $U$ 
and the energy dissipation rate and map the different behaviours observed in the parameter space
making a phase space diagram.

\section{Inverse transfers and saturation of condensates.} 
\label{theory} 

In this section we present some theoretical estimates for the saturation amplitude of the velocity $U$ and the energy dissipation rate $\epsilon$.
As a first step we consider a fixed energy injection rate $\epsilon$ and use ($\Red,\Rod$) as control parameters. 
We relax this assumption later in the text where we extend these considerations to the case of fixed forcing amplitude.
%

For weak rotating and non-rotating systems ($Ro \to \infty$) the cascade is strictly forward.
The external forcing is balanced either by the viscous forces when $Re$ is small, or by the nonlinearities
that transfer the injected energy to the small scales where viscosity is again effective. 
These considerations lead to the classical scaling for laminar and turbulent flows
between the velocity $U$ and the energy injection rate $\epsilon$, 
\beq
U^2 \propto  \epsilon \frac{ \ell_f^2  }{\nu} \quad \mathrm{for} \quad \Red \to 0 \quad \mathrm{and}  \quad 
U^2 \propto        (\epsilon \ell_f) ^{2/3}   \quad \mathrm{for} \quad \Red \to \infty.
\label{UELT}
\eeq
Note that both of these scalings are independent of the domain size $L$ and the rotation rate $\Omega$. 
Using these scalings one can show that for $\Red \to \infty$ all the definitions of $Re$ given in the previous section are equivalent up to a pre-factor so that
$\Red\sim\Reu\sim\Ref$ and $\Rod\sim\Rou\sim\Rof$.

\begin{figure}
   \centering
   \centerline{
   \includegraphics[width=0.4\textwidth]{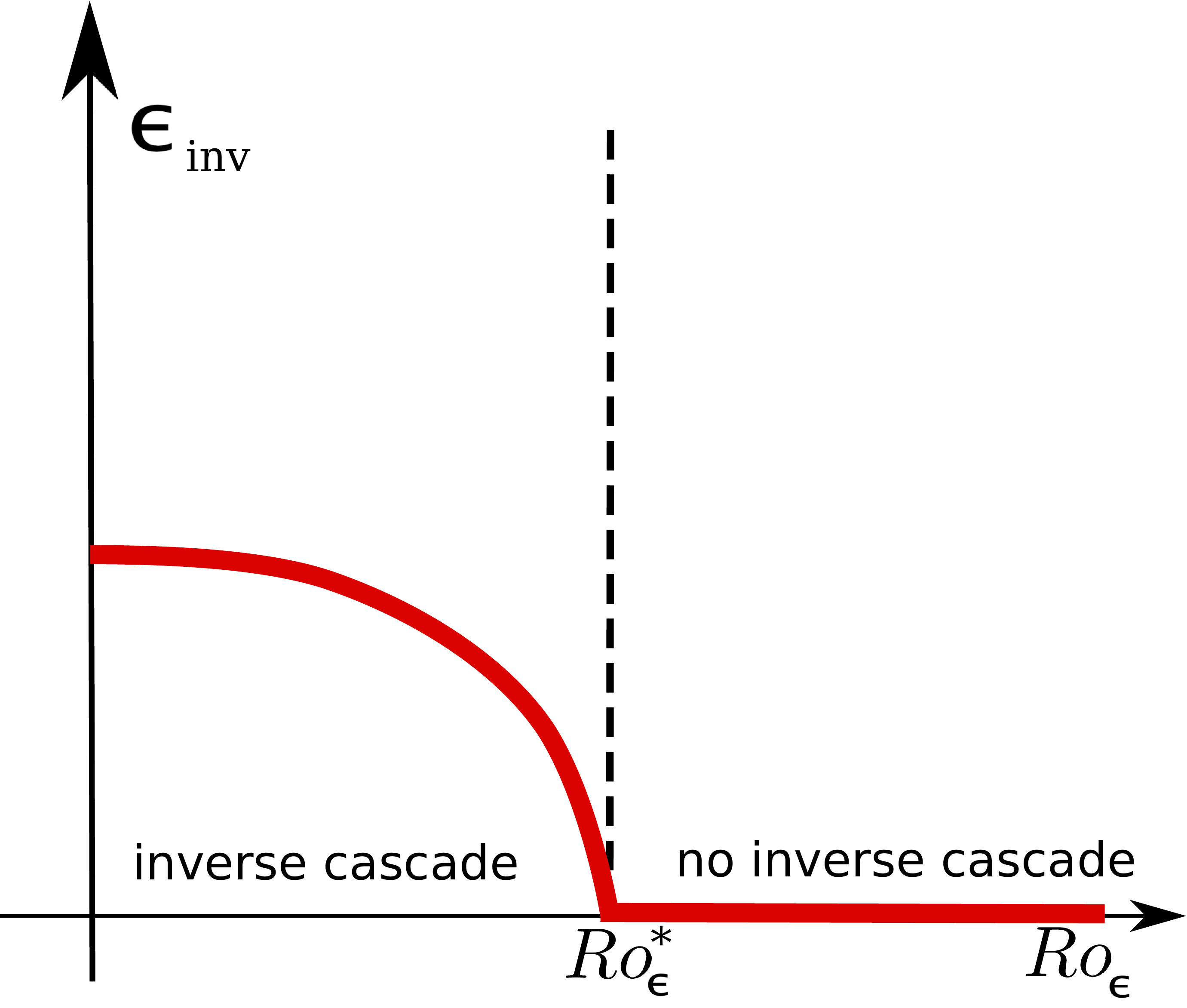} \,\,
   \includegraphics[width=0.4\textwidth]{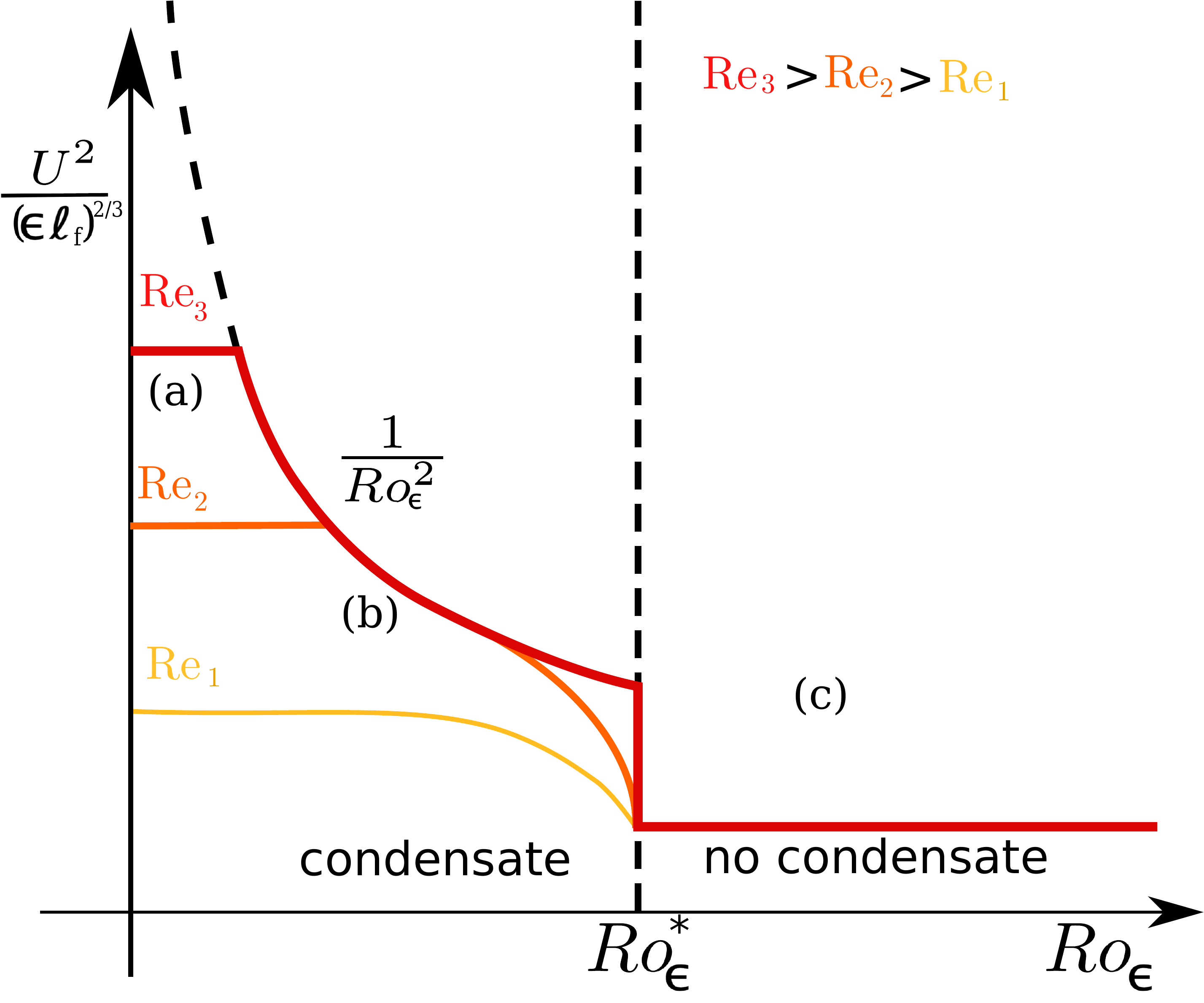} }
   \caption{
   Left panel:
   The figure shows an illustration of the expected dependence of the amplitude of the inverse flux $\einv$ 
   at the early stages of the inverse cascade as a function of the Rossby number $\Rod$.
   Right panel:
   Figure shows an illustration of the total energy $U^2$ as a function of $\Rod$ at the steady state regime 
   for different values of $\Red$.
   The dashed vertical line indicates the transition from a flow with no inverse cascade to a flow with an inverse cascade. 
   The dashed curved line shows the scaling $1/\Rod^2$ that reflects the $U^2\propto \Omega^2L^2$ scaling. }
  \label{fig:sketch}
\end{figure}

In the presence of an inverse cascade however the involved mechanisms for saturation become considerably different, altering these scaling relations.
At late times, in order for the system to reach a steady state and saturate the initial increase of the large scale energy,
it has to either suppress the rate that energy cascades inversely $\einv$ or 
to reach sufficiently high amplitudes so that the energy can be dissipated by viscosity.
If indeed the transition from forward to an inverse cascade has a critical behaviour,
the amplitude of the inverse cascade will depend as a power law on the deviation from criticality $\Rod^*$,
\beq
 \einv = C_1 \left(\frac{\Rod^*-\Rod}{\Rod^*}\right)^\gamma \epsilon \qquad \mathrm{for} \quad 0 < \Rod^*-\Rod \ll \Rod^*,
 \label{crit1}
\eeq
while away from criticality it is expected that,
\beq
 \einv = C_2 \, \epsilon \qquad \mathrm{for} \quad 0 < \Rod \ll \Rod^*.
 \label{away1}
\eeq
Here $\Rod^*$ denotes the critical value of Rossby for which the inverse cascade starts.
Note that $\Rod^*$ is found to depend on the height of the box but not on the horizontal dimensions.
A sketch of the dependence of $\einv$ on $\Rod$ is shown at the left panel of figure \ref{fig:sketch}.

The pre-factors $C_1$ and $C_2\le 1$ and the exponent $\gamma$ have not yet been determined neither by DNS nor by experiments.
In fact, even the conjecture of criticality is very hard to verify with DNS. Although it seems to be plausible,
it has  been demonstrated with some accuracy only for two dimensional models
(see \cite{benavides2017critical,seshasayanan2014edge,seshasayanan2016critical}). The reason is that close to the 
transition point, finite size and finite Reynolds effects become important that tend to smooth out the transition.
To demonstrate this criticality, ever increasing box sizes and Reynolds numbers need to be considered and
this is extremely costly for three dimensional simulations.
We thus do expect that the transition might not appear as sharp as eq. \eqref{crit1} might suggest and we will be
rather dominated by finite size effects that will smooth the transition.

For $\Rod < \Rod^*$  the energy that arrives at the domain size piles up 
forming condensates.  In rotating turbulence such condensates can saturate by two possible mechanisms.
First, just like in the case of 2D turbulence, saturation comes from viscous forces:
the amplitude of the large scale condensate $U_{_{2D}}$ becomes so big that viscous dissipation at 
large scale balances the rate $\einv$ that energy arrives at the large scales by the inverse cascade. 
Thus the balance 
$
\epsilon_{inv} \propto\nu \frac{U_{_{2D}}^2}{L^2} 
$
is reached. 
The scaling for the amplitude of the condensate close to the transition point $\Rod^*$ thus follows,
\begin{align}
U^2 \propto C_1 \frac{ \epsilon L^2}{\nu} \left(\frac{\Rod^*-\Rod}{\Rod^*} \right)^\gamma,  \qquad \mathrm{for} \quad 0 < \Rod^*-\Rod \ll \Rod^*.
\label{UV}     
\end{align}
This argument indicates that if the injection rate $\epsilon$ is fixed, 
the amplitude of the condensate $U$ scales super-critically with $\Omega$ with an exponent $\gamma/2$. 
For strong rotations away from criticality $\Rod \ll \Rod^*$, 
we expect the scaling for the condensate of 2D turbulence,
\begin{align}
U^2 \propto C_2 \frac{ \epsilon L^2}{\nu},  \qquad \mathrm{for} \quad 0 < \Rod \ll \Rod^*. 
\label{UV2}
\end{align}
We will refer to the condensate in this case as a {\it viscous condensate} because it is the viscosity that saturates the growth of energy at the large scales.


A different way to saturate the inverse cascade for fast rotating flows is by breaking the conditions that make the flow quasi-2D.
This can happen in domains with  periodic boundary conditions where 
due to the conservation of vorticity flux the shape of the condensate takes the form of a dipole with one co-rotating vortex and a counter rotating vortex.
Saturation of the inverse transfer of energy can then happen when the counter rotating vortex 
cancels locally the rotation rate and energy cascades forward again (see \cite{Bartello1994,alexakis2015rotating} ). 
This balance is achieved when eddy turn over time of the condensate $L/U$ becomes comparable to the rotation $\Omega$.
This leads to the scaling,
\begin{align}
U^2 \propto \Omega^2 L^2.
\label{UO}
\end{align}  
This scaling was realized in simulations of rotating Taylor-Green flows, see \cite{alexakis2015rotating}.
Note that this scaling is independent of the amplitude of the inverse cascade, and thus independent from the deviation from criticality,
that suggest that the transition will be sub-critical. This was indeed found to be the case in \cite{alexakis2015rotating}. Further
more, recently \cite{2017arXiv170108497Y} were able to follow the hysteresis diagram of the subcritical bifurcation.
Finally we also note that in this regime a strong asymmetry between co-rotating and counter rotating vortexes is 
expected, (see for example  \cite{hopfinger1982turbulence,Bartello1994,morize2006energy,bourouiba2007intermediate,
sreenivasan2008formation,van2008refined,Staplehurst2008,moisy2011decay,Gallet2014Scale}). 
We will refer to the condensate in this case as a  {\it rotating condensate} 
because the energy at the large scales depends on the rotation rate.

From the two mechanisms the one that predicts a smaller value of $\UTD$ is going to be more effective. 
As the Rossby number $\Rod$ is varied slightly below the critical value, we expect that due to the small amplitude of the inverse cascade,
viscosity will be effective in saturating the inverse cascade and the saturation amplitude will be given by eq. \eqref{UV}. 
Away from criticality however, the breaking of the quasi-2D condition becomes more effective as
the amplitude predicted by \eqref{UO} will become smaller that \eqref{UV}, and the saturation amplitude will depend on rotation as in eq.
\eqref{UO}.
The region for which the first scaling \eqref{UV} holds becomes smaller as $\Red$ increases. 
Thus in the limit of large $\Red$ the transition will become discontinuous.  
Viscosity will become effective again at very small $\Rod$ where the saturation amplitude will be governed by equation \eqref{UV2}.
The value of $\Rod$ at which the behaviour transitions from the scaling \eqref{UO} to the scaling \eqref{UV2} can be obtained by equating the two predictions. 
This leads to 
\beq \Rod \propto \Red^{-1/2}  \label{Rotran} \eeq
which implies that the transition from a rotating condensate to a viscous condensate occurs when 
the micro-Rossby number is of order unity $\Rol=\mathcal{O}( 1)$.

The right panel of figure \ref{fig:sketch} shows a sketch of these expected transitions. The parameter space is thus split in three regions 
(a) one where a condensate forms that is balanced by viscosity for $\Rod \ll \Red^{-1/2} \ll \Rod^*$, (b) a second in which the condensate that forms equilibrates to a steady state 
by the counter rotating vortex cascading energy back to the small scales for $\Red^{-1/2} \ll \Rod < \Rod^* $,
and finally (c) where there is no inverse cascade and the system is close to isotropy for $\Rod > \Rod^*$.
We stress that based on these arguments the behaviour of the flow at large $\Red$ and low $\Rod$ depends on
precise order in which the limits $\Rod \to 0$ and $\Red \to \infty$ are taken.

We now relax the assumption of fixed energy injection rate and consider the case that 
the system is forced by a constant in time forcing of fixed amplitude as in our simulations.
For weak rotation the relation between the forcing amplitude and energy injection rate if the Reynolds number is small is given by:
\beq
\epsilon \propto   \frac{f_0^2\ell_f^2}{\nu} \quad \mathrm{for} \,\, \Ref \ll 1,   \quad         (\mathrm{laminar \,\,scaling}),
\label{LamF}
\eeq  
while for large Reynolds numbers we have a viscosity independent scaling:
\beq
\epsilon \propto   f_0^{3/2} \ell_f^{1/2} \quad \mathrm{for} \,\, \Ref \gg 1,                \quad                    (\mathrm{turbulent \,\,scaling}).
\label{TurbF}
\eeq 
For high rotation rates however the injection rate can depend on $\Omega$
if the forcing is not invariant along the axis of rotation. This was shown for the Taylor-Green forcing where the flow was shown to re-laminarize 
at high rotation rates \citep{alexakis2015rotating}. This effect will not take place in the present investigation for which the forcing is $z$ independent
and we thus expect that the scaling in eq. \eqref{TurbF} remains valid, that along with eq.\eqref{UV2}  leads to the prediction,
\begin{align}
U^2 \propto U_f^2 \Ref (k_fL)^2         \quad            \mathrm{for } \,\, \Rof \ll 1,
\label{UF}
\end{align}
for the amplitude of the condensate. We note that in the presence of of large scale separation this relation is altered to the weaker scaling 
$U^2 \propto U_f^2 \Ref^{2/3}(k_fL)^{4/3}$ due to the effect of sweeping (see \cite{shats2007suppression, xia2008turbulence, tsang2009forced, gallet2013two}).  
Such an effect however is not expected to be present in our case for which $k_f L=4$.
For moderate values of $\Rof$ such that
the saturation comes from the cancelling of the quasi-2D condition of the counter rotating vortex, $U^2$ is independent of the energy injection rate and thus from the forcing amplitude.
and it is thus given by eq. \eqref{UO}.
%
%
Thus, a qualitative difference between the constant injection of energy and constant forcing amplitude 
is only expected for viscous condensates and only alters the dependence of the saturation amplitude on $\Ref$
and not on $\Rof$.

\section{Simulation Results} 
\label{data} 

\begin{figure}
\begin{subfigure}[b]{0.5\textwidth}
\includegraphics[scale=0.335]{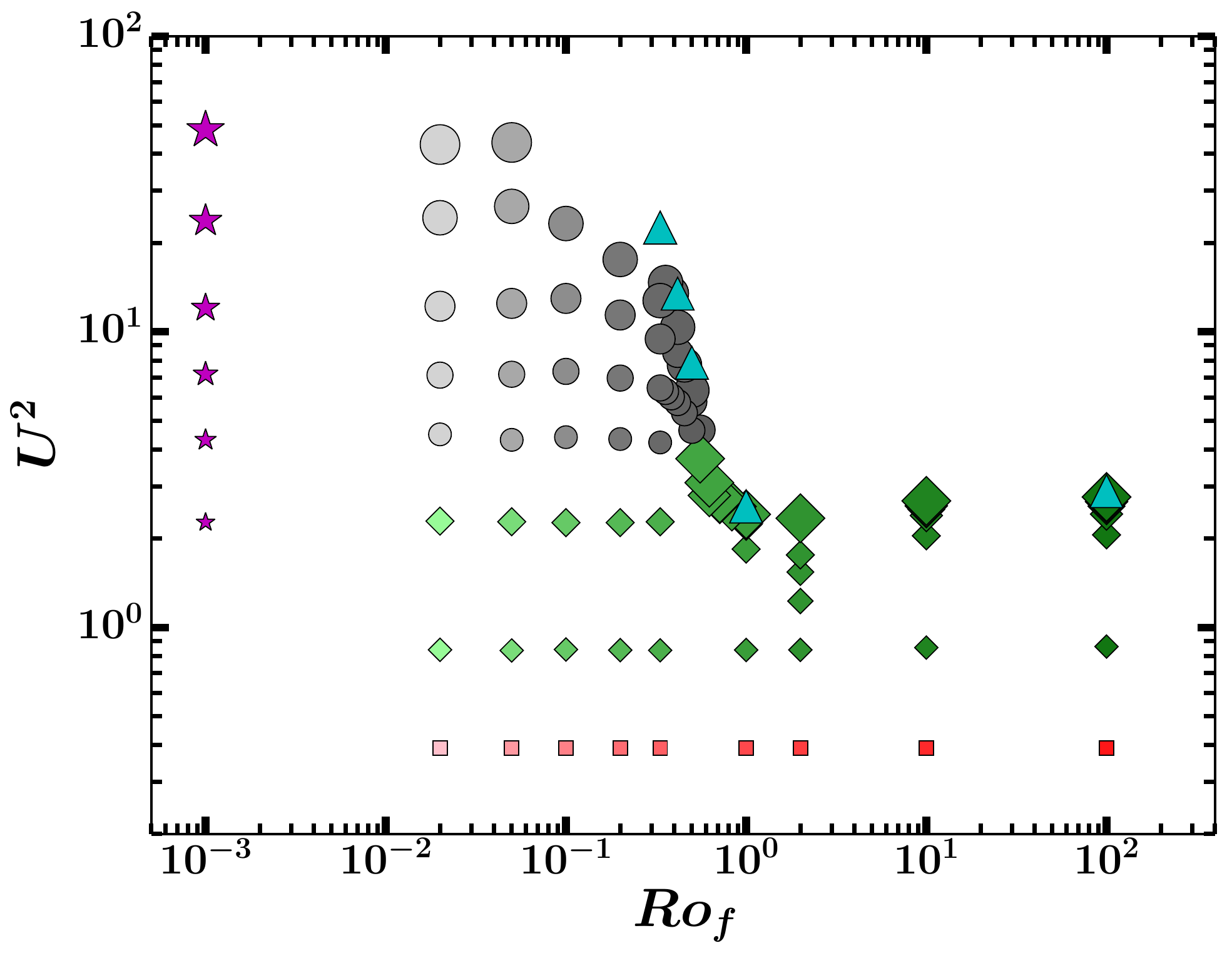}
\caption{}
\label{fig:U1a}
\end{subfigure} 
\begin{subfigure}[b]{0.5\textwidth}
\includegraphics[scale=0.335]{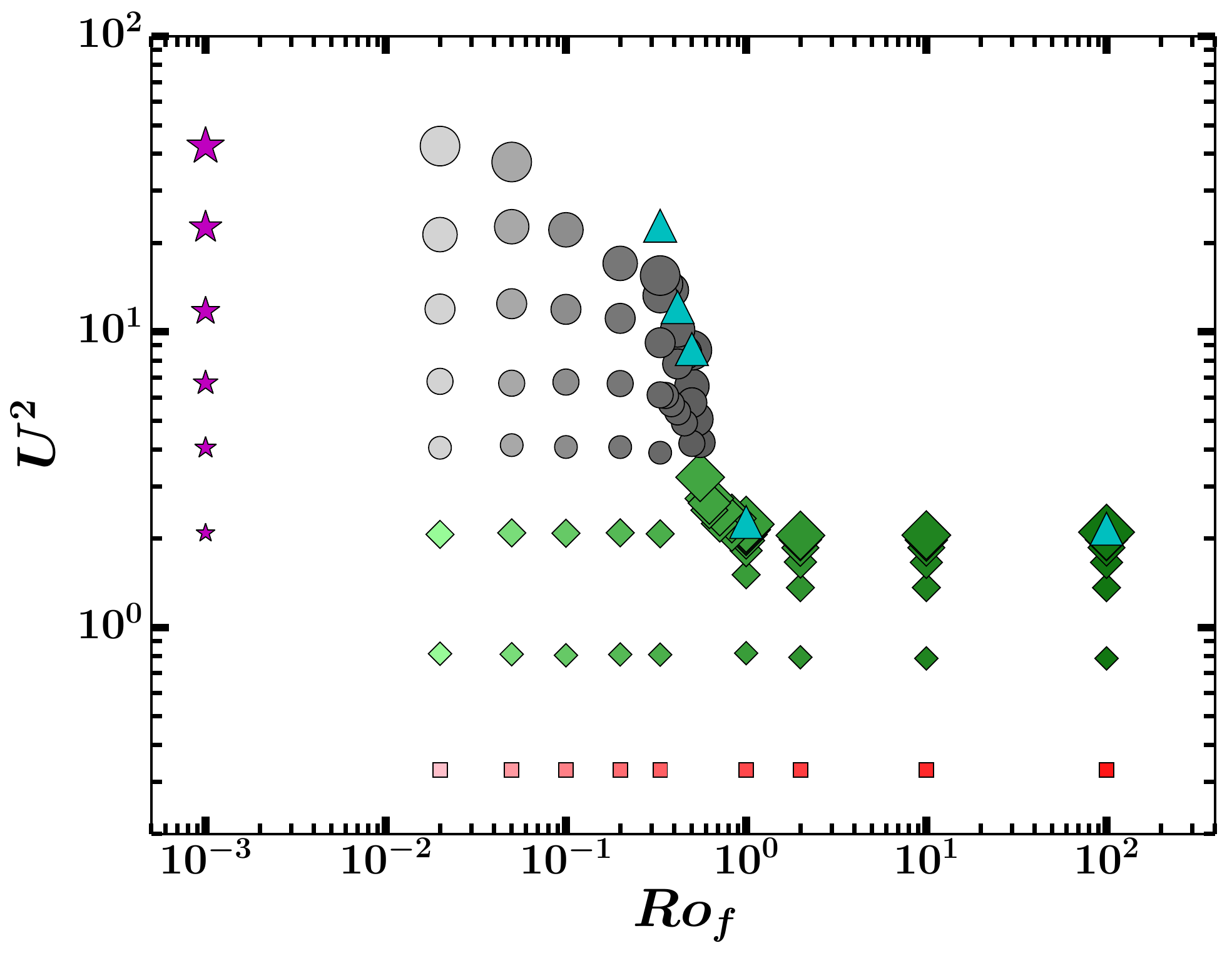}
\caption{}
\label{fig:U1b}
\end{subfigure}
\caption{The figures show the total energy $U^2$ (in units of $U_f^2$) as a function of $Ro_f$ for the examined numerical runs for a) helical flow and b) non-helical flow. Larger symbols denote larger values of $Re_f$ and lighter symbols correspond to smaller values of $Ro_f$. Different symbols correspond to different behaviour of the flow. }
\label{fig:U1}
\end{figure}

We begin by plotting in figure \ref{fig:U1} the square of the velocity saturation amplitude $U^2$
(in units of $U_f^2$) as a function of the Rossby number for the entirety of our data points for the helical (left panel) and the non-helical (right panel) runs.
For both cases the velocity amplitude increases rapidly as $\Rof$ decreases beyond a critical value $\Rof^*=\mathcal{O}(1)$. 
This increase appears to become stronger  for larger values of $\Ref$ (larger symbols). 
For larger values of $\Rof$ (weakly rotating runs), $U_f^2$ quickly saturates to a $\Rof$ and $\Ref$ independent value
provided $\Ref$ is sufficiently above the laminar instability threshold. 
%

\begin{figure}
\begin{subfigure}[b]{0.5\textwidth}
\includegraphics[scale=0.335]{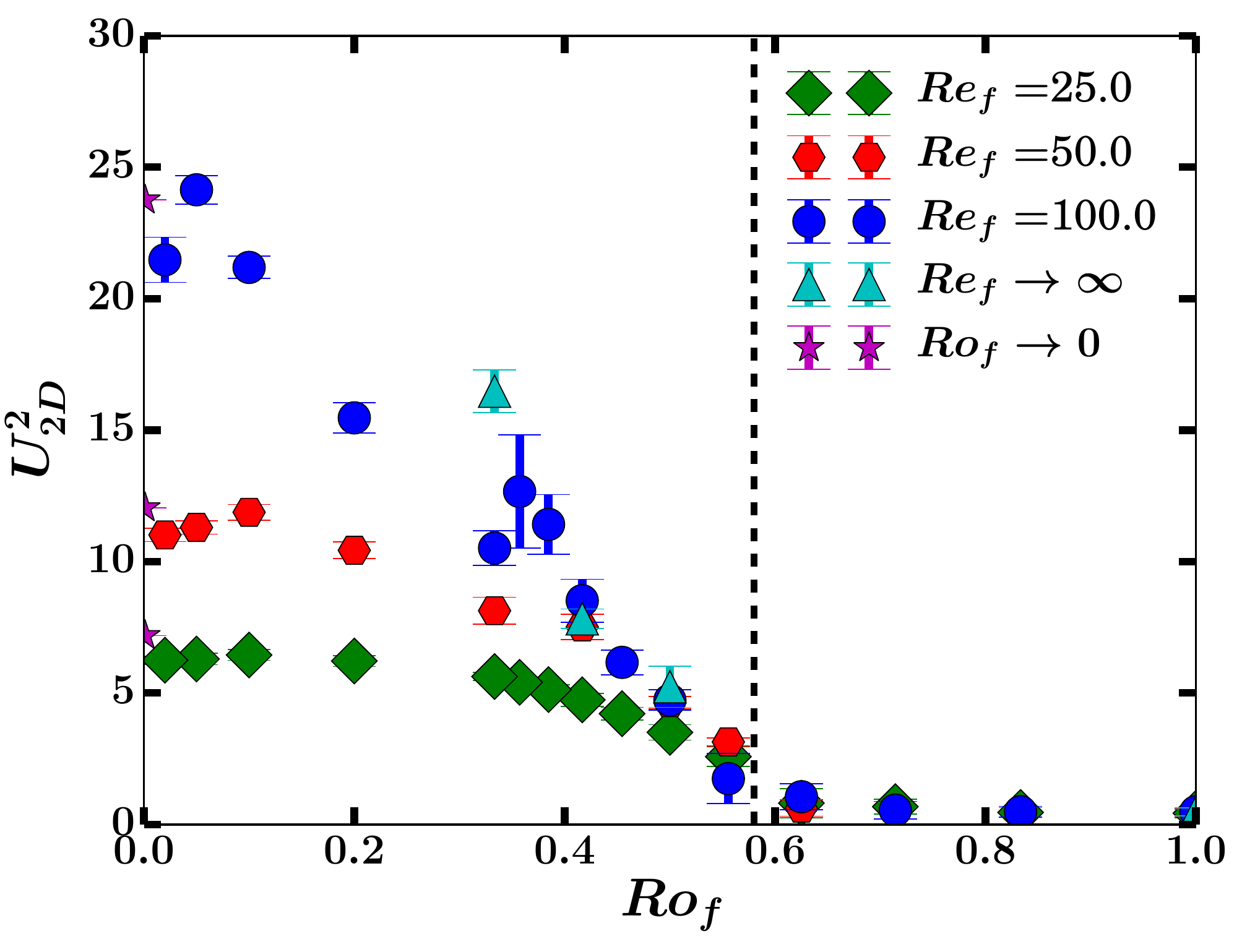}
\caption{}
\end{subfigure} 
\begin{subfigure}[b]{0.5\textwidth}
\includegraphics[scale=0.335]{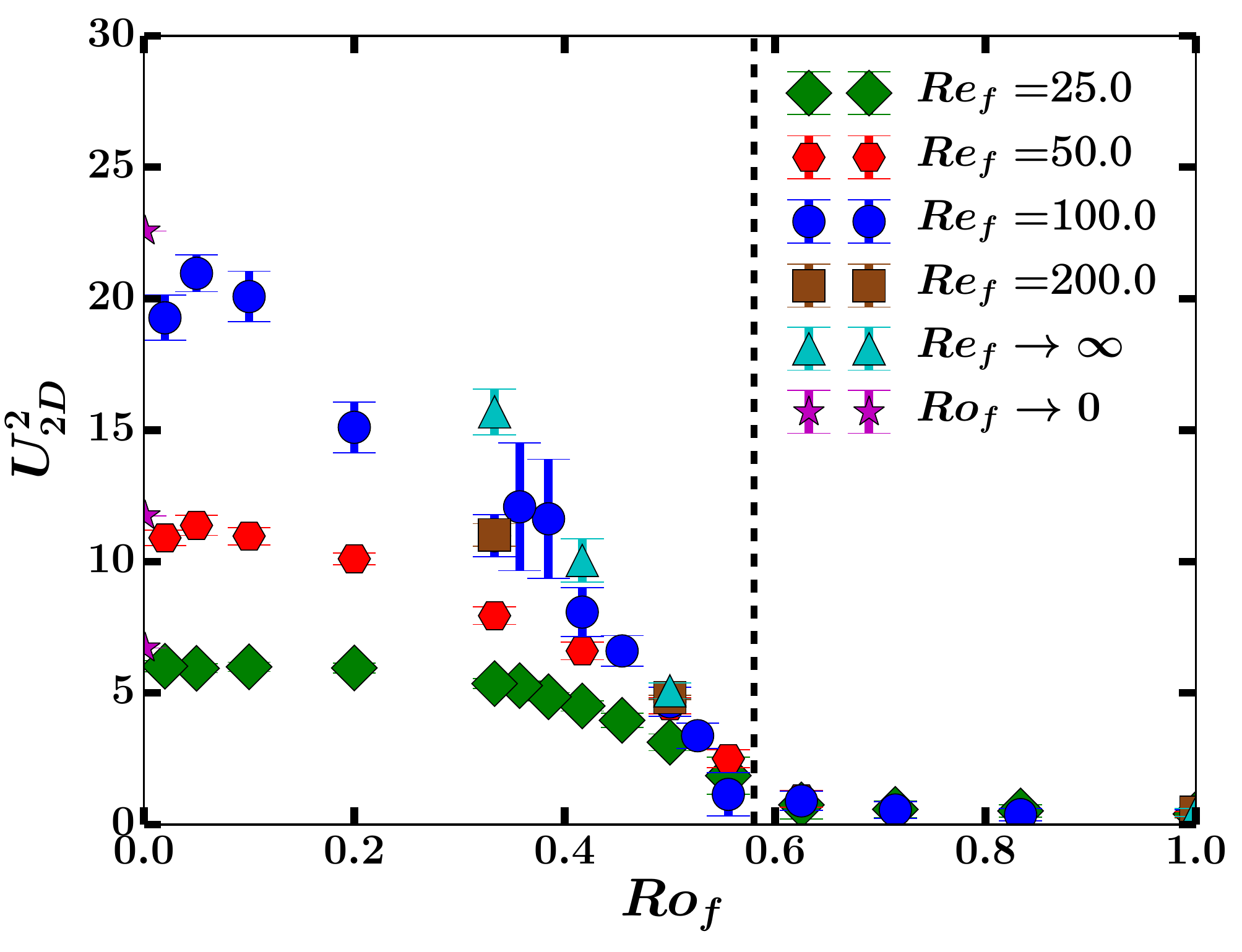}
\caption{}
\end{subfigure}
\caption{ The figures show the energy at large scales $\UTD$ as a function of $\Rof$  (in units of $U_f^2$) for a few different values of $Re_f$ for 
a) the helical flow and b) the nonhelical flow. 
The vertical dashed line at $\Rof \sim 0.6$ denotes the critical Rossby number $\Rof^*$ 
at which the system transitions to a condensate. }
\label{fig:U2}
\end{figure}

The large increase of $U^2$ indicates the formation of a condensate at large scales.
This is clear for large $\Ref$ and strong rotation where $\Rof$ is much smaller than the critical value. 
However for values of the rotation close to the critical value $\Rof^*$
or for small $\Ref$ for which the condensate does not obtain such large values, a better indicator for a condensate formation is the energy $U_ {_{2D}}^2$ contained
in the largest Fourier mode $|{\bf k }|=1$,  or in terms of the energy spectrum $E_k$ we have $\UTD=E_{1}$ where
\beq
E_k= \left\langle \int |{\bf \hat u(q)} |^2 \delta(|{\bf q}| - k ) d{q}^3 \right\rangle \,\, \quad \mathrm{with} \quad \,\,  {\bf\hat u(q)} \equiv \frac{1}{2\pi^{3/2}} \int {\bf u} e^{i{\bf q}\cdot {\bf x}} d{ x}^3
\eeq
Figure \ref{fig:U2} shows $U_{_{2D}}^2$ as a function of $\Rof$ for few different values of $\Ref$
for the helical and the nonhelical flow.  From this figure the critical value $\Rof^*$ is estimated to be $\Rof^* \simeq 0.6$ for both flows.
The value of $\Rof^*$ is denoted by a vertical dashed line in the figures. For values above $\Rof^*$ the large scale energy remains close to zero. 
Below $\Rof^*$ the energy $U_ {_{2D}}^2$ increases as $\Rof$ decreases further from  $\Rof^*$ and asymptotes to
a finite value as $\Rof\to 0 $ is approached. 

Close to the onset the transition to the condensate appears to be supercritical, and $\UTD$
can be fitted to a function of the form $\UTD \propto C_3 (\Rof^*-\Rof)^\gamma$.
From the present data we cannot measure with any significant accuracy the exponent $\gamma$.
We note that increasing $\Ref$ increases the saturation amplitude of $\UTD$
indicating that the prefactor $C_3$ depends on the Reynolds number. 
But in the large $Re_f$ limit, we see that the data points converge for $Ro_f$ close to $Ro_f^*$. This shows that unlike the discussion in the section \ref{theory}, and the results of \cite{alexakis2015rotating,2017arXiv170108497Y} the transition at large $\Ref$ is supercritical. 

\begin{figure}
\begin{subfigure}[b]{0.5\textwidth}
\includegraphics[scale=0.335]{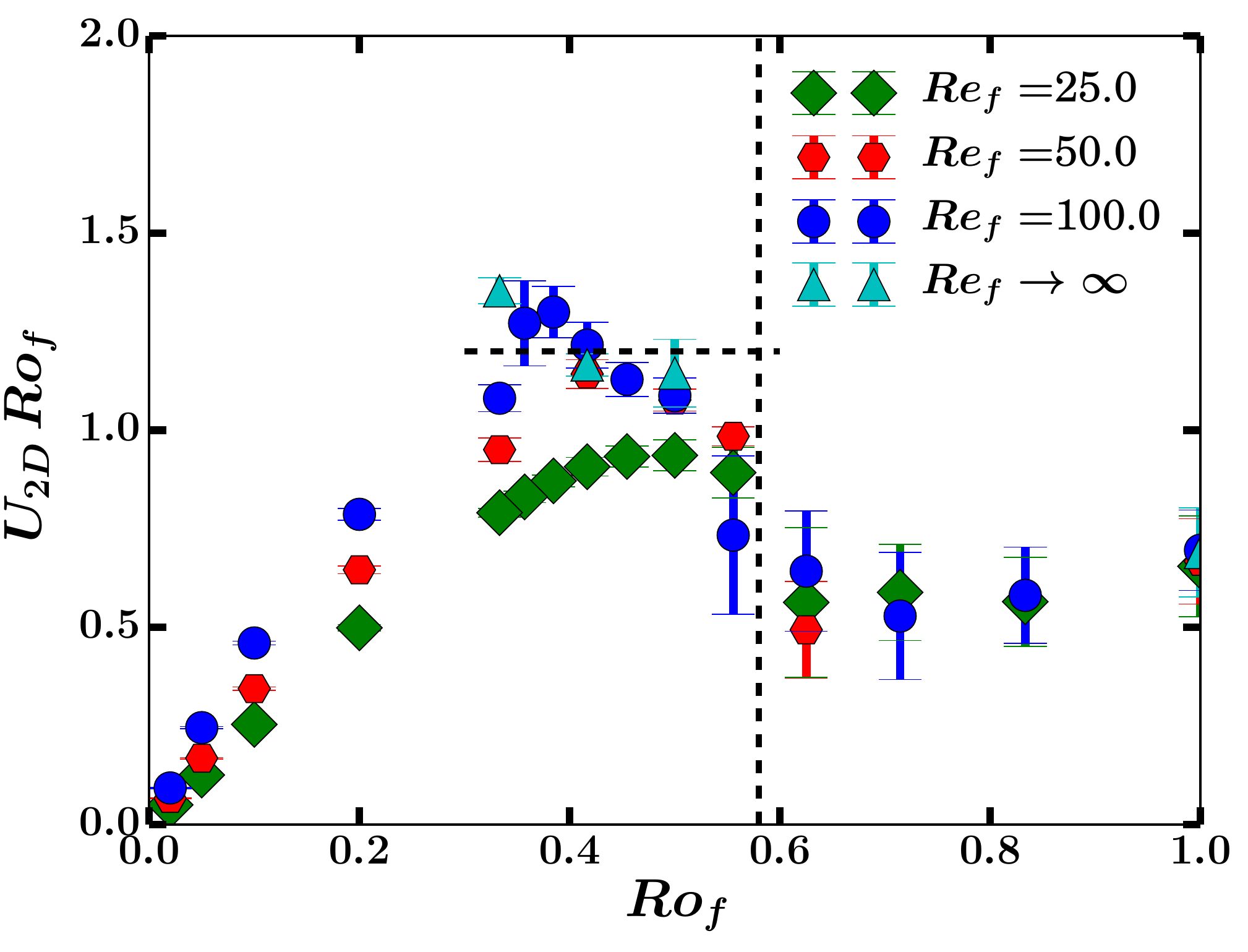}
\caption{}
\end{subfigure} 
\begin{subfigure}[b]{0.5\textwidth}
\includegraphics[scale=0.335]{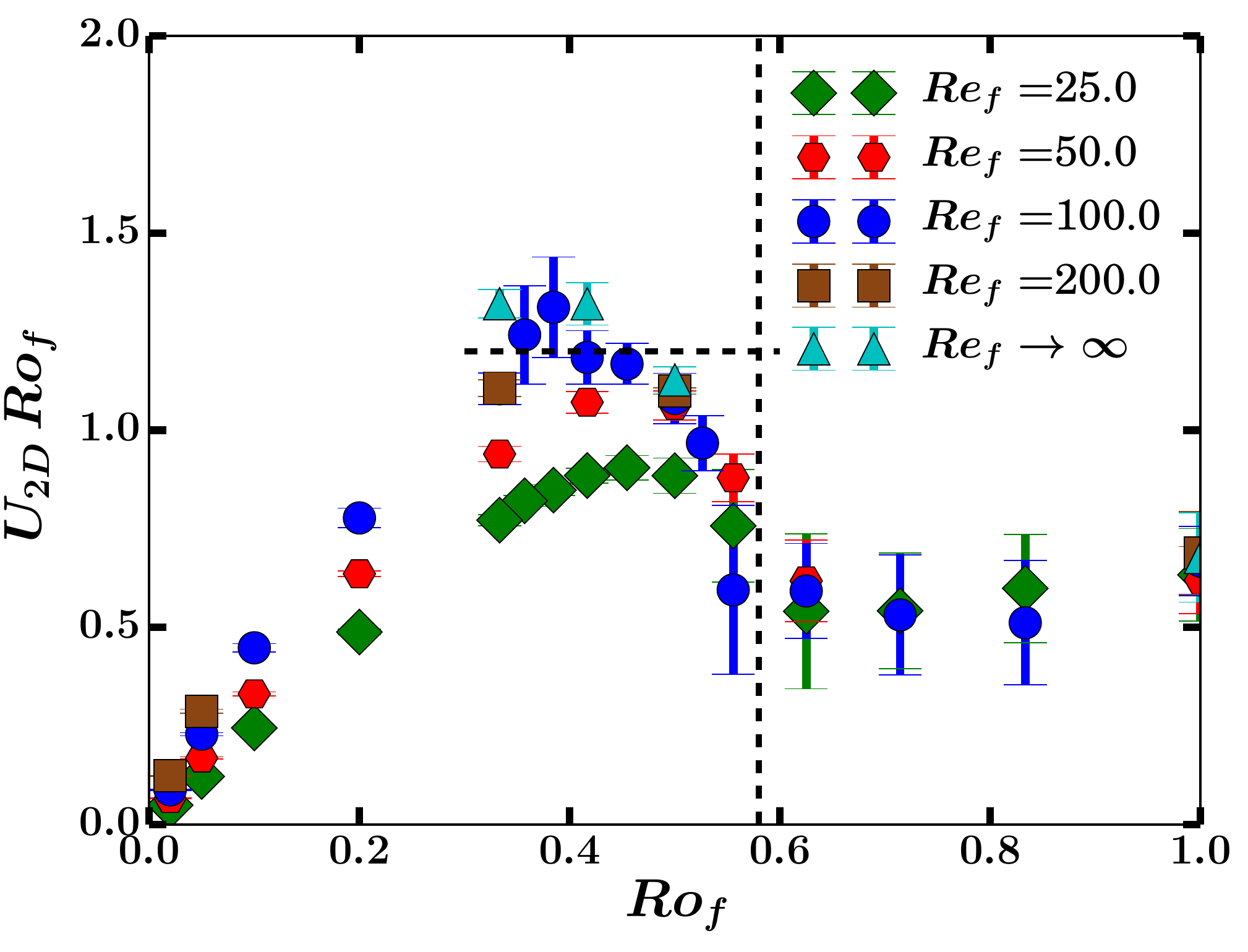}
\caption{}
\end{subfigure}
\caption{ The figures show the product $U_{_{2D}} Ro_f$ (in units of $U_f$) as a function of $Ro_f$ for a few different values of $Re_f$ for a) the case of helical flow and b) the case of the nonhelical flow. The vertical dashed line at $Ro_f \sim 0.6$ denotes the critical Rossby number for the transition to condensates. The horizontal dashed line denotes the scaling $U_{_{2D}} Ro_f \sim 1$.}
\label{fig6}
\end{figure}
For intermediate values of $\Rof$ and for sufficiently large values of $\Ref$, we are expecting that $\UTD$  will 
saturate to values that follow the scaling of the rotating condensates $U_{_{2D}}\propto \Omega L$ (eq. \eqref{UO}), 
that implies that the saturation amplitude is such that $U_{_{2D}} \Rof \simeq 1$.
To test this expectation  we plot in figure \ref{fig6}, $ U_{_{2D}} \Rof $ as a function of $\Rof$ for different values of 
$\Ref$. Indeed in the region $0.3> \Rof > \Rof^*$ the $U_{_{2D}}\Rof$ appears to converge to an order one value as $\Ref$ is increased,   
independent of  $\Rof$. We note that the largest $\Ref$ points are close to the hyper viscous results and this implies independence on $\Ref$ has been reached.
Although the results indicate that the saturation mechanism leading to eq. \eqref{UO} is plausible, 
the range of validity is very small to claim that the scaling has been demonstrated. 
To extend the range of validity to smaller values of $\Rof$ we need to extend our simulations to larger values of $\Ref$.
This however becomes numerically very costly not only because it implies an increase of resolution but also because
the saturation amplitude of the condensate becomes large, and the time-scale to reach saturation increases. 
 As an example we mention that if we would like to extend the range of the rotating condensate to a value of $\Rod$ twice smaller,
it will require to achieve a $\Red$ that it is four times as big as the one used now. This would require a spatial grid that is $4^{3/4}$ bigger in each direction.
If we take in to account the computational cost increase due to the CFL condition \citep{CFL} by a factor of $2 \times 4^{3/4}$ (due to the finer grid and twice larger $U$)
and the twice longer duration of the run  we arrive at a computational cost that is $2^8$ more expensive than the present computations.

\begin{figure}
\begin{subfigure}[b]{0.5\textwidth}
\includegraphics[scale=0.335]{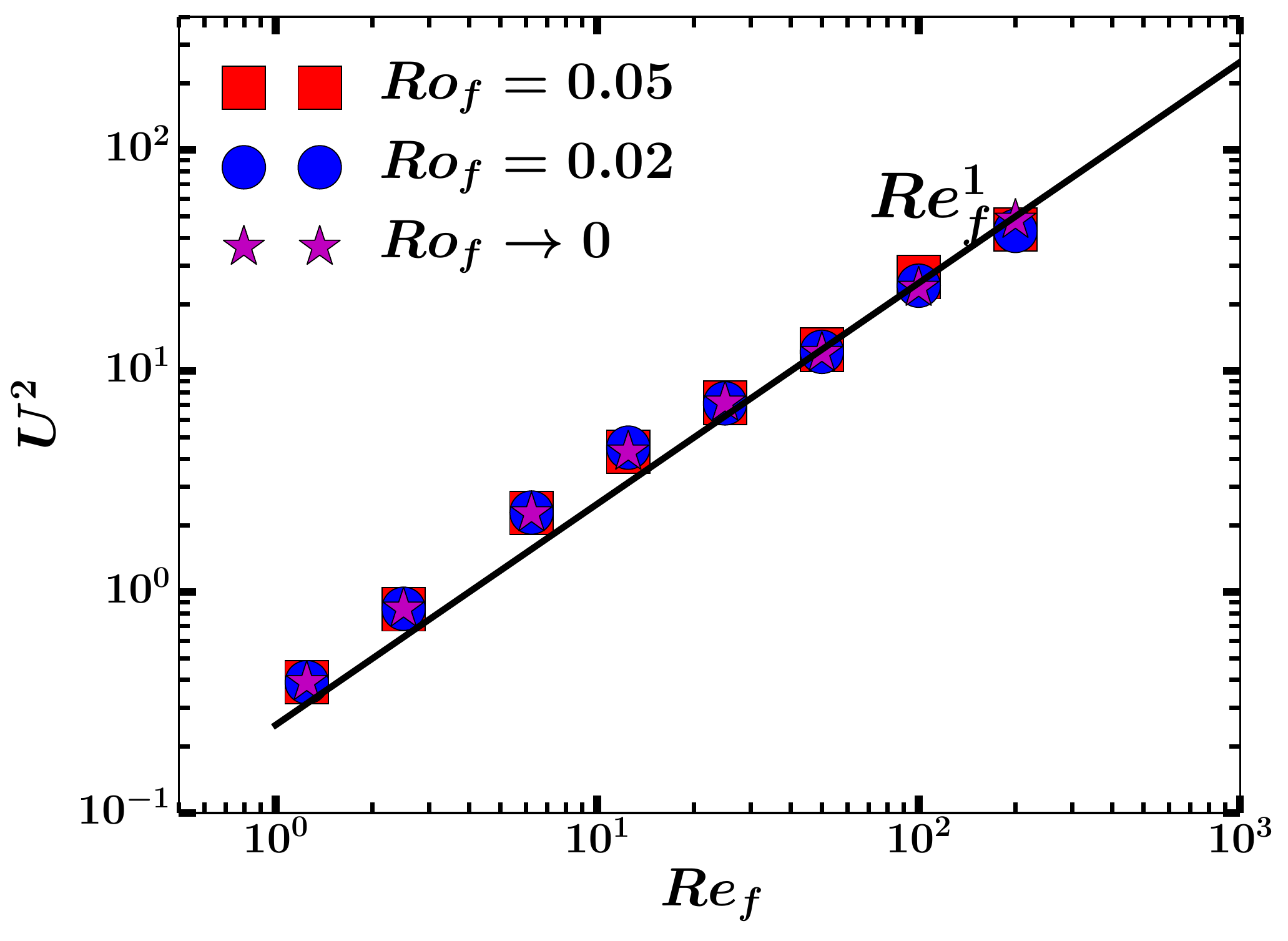}
\caption{}
\end{subfigure} 
\begin{subfigure}[b]{0.5\textwidth}
\includegraphics[scale=0.335]{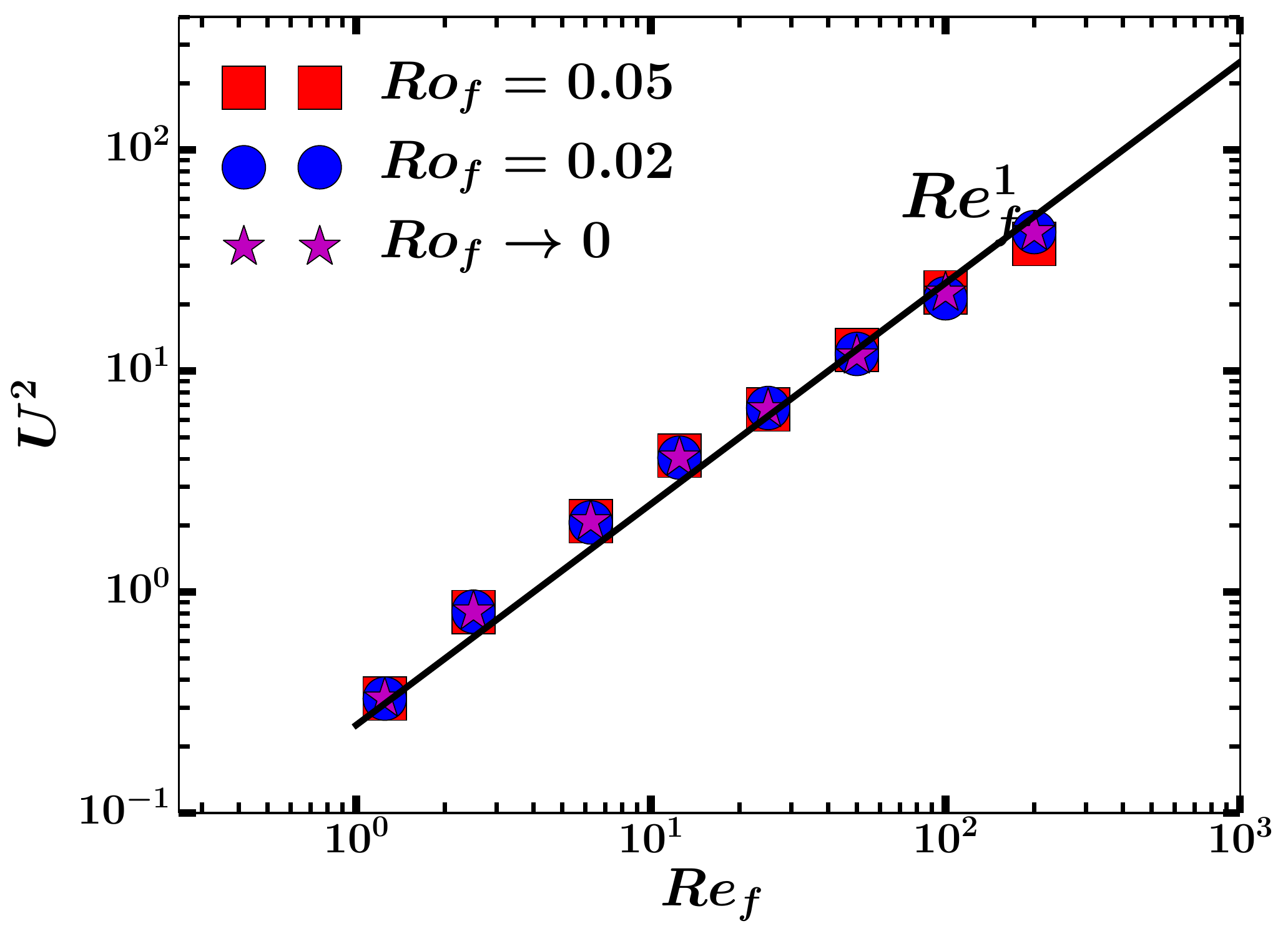}
\caption{}
\end{subfigure}
\caption{ The figures show the energy $U^2$ (in units of $U_f^2$) as a function of $\Ref$ for a few different values of $\Rof\ll1$ for a) the case of helical flow and b) the case of the nonhelical flow. The thick line denotes the linear scaling with $\Ref$.  }
\label{fig7}
\end{figure}

Finally in figure \ref{fig:U2}, for very small values of $\Rof$ the energy $\UTD$ asymptotes to a finite value.
This value matches the results obtained from the $2D$ simulations using 
equation \eqref{R2D} that are marked by a star, indicating that the flow has become two dimensional.
The saturation amplitude $\UTD$ at the $\Rof=0$ limit however depends on the value of $Re$. 
In figure \ref{fig7} we plot $U^2$ for the smallest values of $\Rof$ examined as a function of $\Ref$ along with the results
from the system \eqref{R2D}. The data scale linearly with $\Ref$ in agreement with the prediction given in eq. \eqref{UF} for the viscous condensate.
Equating the two results shown in figure \ref{fig6} and in figure \ref{fig7} we obtain that the transition from the rotating condensate regime
to the viscous condensate regime occurs when $ \Rof^{-2} \sim \Ref$ as seen in eq. \eqref{Rotran}.  

\begin{figure}
\begin{subfigure}[b]{0.48\textwidth}
\includegraphics[scale=0.3]{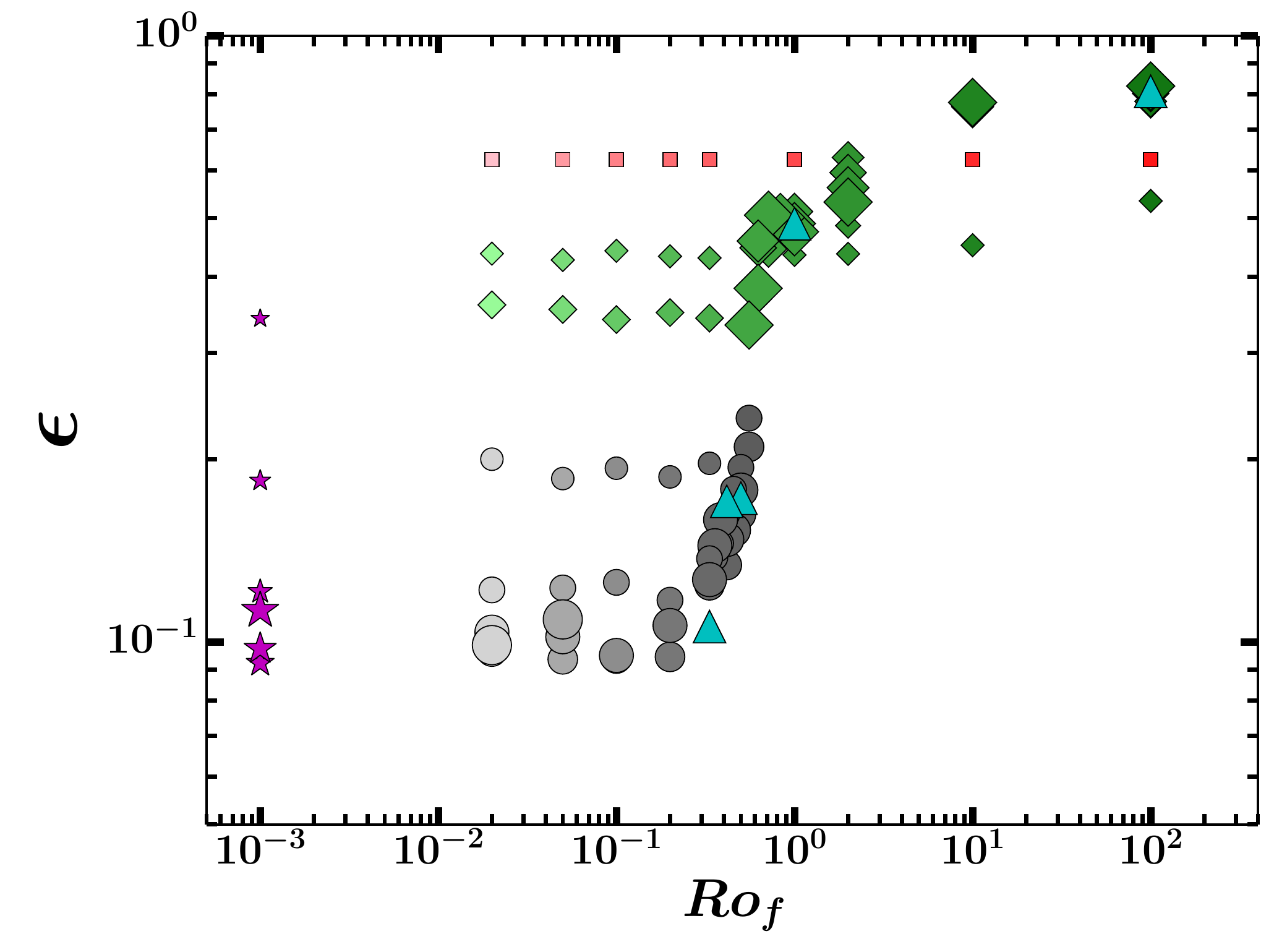}
\caption{}
\end{subfigure} 
\begin{subfigure}[b]{0.48\textwidth}
\includegraphics[scale=0.3]{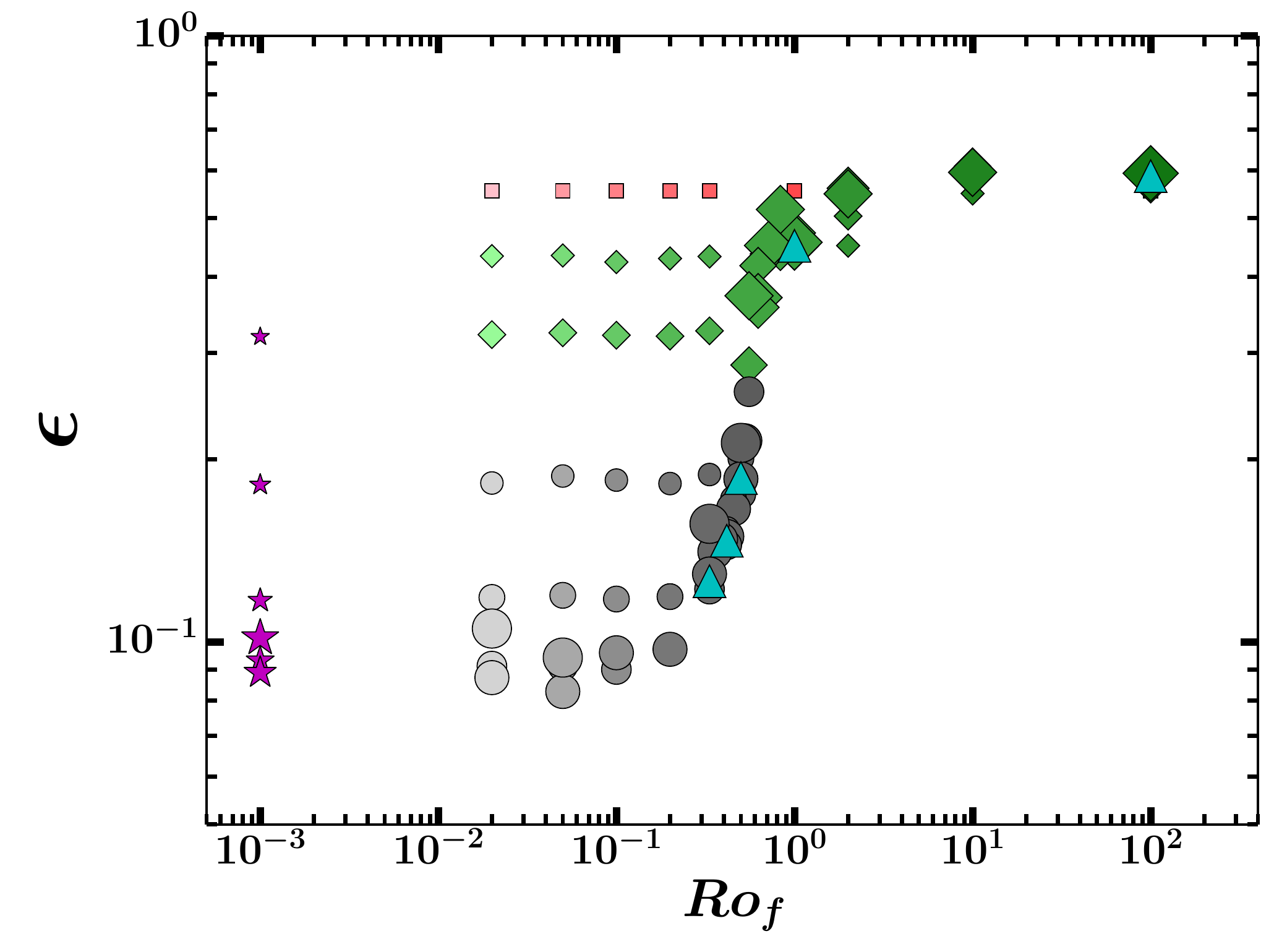}
\caption{}
\end{subfigure}
\caption{ The figures show the injection/dissipation rate $\epsilon$ (in units of $U_f^3k_f$) as a function of $\Rof$ for the examined numerical runs for a) the helical flow and b) the nonhelical flow. Larger symbols denote larger values of $Re_f$ and lighter symbols correspond to smaller values of $Ro_f$. Different symbols correspond to different behaviour of the flow.  }
\label{fig:E1}
\end{figure}

\begin{figure}
\begin{subfigure}[b]{0.5\textwidth}
\includegraphics[scale=0.335]{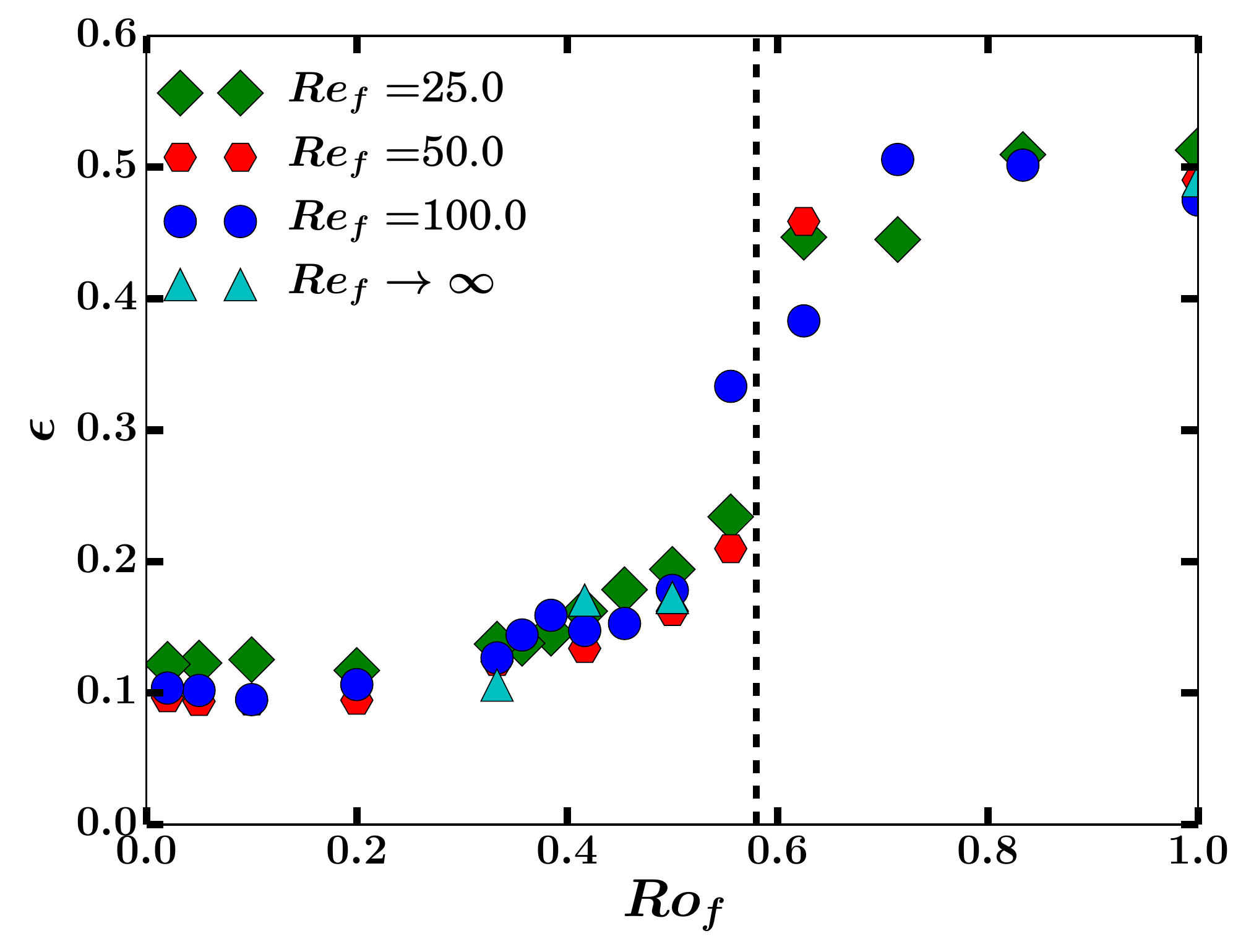}
\caption{}
\label{fig:chap4_fig1a}
\end{subfigure} 
\begin{subfigure}[b]{0.5\textwidth}
\includegraphics[scale=0.335]{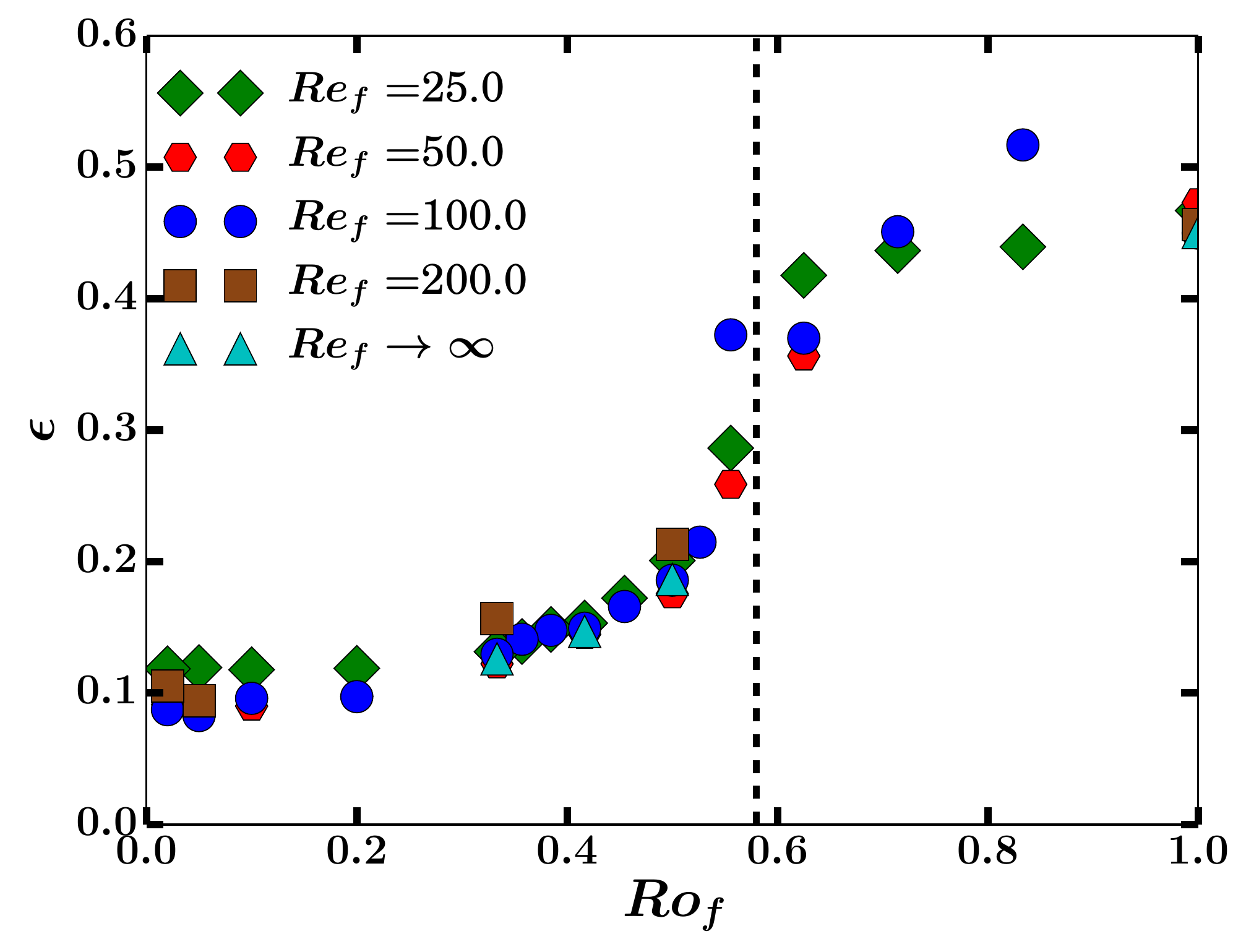}
\caption{}
\label{fig:chap4_fig1b}
\end{subfigure}
\caption{ The figures show the dissipation rate $\epsilon$ (in units of $U_f^3k_f$) as a function of $Ro_f$ for a few different values of $Re_f$ for a) the case of helical flow and b) the case of the nonhelical flow. The vertical dashed line at $Ro_f \sim 0.6$ denotes the critical Rossby number for the transition to condensates.}
\label{fig:E2}
\end{figure}

We now focus on the effect of rotation on the energy injection rate in the system.
In figure \ref{fig:E1} we plot the energy injection rate $\epsilon$
(in units of $f_0^{3/2}k_f^{1/2}$) as a function of $\Rof$ for the entirety of our data points for the helical (left panel) and the non-helical (right panel) runs.
We remind the reader that smaller symbols indicate smaller Reynolds numbers as in figure \ref{map1}.
The energy dissipation as $\Ref$ increases (from small to large symbols), saturates to a $\Ref$ independent value. 
This value however is different for small and large values of $\Rof$
with the transition occurring over a thin region close to $\Rof=\Rof^*$.
This is seen more clearly in figure \ref{fig:E2} where we have concentrated to five largest values of $\Ref$
and plotted the data in linear scale close to $\Rof^*$.
For these values of $\Ref$, the energy injection rate 
is decreased to a five times  smaller value as $\Rof$ is decreased.
The transition from one value to the other occurs very fast when $\Rof$ is close to its critical value $\Rof^*=0.6$.
The transition by this sudden jump at $\Rof^*$ indicates that possibly close to the critical point the dependence
of $\epsilon$ on $\Rof$ could be discontinuous or an other possibility is that it is 
continuous but with diverging derivatives. Similar behaviour 
has been observed close to the transition to an inverse cascade for a 2DMHD flow where the low dimensionality of the system
allowed a much closer investigation.
In any case the investigation of the energy injection close to critical rotation rate 
is very interesting but would require long runs that are expensive for numerical simulations but could be 
addressed more easily with experiments.

\begin{figure}
\begin{subfigure}[b]{0.5\textwidth}
\includegraphics[scale=0.335]{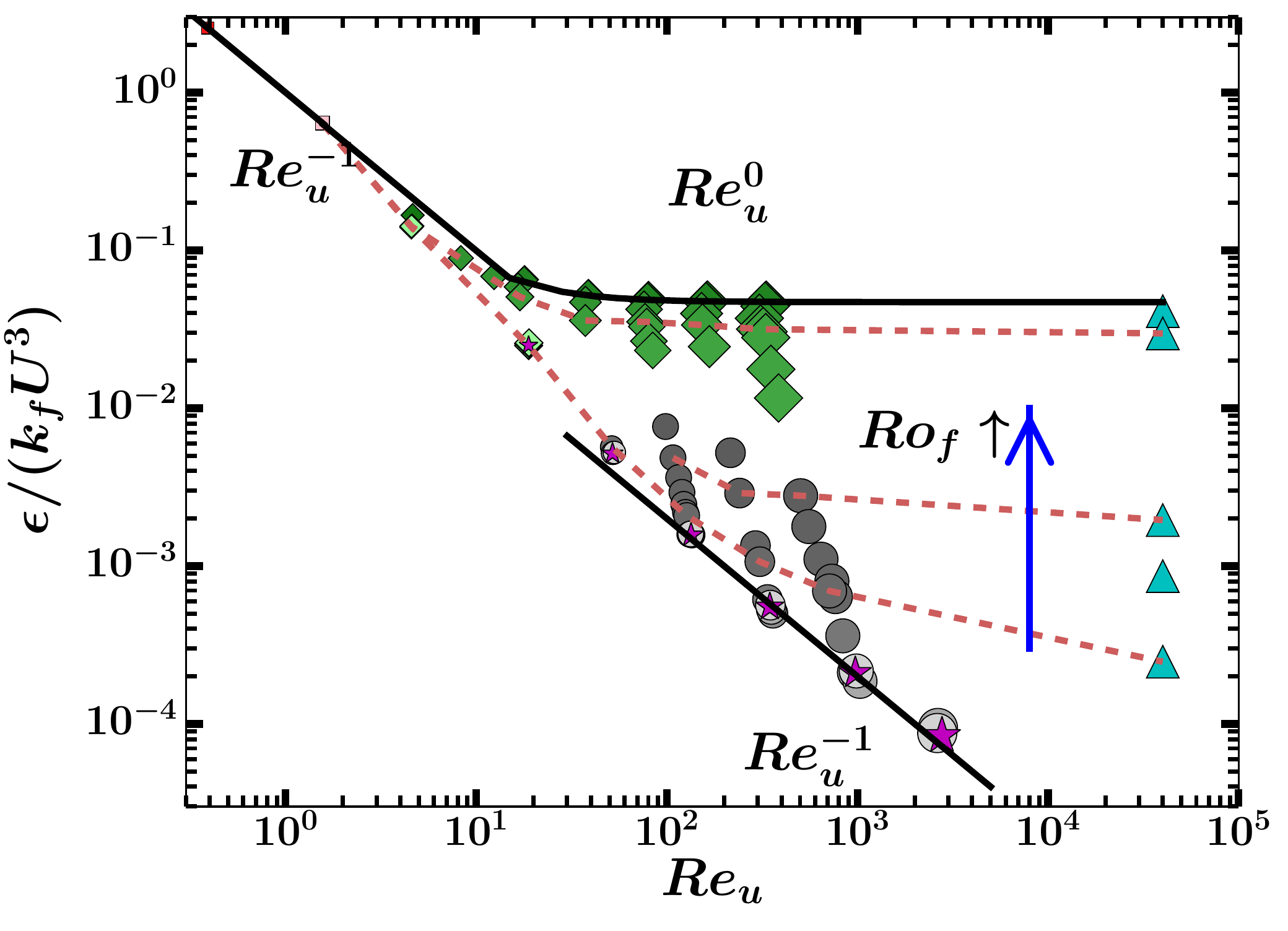}
\caption{}
\label{fig:chap4_fig1a}
\end{subfigure} 
\begin{subfigure}[b]{0.5\textwidth}
\includegraphics[scale=0.335]{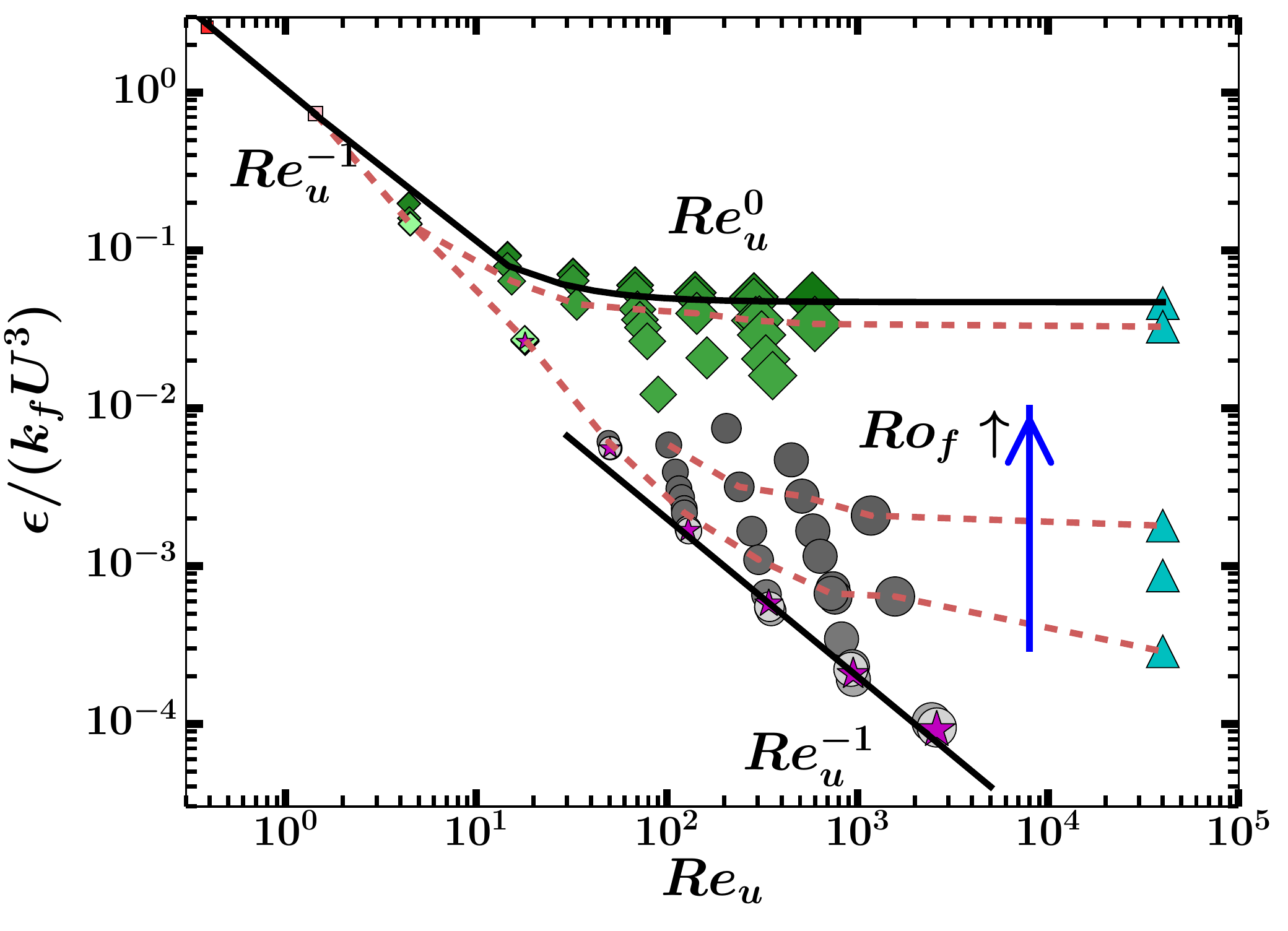}
\caption{}
\label{fig:chap4_fig1b}
\end{subfigure}
\caption{ The figures show the normalized dissipation rate $\epsilon/(U^3 k_f)$ as a function of $Re_u$ for the examined numerical runs for a) the helical case and b) the nonhelical case. Larger symbols denote larger values of $Re_f$ and lighter symbols correspond to smaller values of $Ro_f$. Different symbols correspond to different behaviour of the flow. The thick lines denote the laminar
scaling $Re_u^{-1}$ and the turbulent scaling $Re_u^{0}$. 
 The blue vertical arrow indicates the direction of increasing $Ro_f$ (decreasing $\Omega$). 
 The three dashed brown lines connect the data points of three values $Ro_f = 1.0,\, 0.5,\, 0.33$ as we move from top to bottom. }
\label{fig:E3}
\end{figure}

We conclude this section by considering the ratio $\epsilon/(U^3 k_f)$.
The quantity $\epsilon/(U^3 k_f)$ is sometimes referred as the drag coefficient. 
For laminar flows it scales like $1/\Reu$ while it tends to a non-zero constant for strongly turbulent flows at large $\Reu$. 
The finite asymptotic value of this ratio at large $\Reu$ gives one of the fundamental assumptions of turbulence theory, 
that of finite dissipation at the zero viscosity limit.
This has been clearly demonstrated in experiments of non rotating turbulence and large scale numerical simulations, see \cite{sreenivasan1984scaling,kaneda2003energy,ishihara2016energy}.
In rotating turbulence experiments it has been investigated in \cite{campagne2016turbulent}, where the drag coefficient has been shown to scale as $\Rou$  for sufficiently small $\Rou$.
We note that in their experimental set-up it was the velocity of the propellers that were used to define $\Rou$.
%
In figure \ref{fig:E3} we plot the ratio $\epsilon/(U^3 k_f)$
as a function of $\Reu$ for different Rossby numbers. {The arrow indicates the direction that $\Rof$ is increased (ie rotation is decreased).}
The dashed lines connect points with the same value of $\Rof$ for three different values of $Ro_f = 1.0,\, 0.5,\, 0.33$ as we move from top to bottom.  

For rotation rates such that $\Rof > \Rof^* $ (diamonds), the data show a $\Reu^{-1}$ scaling at low $\Reu$ that transitions to a constant at large $\Reu$
demonstrating a finite dissipation at infinite $\Reu$. This asymptotic value decreases slightly with $\Rof$. 
For the runs with $\Rof < \Rof^* $ (circles) on the other hand, the region of the laminar scaling $\Reu^{-1}$ appears to extend to larger values of 
$\Reu$. 
The very fast rotating runs (circles with light colours) and the $2D$ simulations from eq. \eqref{R2D} show a $\Reu^{-1}$ scaling
through out the examined range. The pre-factor in front of $\Reu^{-1}$ has decreased at the condensate regime because the laminar vortices 
are at the scale of the forcing (eq. \eqref{UELT}) while the viscous condensate vortices are at the scale of the box size eq. \eqref{UV2}.
However, for fixed $\Rof$ (dashed lines),  as the Reynolds number is increased  the $\Reu^{-1}$ scaling appears to flatten to a $\Reu$ independent scaling. 
This occurs for the flows that are in the rotating condensate regime.
This suggests that even for the rotating runs, the ratio $\epsilon/(U^3 k_f)$ will reach an asymptotic non-zero value at $\Reu\to\infty$ 
(for fixed $\Rof$) matching the one obtained by the hyper-viscous simulations. 
This asymptotic value however is different for different values of $\Rof$. Indicating that the value of the drag coefficient depends on the Rossby number.

\begin{figure}
\begin{subfigure}[b]{0.5\textwidth}
\includegraphics[scale=0.335]{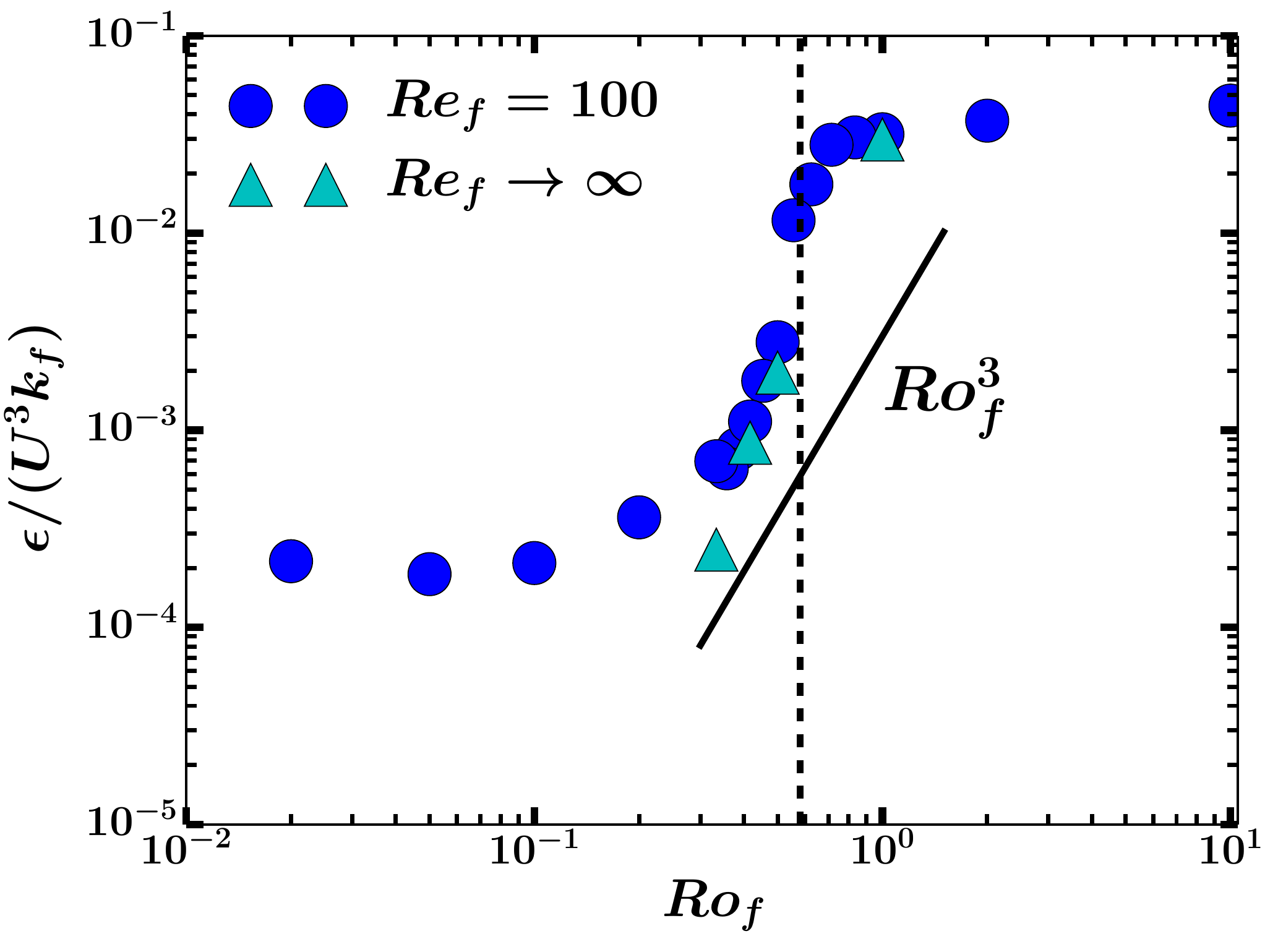}
\caption{}
\label{fig:chap4_fig1a}
\end{subfigure} 
\begin{subfigure}[b]{0.5\textwidth}
\includegraphics[scale=0.335]{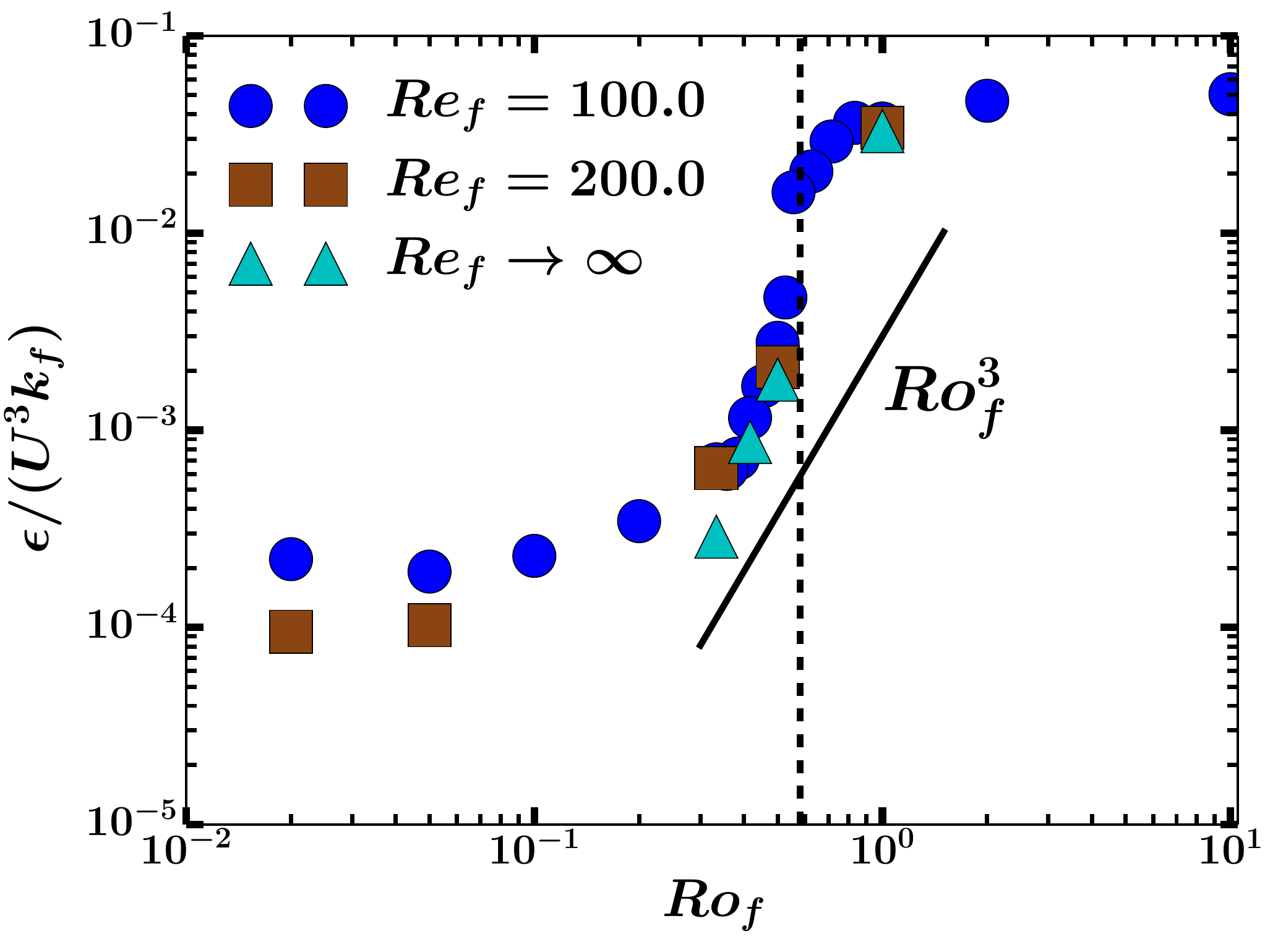}
\caption{}
\label{fig:chap4_fig1b}
\end{subfigure}
\caption{ The figures show the normalized dissipation rate $\epsilon/(U^3 k_f)$ as a function of $Ro_f$ for a few different values of $Re_f\gg1$ for a) the case of helical flow and b) the case of the nonhelical flow. The thick line denotes the scaling $Ro_f^3$ and the vertical dashed line denotes the critical Rossby number for the transition to condensates. }
\label{fig:E4}
\end{figure}

The values of this asymptotic behaviour along with the results of the hyper viscous runs are shown in figure \ref{fig:E4} 
where they are compared with the scaling $\epsilon/(U^3 k_f) \propto \Rof^3$ that is the scaling obtained if assuming the saturation amplitude follows $U\propto \Omega L$.
The data appear to be slightly steeper. 
Perhaps this is not surprising considering the small range of $\Rof$ that the scaling $U\propto \Omega L$ was shown to hold in figure \ref{fig6}.
Note that a weak turbulence scaling would predict $\epsilon/(U^3 k_f)\propto \Rod$ that is clearly not obtained here.


\section{Structures, Spectra and Dynamical behaviour}  
\label{Dyn} 

In this section we try to obtain an understanding of the results in the previous section 
by visualizing the structures involved and examining their spectral and temporal behaviour.
%
\begin{figure}
\begin{subfigure}[b]{0.3\textwidth}
\includegraphics[scale=0.135]{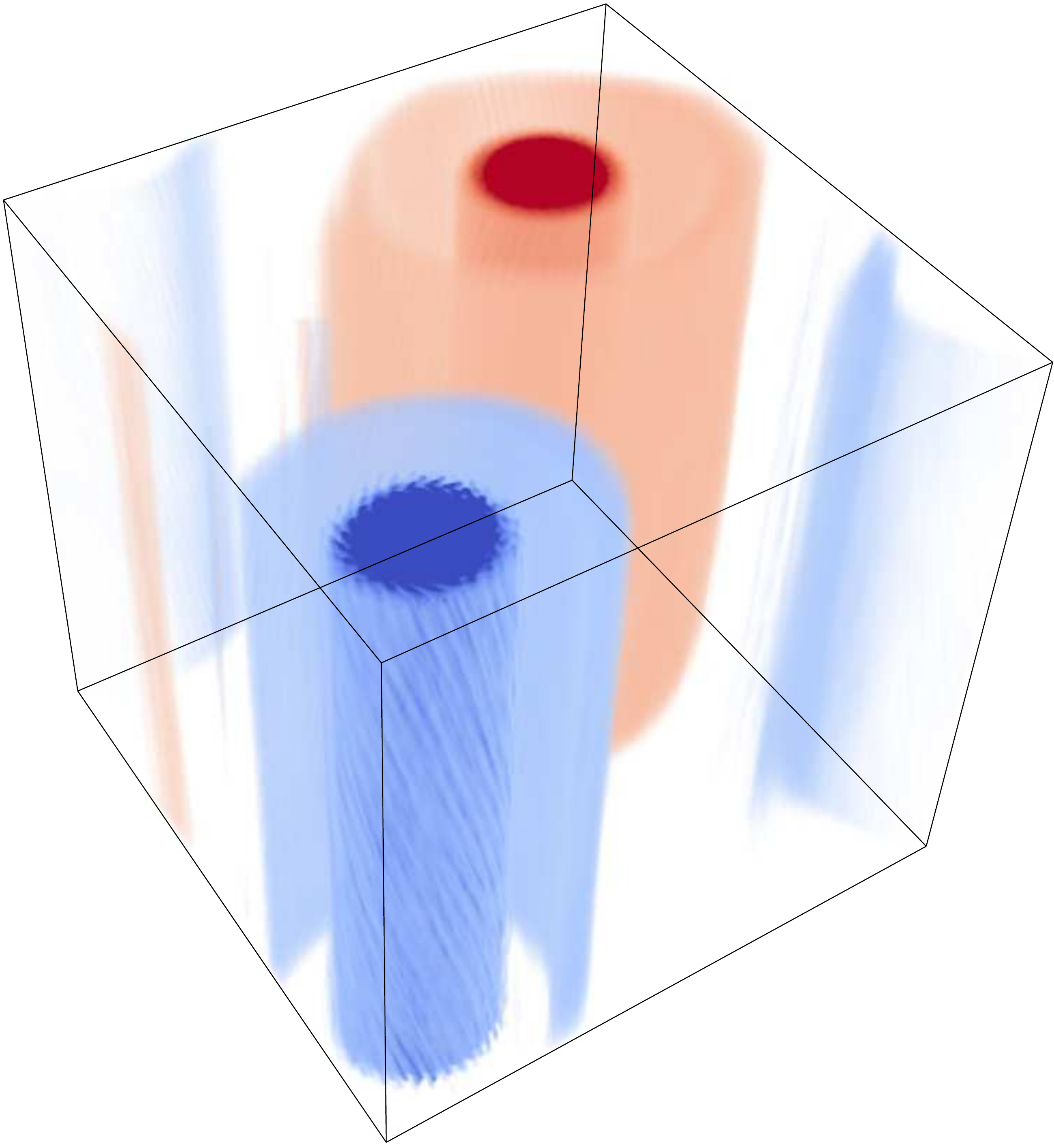}
\caption{ }
\end{subfigure} 
\begin{subfigure}[b]{0.3\textwidth}
\includegraphics[scale=0.135]{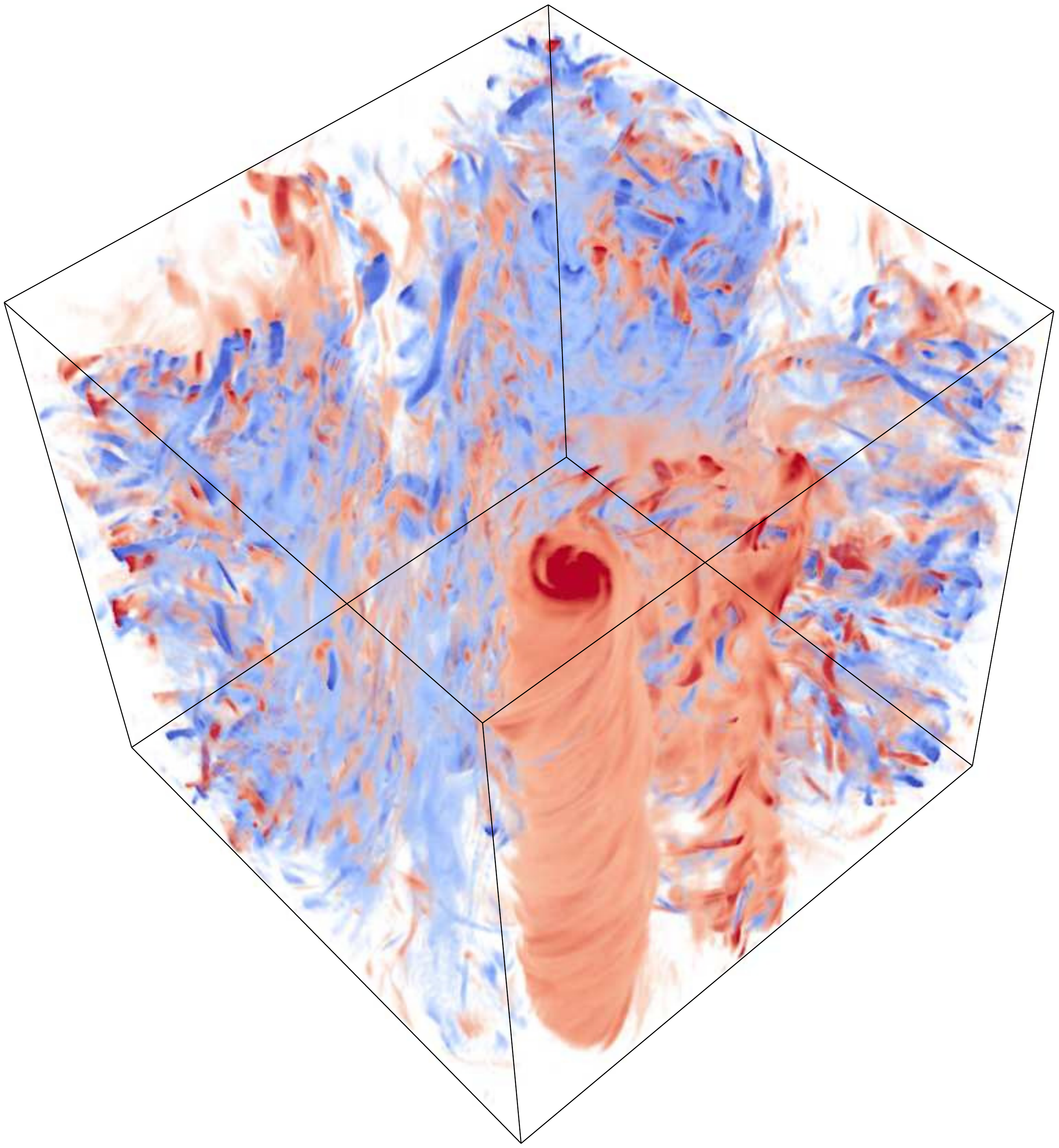}
\caption{ }
\end{subfigure}
\begin{subfigure}[b]{0.3\textwidth}
\includegraphics[scale=0.135]{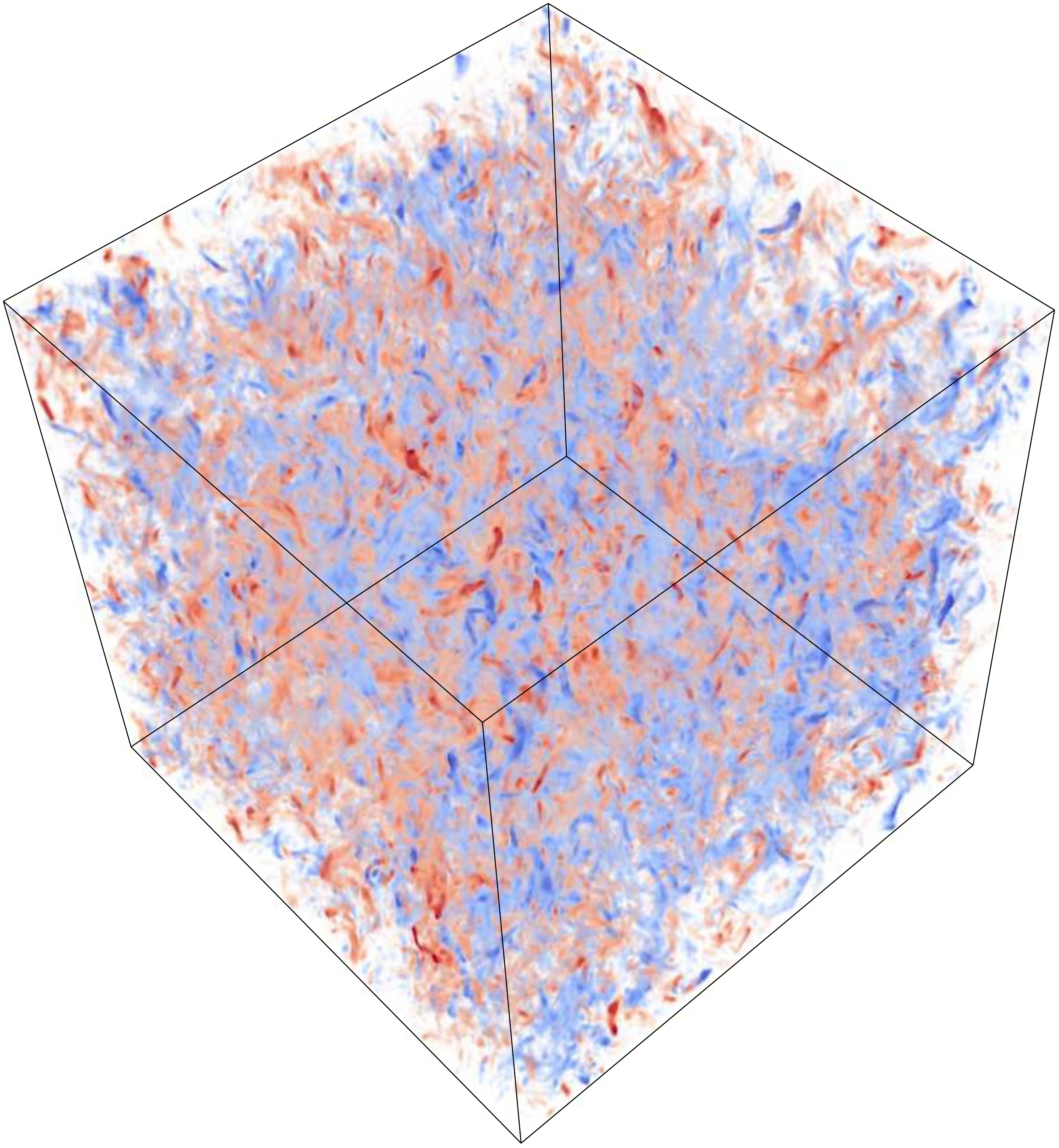}
\caption{ }
\end{subfigure}
\caption{ The figures show the contours of the vertical vorticity $\omega_z$ for a) $Ro_f = 0.02, Re_f = 100$, b) $Ro_f = 0.5, Re_f = 100$, c) $Ro_f = \infty, Re_f = 100$. The red colour corresponds to positive vorticity and the blue colour corresponds to negative vorticity. }
\label{fig:VIS1}
\end{figure}
%
We start by the visualization of the flows. 
Figure \ref{fig:VIS1} shows colour coded visualizations of the vertical vorticity field.
The red colour corresponds to vorticity parallel to rotation while the blue colours
correspond to vorticity anti parallel to rotation. The three images have been constructed  
from numerical simulations corresponding to the three regimes discussed in the previous section, 
(a) the viscous condensate, (b) the rotating condensate with the counter rotating vortex cascading energy 
back to the small scales and (c) weakly rotating (or non-rotating) turbulence. In the first case (a) the flow looks
very close to a 2D state with no visible variations along the $z$ direction and no observed asymmetry between co-rotating and counter-rotating vortex.
In the second case (b) a condensate is also formed but only clearly observed for the co-rotating vortex.
The counter rotating vortex, although present, is infested with small scale eddies that extract energy from it.
Finally in case (c) no large scale condensate is observed and the flow looks isotropic.

\begin{figure}
\begin{center}
\includegraphics[scale=0.30]{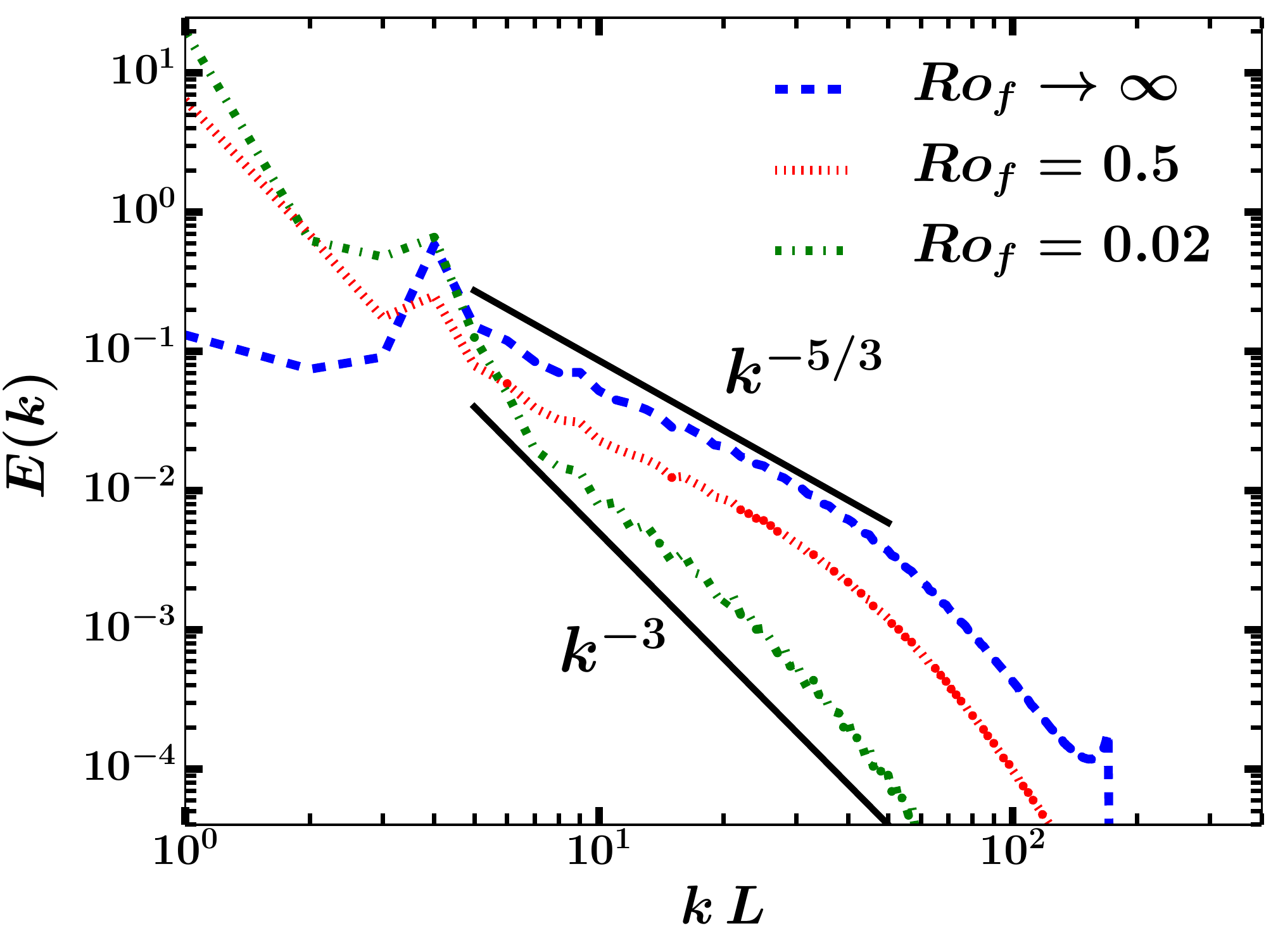}
\end{center}
\caption{The figure show the energy spectra for a few different values of $Ro_f$ with $Re_f = 100$. The black thick lines denote the scaling $k^{-5/3}, k^{-3}$.}
\label{fig:spec_3cas}
\end{figure}

The spectra for the three cases are shown in the figure \ref{fig:spec_3cas}. The spectrum for the flow in the
viscous condensate regime (a) is shown with a green dash dot line. 
The energy is concentrated at the smallest wavenumber $kL=1$, with the energy for wave numbers above $k_fL=4$
dropping very fast. In the non rotating case (c), shown by a dashed line,
energy is concentrated at the forcing wavenumber $k_fL=4$ that is followed
by a power-law spectrum close to $k^{-5/3}$. 
Finally, the case in the intermediate regime (b) shown by the dotted line, shows signs of both behaviours:
the energy is concentrated at the largest scale $kL=1$ as in case (a) but the 
spectrum at the small scales follows a  $k^{-5/3}$ power-law as in the non-rotating case. 
Thus the spectrum for the rotating condensate is in agreement with the co-existence of a condensate along with a forward cascade.  

We next examine the behaviour of the flow close to the transition point $\Rof^*$.
The arguments made in section \ref{theory} suggested that at large $\Ref$ this transition would become discontinuous (subcritical)
which was what was found for the Taylor-Green forcing \cite{alexakis2015rotating,2017arXiv170108497Y}. The results in the previous section
however showed that even at large $\Ref$ the transition remains supercritical.

\begin{figure}
\begin{subfigure}[b]{0.5\textwidth}
\includegraphics[scale=0.275]{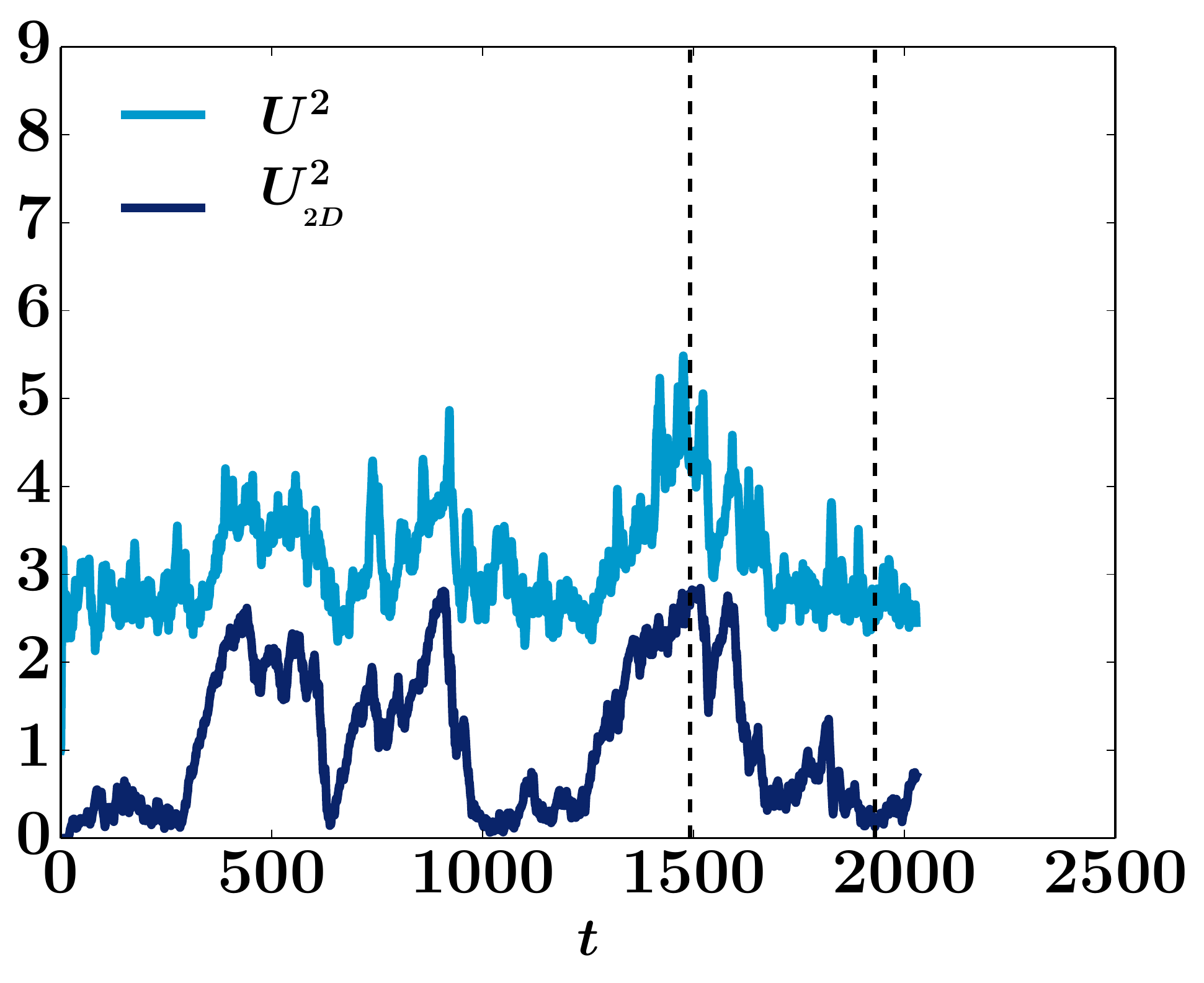}
\caption{}
\label{fig:bistability1}
\end{subfigure} 
\begin{subfigure}[b]{0.5\textwidth}
\includegraphics[scale=0.275]{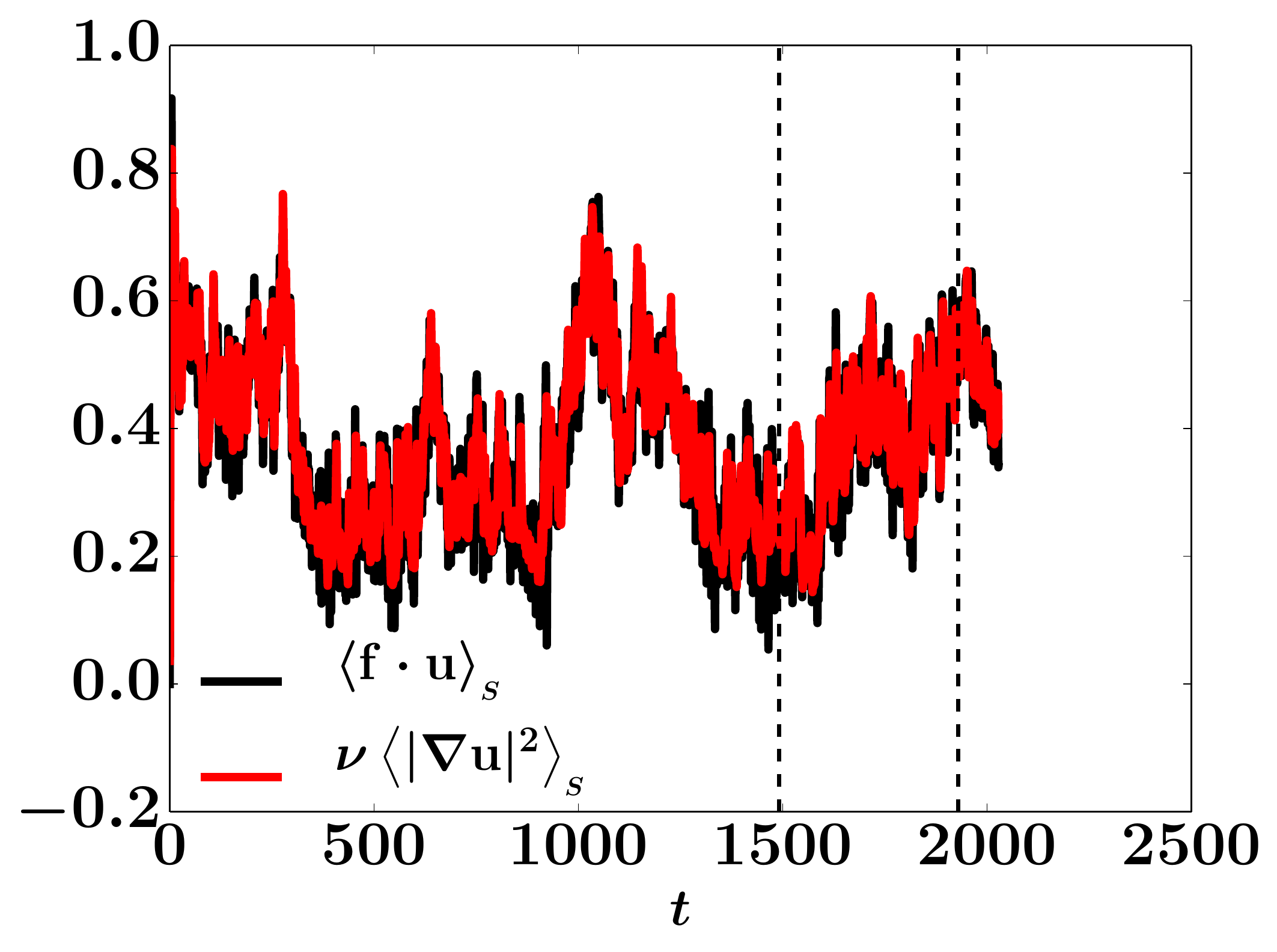}
\caption{}
\label{fig:bistability2}
\end{subfigure}
\caption{The figure a) show the time series of the total energy $U^2$ and the energy at the large scales $U_{_{2D}}^2$ for the case of $Ro_f = 0.556, Re_f = 100$ which is close to $Ro_f^*$. The vertical dashed lines denote the time instances at which the visualizations \ref{fig:VIS2} are taken. Figure b) shows the spatial averaged energy injection rate $\left\langle {\bf f} \cdot {\bf u} \right\rangle_{_S}$ and the dissipation rate $\nu \left\langle |{\bm \nabla} {\bf u}|^2 \right\rangle_{_S}$ for the same run. }
\label{fig:bistability}
\end{figure}
\begin{figure}
\begin{subfigure}[b]{0.5\textwidth}
\includegraphics[scale=0.135]{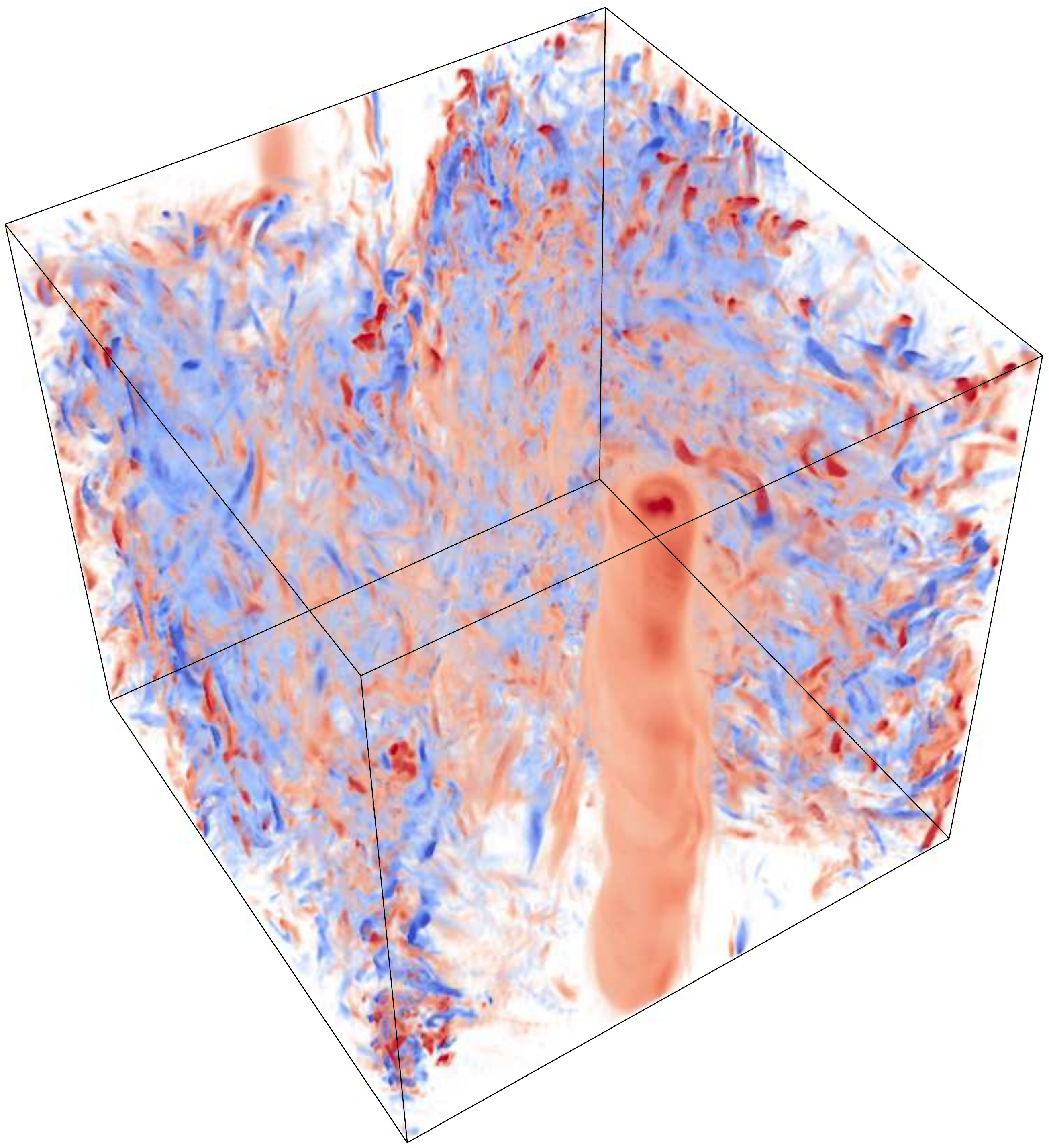}
\caption{}
\end{subfigure} 
\begin{subfigure}[b]{0.5\textwidth}
\includegraphics[scale=0.135]{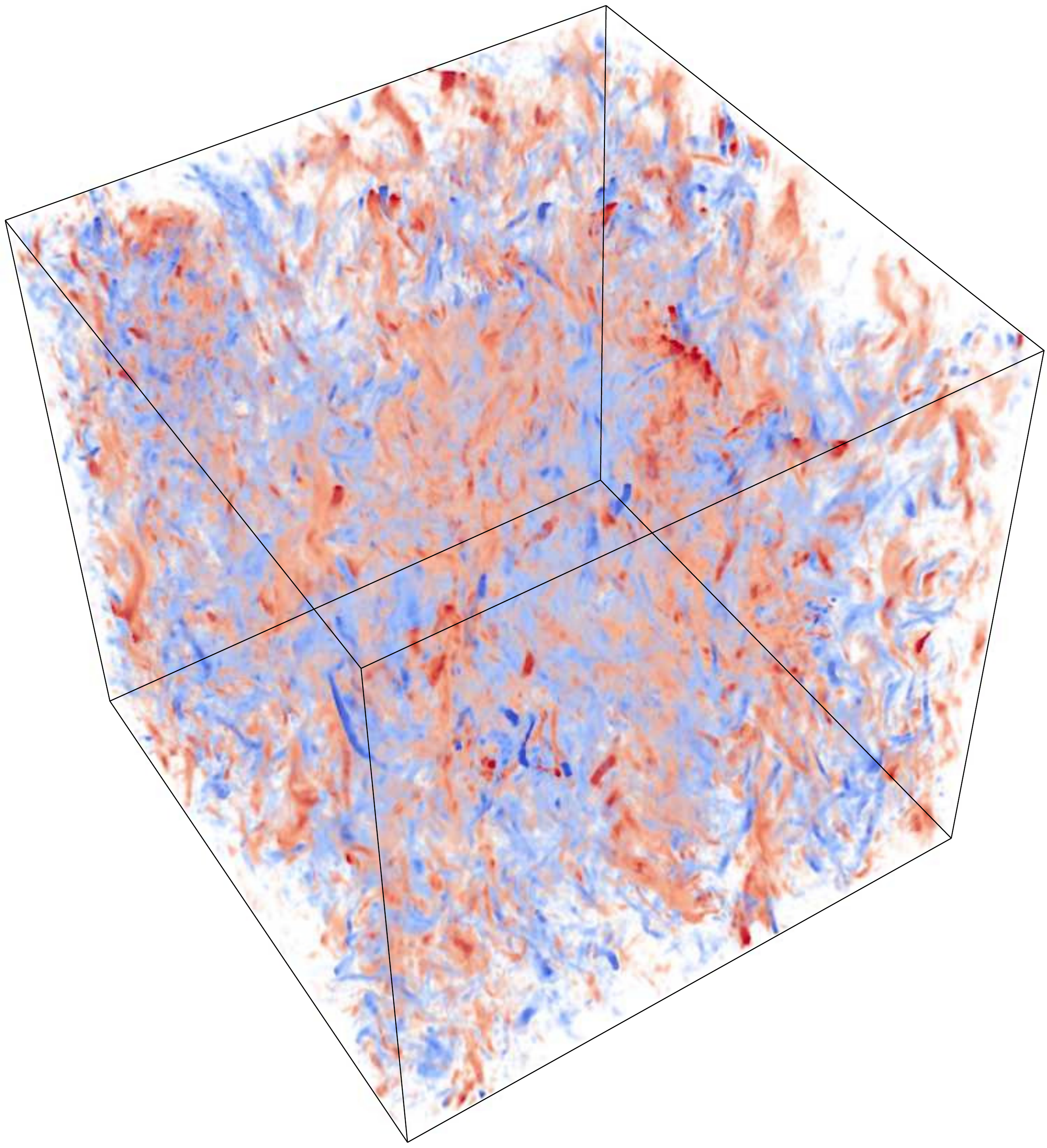}
\caption{}
\end{subfigure}
\caption{The figures show the contours of the vertical vorticity $\omega_z$ for $Ro_f = 0.556, Re_f = 100$ for the two time instances marked in figure \ref{fig:bistability1}. The red colour corresponds to positive vorticity and the blue colour corresponds to negative vorticity. The figure a) shows a co-rotating vortex formed when the system is in the condensate regime, while the figure b) does not have any large scale structure. }
\label{fig:VIS2}
\end{figure}

To understand this discrepancy, in figure \ref{fig:bistability1} we show the time evolution of the total energy $U^2$ with a dark line and the 
energy of the large scales $U_{_{2D}}^2$ for a value of $\Rof=0.556$ close to the critical value
and a relative large $\Ref = 100$. The flow randomly oscillates between two distinct states:
one where the energy of the large scales is weak and most of the energy lies in the forcing scales and one
where the energy of the large scales dominates and accounts for more than 60\% of the total energy.
The energy at the large scales varies by an order of magnitude between these two states
with $\UTD \sim \Omega L$ when $\UTD$ dominates and $ \UTD \ll \Omega L$ at its low values.
In the panel \ref{fig:bistability2} the time series of the spatial averaged energy injection rate 
$\left\langle {\bf f} \cdot {\bf u} \right\rangle_{_{S}}$ and the energy dissipation rate 
$\nu \left\langle |{\bm \nabla} {\bf u}|^2\right\rangle_{_{S}}$ is shown. 
A burst of energy dissipation is observed at the time instances that the flow transitions from the condensate 
to the 3D turbulent state. This correlation between the change of state in the large scales
and the energy dissipation/injection is typical of bimodal systems \citep{mishra2015dynamics}.
%
Visualizations of the vertical vorticity of the flow 
are shown in figure \ref{fig:VIS2} at the two different times
indicated by the vertical dashed lines in figure \ref{fig:bistability1} and \ref{fig:bistability2}. The two figures resemble 
the ones shown in panel (b) and (c) in figure \ref{fig:VIS1} that were obtained for different values of the parameter $\Rof$.

It appears thus that the transition from isotropic turbulence to rotating condensate occurs through 
a bistable regime where both states are realized at different instances of time. 
The two state are distinct i.e. they are separated by finite amount of energy however 
the time the system spends in each one of these states can depend on the deviation from the onset $\Rof^*$,
becoming infinite for the condensate  state for $\Rof$ sufficiently smaller than $\Rof^*$. 
The time averaged quantities displayed in the previous sections thus remain continuous.
This bistable behaviour if it persists at larger $\Ref$ will indicate that the transition will remain supercritical.
Similar behaviour has been observed in experiments in a rotating tank where intermittent switching between 
blocked and large scale zonal patterns have been observed \cite{weeks1997transitions}. This presents an alternate mechanism other than the sub-critical transition discussed in \ref{theory} and observed in \cite{alexakis2015rotating,2017arXiv170108497Y}.

\begin{figure}
\begin{subfigure}[b]{0.5\textwidth}
\includegraphics[scale=0.275]{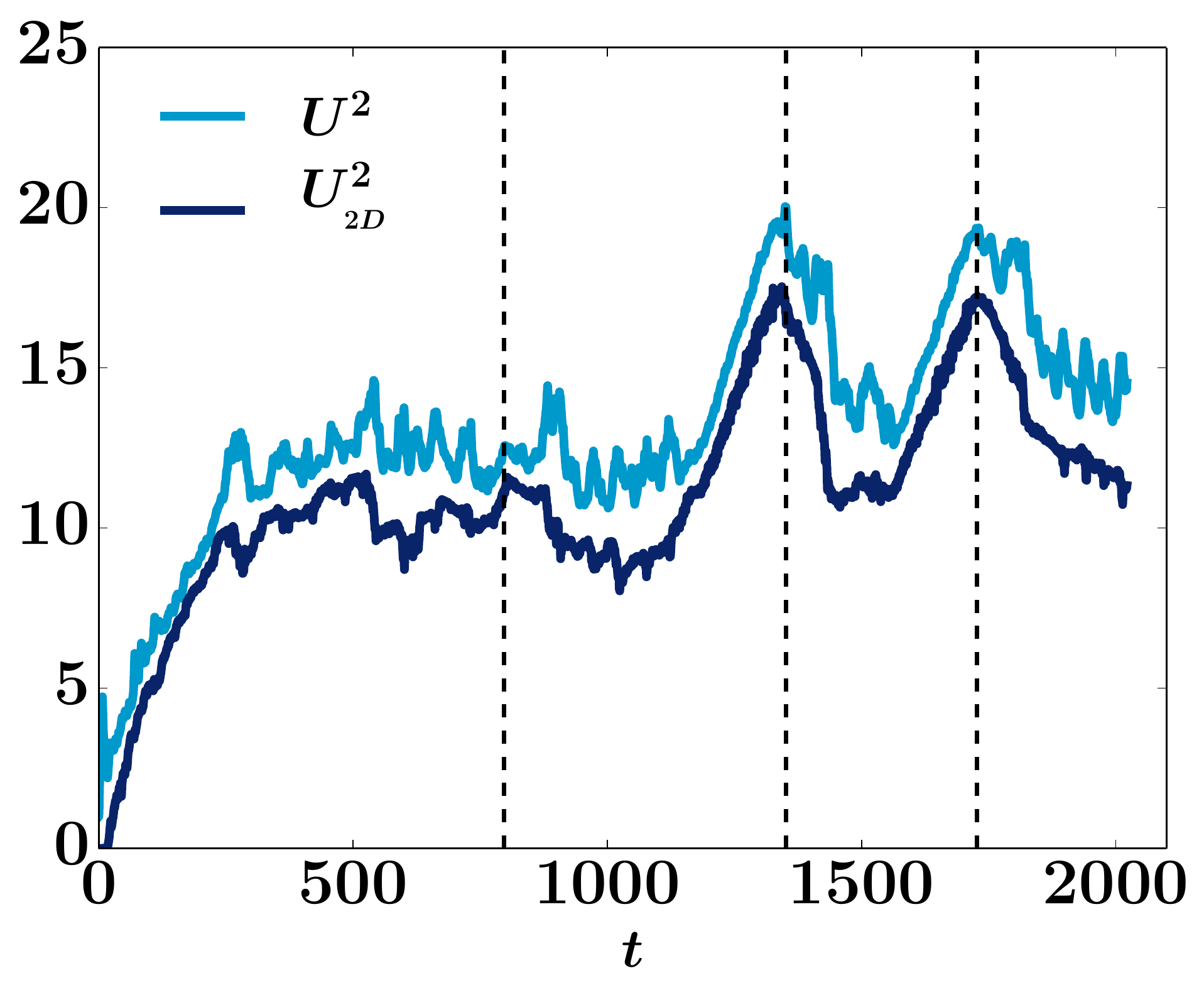}
\caption{}
\label{fig:nobistability1}
\end{subfigure} 
\begin{subfigure}[b]{0.5\textwidth}
\includegraphics[scale=0.275]{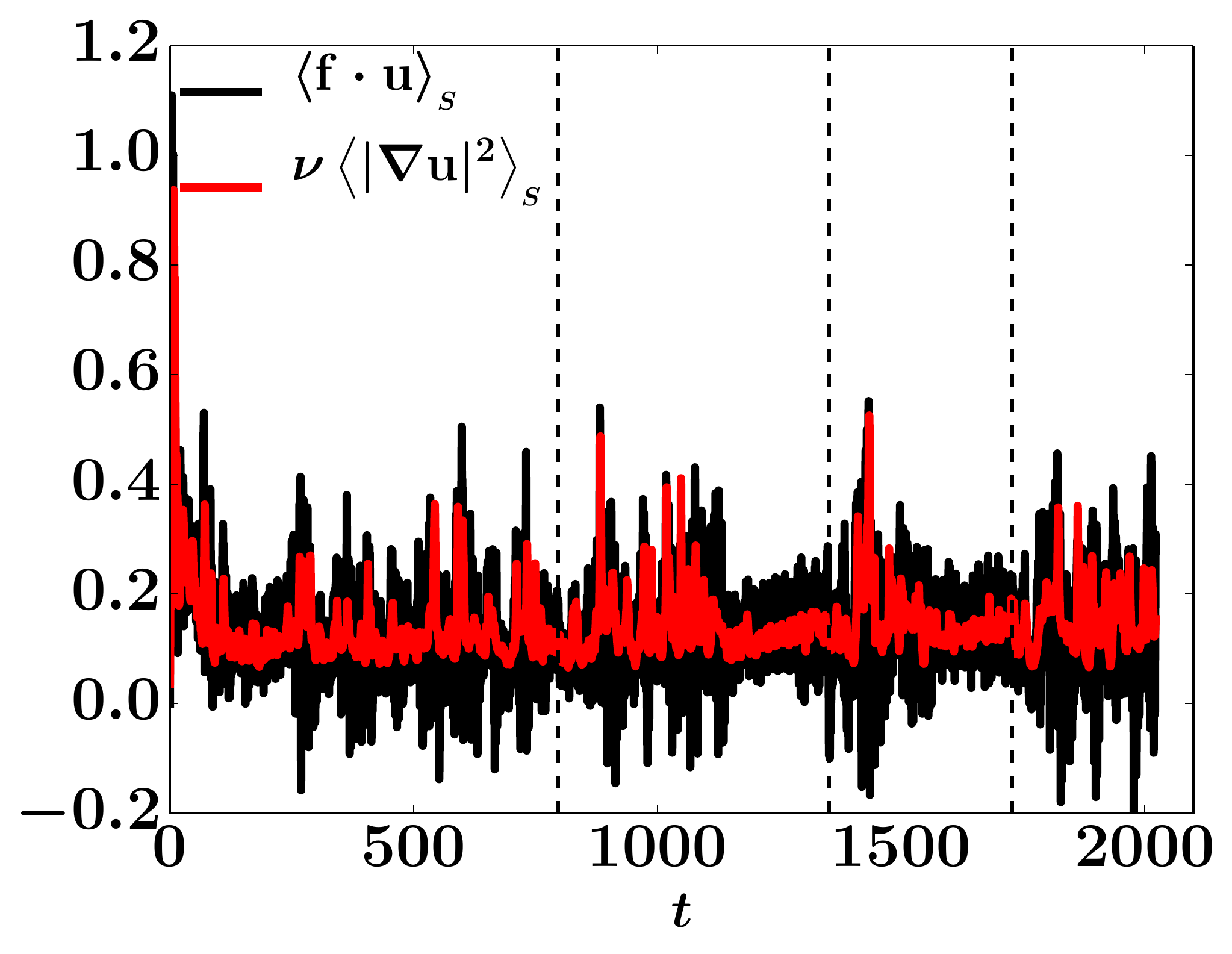}
\caption{}
\label{fig:nobistability2}
\end{subfigure}
\caption{The figure a) show the time series of the total energy $U^2$ and the energy at the large scales $U_{_{2D}}^2$ for the case of $Ro_f = 0.357, Re_f = 100$ which is close to $Ro_f^*$. The vertical dashed lines denote the time instances at which the visualizations \ref{fig:VIS3} are taken. Figure b) shows the spatial averaged energy injection rate $\left\langle {\bf f} \cdot {\bf u} \right\rangle_{_S}$ and the dissipation rate $\nu \left\langle |{\bm \nabla} {\bf u}|^2 \right\rangle_{_S}$ for the same run. }
\label{fig:nobistability}
\end{figure}

\begin{figure}
\begin{subfigure}[b]{0.3\textwidth}
\includegraphics[scale=0.135]{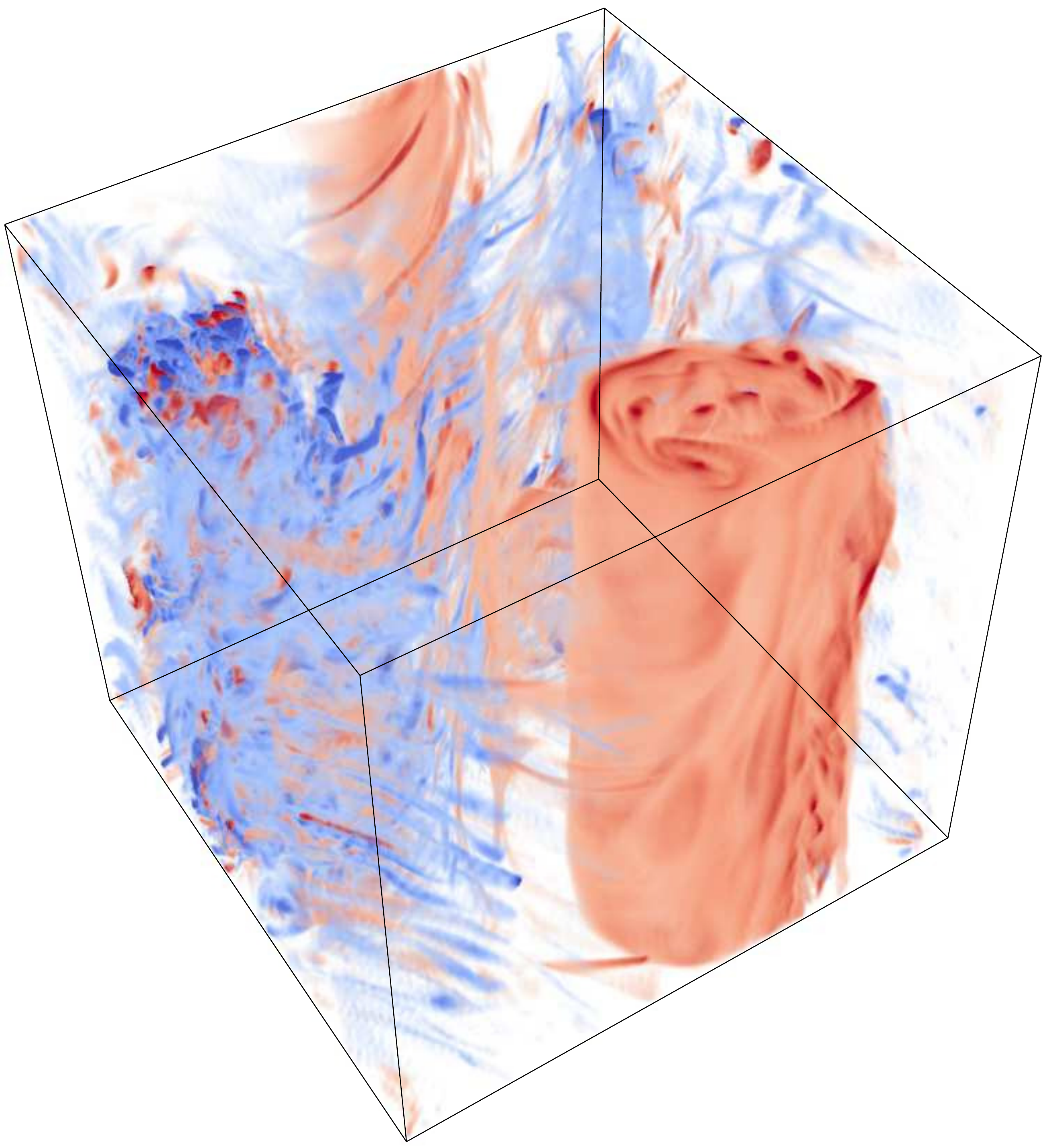}
\caption{}
\end{subfigure} 
\begin{subfigure}[b]{0.3\textwidth}
\includegraphics[scale=0.135]{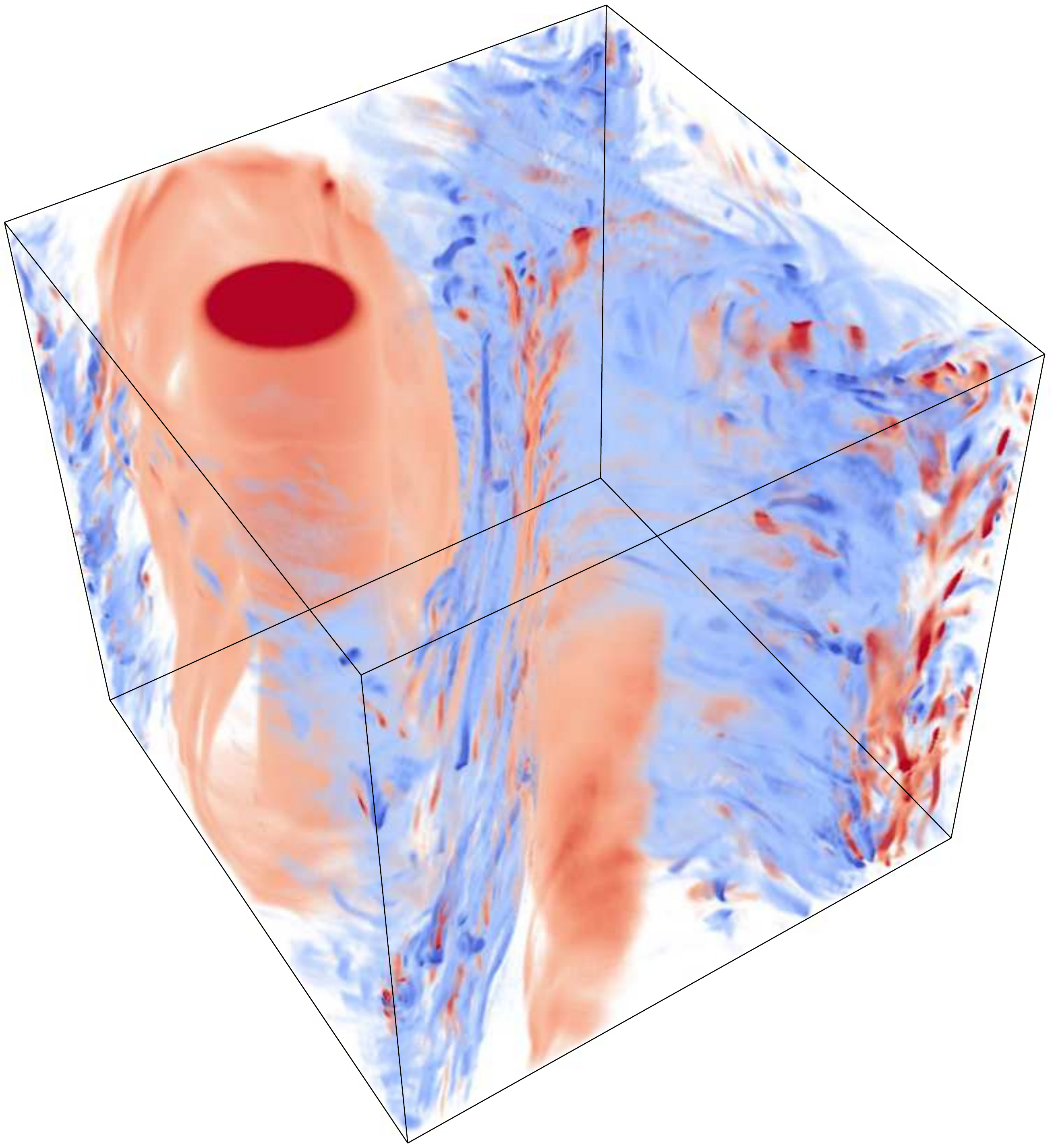}
\caption{}
\label{fig:chap4_fig1b}
\end{subfigure}
\begin{subfigure}[b]{0.3\textwidth}
\includegraphics[scale=0.135]{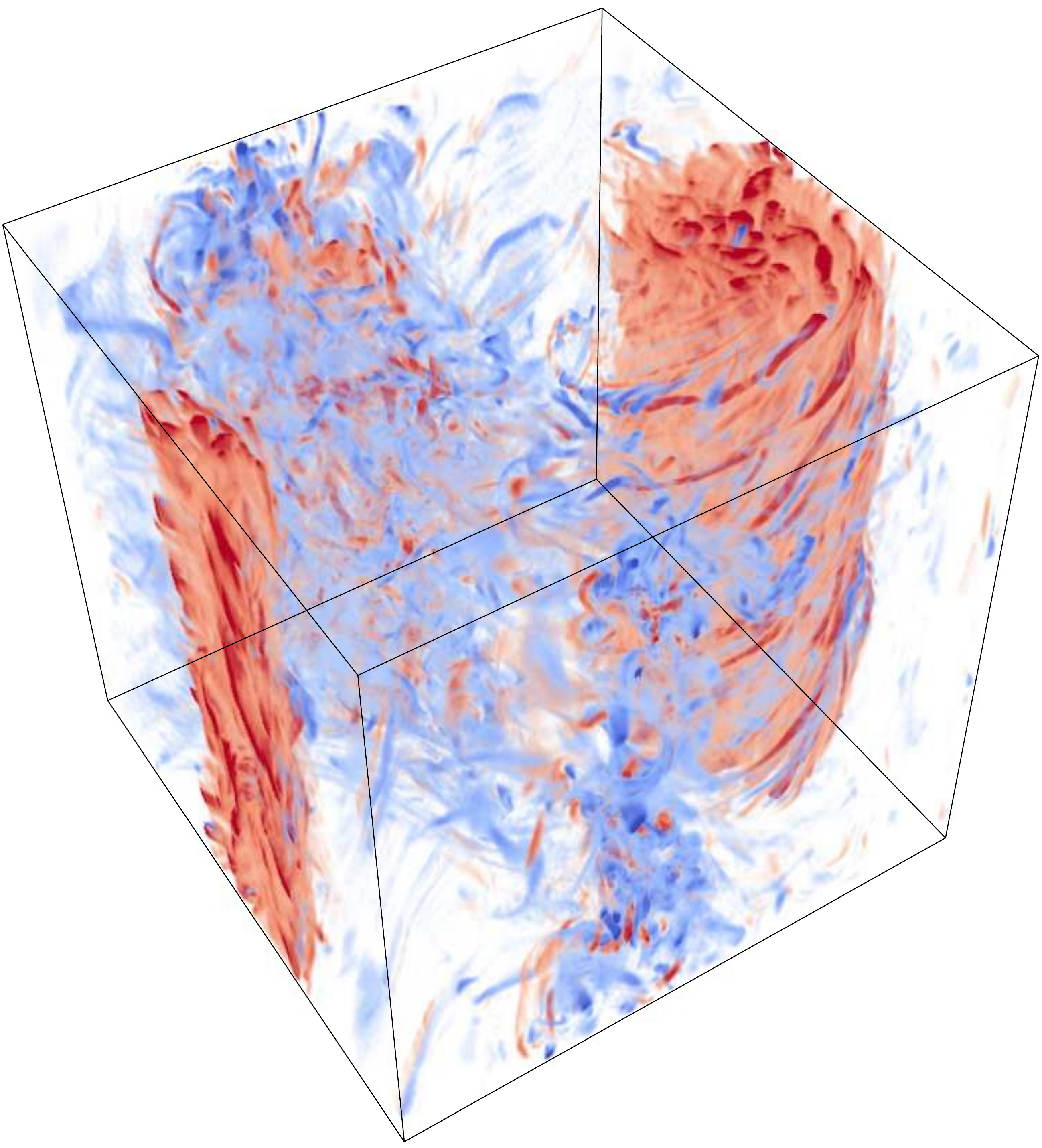}
\caption{}
\end{subfigure}
\caption{ The figures show the contours of the vertical vorticity $\omega_z$ for $Ro_f = 0.357, Re_f = 100$ for the three time instances marked in figure \ref{fig:nobistability1}. The red colour corresponds to positive vorticity and the blue colour corresponds to negative vorticity.}
\label{fig:VIS3}
\end{figure}

A similar oscillating behaviour is observed even further from the onset $\Rof^*$.
In figure \ref{fig:nobistability1} we show the time evolution of $\UTD$ and $U^2$ 
as in figure \ref{fig:bistability1} for a slightly smaller value of $\Rof=0.357$. 
Although, the system never returns to the isotropic case and $\UTD$ is always
dominant,  strong fluctuations are still present. 
Figure \ref{fig:nobistability2} 
shows the time series of the spatially averaged energy injection rate 
$\left\langle {\bf f} \cdot {\bf u} \right\rangle_{_{S}}$ and the energy dissipation rate 
$\nu \left\langle |{\bm \nabla} {\bf u}|^2\right\rangle_{_{S}}$ for the same run as \ref{fig:nobistability1}. 
Again peaks of energy injection/dissipation are correlated with changes in the large scale flow states.
We note that in the condensate regime, even though the dissipation is always positive, the energy injection rate 
takes both negative and positive values. This means that at certain instances of time the forcing 
takes energy out of the system. 
Visualization of the
flows in figure \ref{fig:VIS3}  at different times reveal that these fluctuations correspond to a transition of the
flow from a state that has a co-rotating vortex that is stable 
to states that are unstable to 3D fluctuations that however fail to destroy it. 
It seems thus that the key for understanding the behaviour of this flow lies in understanding the
stability properties of these free evolving vortexes.

\section{Conclusions} 
\label{cons} 

\begin{figure}
\centering
\includegraphics[scale=0.6]{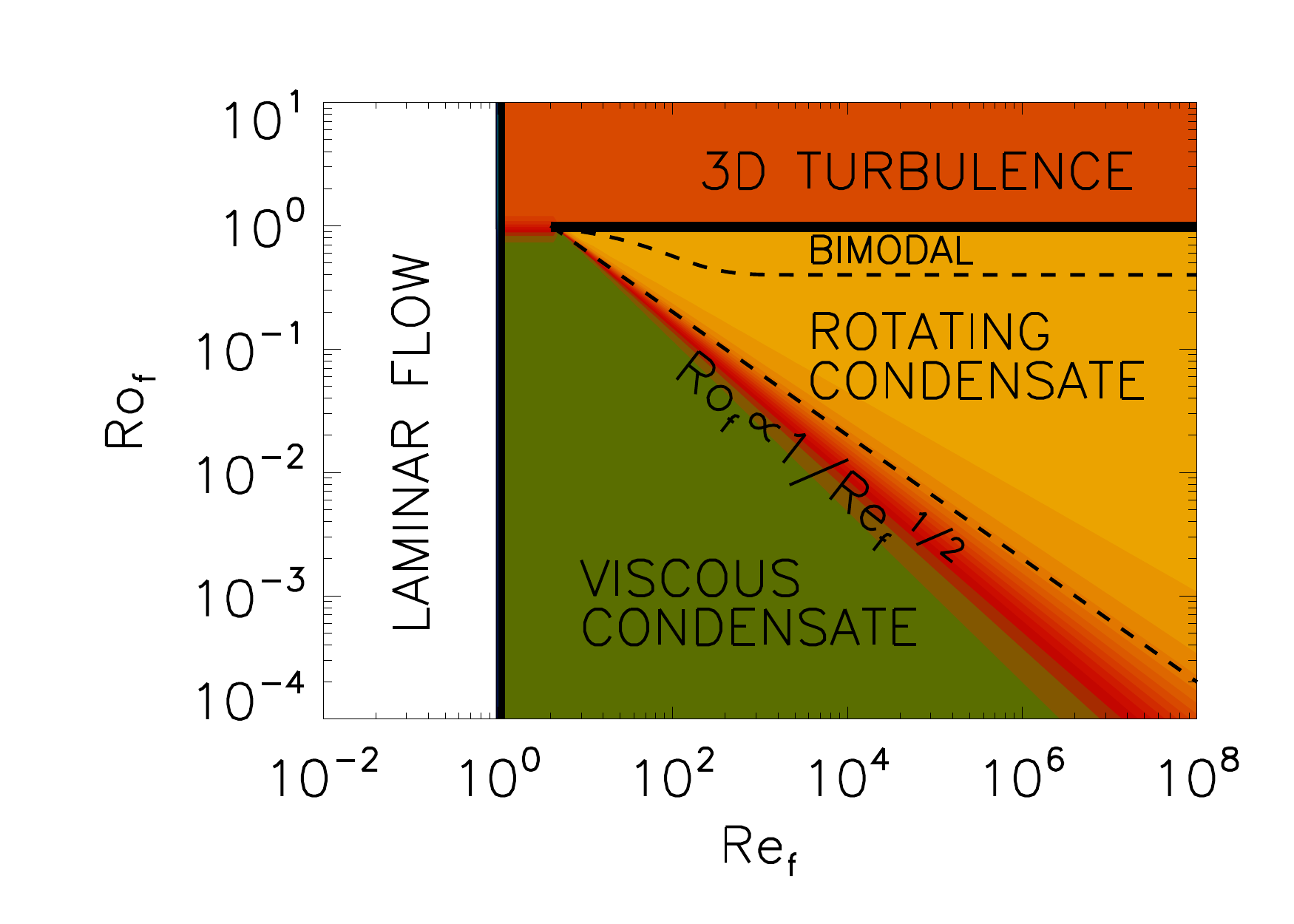}
\caption{The figure shows the phase space diagram with the different flow behaviour marked. Solid lines denote sharp/critical transitions. Dashed line denotes smooth transitions. }
\label{fig:PhaseSpace}
\end{figure}

This work has  given a description of steady state rotating turbulence when the forcing acts directly on the slow manifold,
by mapping to the parameter space the different
behaviours observed and the resulting scaling relations. Our results are concisely summarized in figure 
\ref{fig:PhaseSpace} where the four different phases of the rotating flows examined are shown 
in the parameter space $\Ref,\Rof$: Laminar flow, 3D turbulence flow, rotating condensate flow 
and viscous condensate flow. Solid lines indicate the boundaries where a critical transition take place,
while dashed lines indicate smooth transitions. 

For values of $\Ref$ bellow a critical value $\Ref^*$ that is independent of $\Rof$ the 
flow has a laminar behaviour. At this state the resulting flow is 2D, time independent
and proportional to the inverse Laplacian of the forcing.  The first unstable mode in this 
laminar state is a 2D mode that is not affected by the rotation and thus the instability
boundary does not depend on $\Rof$.   

For $\Ref \gg \Ref^*$ and $\Rof$ above a critical value $\Rof^*$ the flow displays quasi isotropic 3D turbulence.
This regime is described to a good degree by Kolmogorov-Richardson phenomenology \cite{K41,richardson1926atmospheric} 
and classical results of turbulence like the finite energy dissipation at zero viscosity limit and a $k^{-5/3}$ energy spectrum 
appear to hold.

For $\Ref \gg \Ref^*$ and $\Ref^{-1/2} \ll  \Rof \le \Rof^* $ the flow is shown to be in what we refer as a rotating condensate state.
In this state a co-rotating 2D vortex is dominating at the large scales while the counter rotating vortex breaks down to 3D eddies cascading energy back to the small scales. At this state the amplitude of the condensate $U_{_{2D}}$ (in the rather small range examined by our simulations) 
was shown to be proportional to the rotation rate $U_{_{2D}}\propto \Omega L$. Our results also indicated 
(with the help of hyper-viscous simulations) that in this regime the  finite energy dissipation at zero viscosity limit still holds 
but with a drag coefficient that rapidly decreases with $\Rof$. The spectra at the small scales follow a close to $k^{-5/3}$ power-law,
while a large peak appears at the largest scale indicating the presence of the condensate.

The transition from the  quasi isotropic 3D turbulent state to the rotating condensate state was shown to be supercritical, contrary to
the arguments described in the introduction that were predicting that at sufficiently large $\Ref$ right bellow criticality $\Rof^*$ the
system would transition discontinuously to the rotating condensate value $\UTD\propto \Omega^2 L^2$. The reason for this discrepancy is 
in part because the arguments in section \ref{theory} assumed weak dependence of the energy injection rate $\epsilon$ at criticality
while the DNS showed a strong  sensitivity of $\epsilon$ on $\Rof$ close to criticality. 
The second reason is that the system close to criticality 
showed a bimodal behaviour where part of the time it was spending in the 3D turbulence state with $ \UTD\ll\Omega^2L^2$ 
and remaining part was spent in the rotating condensate state with $\UTD\propto \Omega^2 L^2$. Despite the fact that 
these states appeared distinct, the time spent in the condensate state can decrease continuously to zero as $\Rof\to \Rof^*$ 
(from below) leading to a continuous supercritical transition.

For $\Ref \gg \Ref^*$ and  $\Rof  \ll  \Ref^{-1/2} $ the  flow is shown to be in what we refer to as a viscous condensate state.
In this state the flow is close to 2D and both the co-rotating and counter-rotating 2D vortex exist and dominate the large scales. 
The flow has a normalized energy dissipation rate that decreases with $\Reu$ following the laminar scaling $\Reu^{-1}$.
The transition from the rotating condensate regime to the viscous condensate was found to be smooth.
We note however that an other critical value of $\Rof$ is expected for which the flow becomes exactly 2D and all 
3D perturbations decay exponentially \citep{Gallet2015exact}. Such a transition is expected at even smaller values of $\Rof$ and
to observe it we have to focus on deviations from 2D flows which was not done in the present study. 
A similar study in thin layers has shown that this transition is governed by strong intermittent events \cite{benavides2017critical}. 
This is thus an interesting limit that is worth investigating in the future. 

The difference between the parameters $(\Ref,\Rof)$ and $(\Reu,\Rou)$ or $(\Red,\Rod)$ was not found to be as severe 
as in the Taylor-Green flow where discontinuous (sub-critical) transitions were present which resulting in mapping from one set of parameters to the other not to be one to one nor onto. In particular the difference between $(\Ref,\Rof)$ and $(\Red,\Rod)$ was only found to be significant 
close to the critical point $\Rof^*$ where $\epsilon$ was found to change abruptly. The difference between $(\Ref,\Rof)$ and $(\Reu,\Rou)$ 
was stronger and is due the fact that in the rotating condensate regime the scaling $U \propto \Omega L$ merged all values of $\Rou$ to be close to unity. This left all larger values of $\Rou$ to be in the viscous condensate regime. Thus, at the steady state regime at least $\Rou$ does not appear to be a good indicator for the strength of rotation.

We stress the importance of the ordering of the limits when one considers the low Rossby large Reynolds limits.
If one considers the $\Rof\to 0$ limit first and afterwards the $\Ref \to \infty$ one always falls in the viscous condensate regime. 
While if one considers the $\Ref \to \infty$ first one falls in the rotating condensate regime. To distinguish between the two 
one needs to look at the product $\Rof\Ref^{1/2}$ or $Ro_{\lambda} = \Rod\Red^{1/2}$. Referring thus to the large Reynolds number,
small Rossby number limit is ambiguous unless the ordering is specified.

Finally we comment on the effect of boundaries and the realizability of the present results in experiments.
In the present results we considered only the simplest domain that of a triple periodic geometry 
and we should give word of caution in extrapolating them to domains with no slip boundary conditions.
In the presence of no slip boundaries, rotation will introduce Ekman layers, \cite{ekman1905}, that can lead to large scale drag effects 
\citep{caldwell1972laboratory,howroyd1975characteristics,zavala2001ekman,sous2013},
altering in part the energy balance. Nonetheless we do believe that in a carefully prepared experimental setup 
where these effects are accounted for 
some of the presently observed phenomena would carry over to no-slip boundary conditions.
In particular, it would be interesting to investigate the transition to the rotating condensate
regime from 3D turbulence that displayed such rich behaviour. The high numerical cost of 3D simulations at 
this regime limits our runs to relatively short times and does not allow us to study in detail
their statistical behaviour. Experiments in which long signals are much easier attainable can
then address this issue.




\acknowledgments

The authors would like to thank various discussions with Stephan Fauve and the non-linear physics group at ENS.
This work was granted access to the HPC resources of MesoPSL financed by the Region Ile de France 
and the project Equip@Meso (reference ANR-10-EQPX-29-01) of the program Investissements d'Avenir 
supervised by the Agence Nationale pour la Recherche and the HPC resources of GENCI-TGCC-CURIE \& GENCI-CINES-OCCIGEN
(Project No. x2015056421 \& No. x2016056421 \& No. x2017056421) where the present numerical simulations have been performed.

\bibliographystyle{jfm}
\bibliography{refs}

\begin{thebibliography}{81}
\expandafter\ifx\csname natexlab\endcsname\relax\def\natexlab#1{#1}\fi
\def\au#1{#1} \def\ed#1{#1} \def\yr#1{#1}\def\at#1{#1}\def\jt#1{\textit{#1}}
  \def\bt#1{#1}\def\bvol#1{\textbf{#1}} \def\vol#1{#1} \def\pg#1{#1}
  \def\publ#1{#1}\def\arxiv#1{#1}\def\org#1{#1}\def\st#1{\textit{#1}}

\bibitem[Alexakis(2011)]{Alexakis2011two}
{\sc \au{Alexakis, A.}} \yr{2011}  \at{Two-dimensional behavior of
  three-dimensional magnetohydrodynamic flow with a strong guiding field}.
  \jt{Phys. Rev. E}  \bvol{84}~(5),  \pg{056330}.

\bibitem[Alexakis(2015)]{alexakis2015rotating}
{\sc \au{Alexakis, A.}} \yr{2015}  \at{Rotating taylor--green flow}.  \jt{J.
  Fluid Mech.}  \bvol{769},  \pg{46--78}.

\bibitem[Baroud {\em et~al.\/}(2002)Baroud, Plapp, She \&
  Swinney]{baroud2002anomalous}
{\sc \au{Baroud, Charles~N}, \au{Plapp, Brendan~B}, \au{She, Zhen-Su} \&
  \au{Swinney, Harry~L}} \yr{2002}  \at{Anomalous self-similarity in a
  turbulent rapidly rotating fluid}.  \jt{Physical Review Letters}
  \bvol{88}~(11),  \pg{114501}.

\bibitem[Baroud {\em et~al.\/}(2003)Baroud, Plapp, Swinney \&
  She]{baroud2003scaling}
{\sc \au{Baroud, Charles~N}, \au{Plapp, Brendan~B}, \au{Swinney, Harry~L} \&
  \au{She, Zhen-Su}} \yr{2003}  \at{Scaling in three-dimensional and
  quasi-two-dimensional rotating turbulent flows}.  \jt{Physics of Fluids}
  \bvol{15}~(8),  \pg{2091--2104}.

\bibitem[{Bartello} {\em et~al.\/}(1994){Bartello}, {Metais} \&
  {Lesieur}]{Bartello1994}
{\sc \au{{Bartello}, P.}, \au{{Metais}, O.} \& \au{{Lesieur}, M.}} \yr{1994}
  \at{{Coherent structures in rotating three-dimensional turbulence}}.  \jt{J.
  Fluid Mech.}  \bvol{273},  \pg{1--29}.

\bibitem[Benavides \& Alexakis(2017)]{benavides2017critical}
{\sc \au{Benavides, Santiago~Jose} \& \au{Alexakis, Alexandros}} \yr{2017}
  \at{Critical transitions in thin layer turbulence}.  \jt{Journal of Fluid
  Mechanics}  \bvol{822},  \pg{364--385}.

\bibitem[Biferale {\em et~al.\/}(2016)Biferale, Bonaccorso, Mazzitelli, van
  Hinsberg, Lanotte, Musacchio, Perlekar \& Toschi]{biferale2016coherent}
{\sc \au{Biferale, Luca}, \au{Bonaccorso, Fabio}, \au{Mazzitelli, Irene~M},
  \au{van Hinsberg, Michel~AT}, \au{Lanotte, Alessandra~S}, \au{Musacchio,
  Stefano}, \au{Perlekar, Prasad} \& \au{Toschi, Federico}} \yr{2016}
  \at{Coherent structures and extreme events in rotating multiphase turbulent
  flows}.  \jt{Physical Review X}  \bvol{6}~(4),  \pg{041036}.

\bibitem[Bourouiba \& Bartello(2007)]{bourouiba2007intermediate}
{\sc \au{Bourouiba, Lydia} \& \au{Bartello, Peter}} \yr{2007}  \at{The
  intermediate rossby number range and two-dimensional--three-dimensional
  transfers in rotating decaying homogeneous turbulence}.  \jt{Journal of Fluid
  Mechanics}  \bvol{587},  \pg{139--161}.

\bibitem[Caldwell {\em et~al.\/}(1972)Caldwell, Van~Atta \&
  Helland]{caldwell1972laboratory}
{\sc \au{Caldwell, D.~R.}, \au{Van~Atta, C.~W.} \& \au{Helland, K.~N.}}
  \yr{1972}  \at{A laboratory study of the turbulent ekman layer}.
  \jt{Geophys. \& Astrophys. Fluid Dyn.}  \bvol{3}~(1),  \pg{125--160}.

\bibitem[{Campagne} {\em et~al.\/}(2014){Campagne}, {Gallet}, {Moisy} \&
  {Cortet}]{Moisy2014direct}
{\sc \au{{Campagne}, A.}, \au{{Gallet}, B.}, \au{{Moisy}, F.} \& \au{{Cortet},
  P.-P.}} \yr{2014}  \at{{Direct and inverse energy cascades in a forced
  rotating turbulence experiment}}.  \jt{Physics of Fluids}  \bvol{26}~(12),
  \pg{125112}.

\bibitem[Campagne {\em et~al.\/}(2015)Campagne, Gallet, Moisy \&
  Cortet]{campagne2015disentangling}
{\sc \au{Campagne, A.}, \au{Gallet, B.}, \au{Moisy, F.} \& \au{Cortet, P.-P.}}
  \yr{2015}  \at{Disentangling inertial waves from eddy turbulence in a forced
  rotating-turbulence experiment}.  \jt{Phys. Rev. E.}  \bvol{91}~(4),
  \pg{043016}.

\bibitem[Campagne {\em et~al.\/}(2016)Campagne, Machicoane, Gallet, Cortet \&
  Moisy]{campagne2016turbulent}
{\sc \au{Campagne, Antoine}, \au{Machicoane, Nathana{\"e}l}, \au{Gallet,
  Basile}, \au{Cortet, Pierre-Philippe} \& \au{Moisy, Fr{\'e}d{\'e}ric}}
  \yr{2016}  \at{Turbulent drag in a rotating frame}.  \jt{Journal of Fluid
  Mechanics}  \bvol{794},  \pg{R5}.

\bibitem[{Celani} {\em et~al.\/}(2010){Celani}, {Musacchio} \&
  {Vincenzi}]{Celani2010turbulence}
{\sc \au{{Celani}, A.}, \au{{Musacchio}, S.} \& \au{{Vincenzi}, D.}} \yr{2010}
  \at{{Turbulence in More than Two and Less than Three Dimensions}}.
  \jt{Physical Review Letters}  \bvol{104}~(18),  \pg{184506}.

\bibitem[{Chen} {\em et~al.\/}(2005){Chen}, {Chen}, {Eyink} \&
  {Holm}]{Chen2005}
{\sc \au{{Chen}, Q.}, \au{{Chen}, S.}, \au{{Eyink}, G.~L.} \& \au{{Holm},
  D.~D.}} \yr{2005}  \at{{Resonant interactions in rotating homogeneous
  three-dimensional turbulence}}.  \jt{J. Fluid Mech.}  \bvol{542},
  \pg{139--164}.

\bibitem[{Courant} {\em et~al.\/}(1928){Courant}, {Friedrichs} \& {Lewy}]{CFL}
{\sc \au{{Courant}, R.}, \au{{Friedrichs}, K.} \& \au{{Lewy}, H.}} \yr{1928}
  \at{{\"Uber die partiellen Differenzengleichungen der mathematischen
  Physik.}}  \jt{{Math. Ann.}}  \bvol{100},  \pg{32--74}.

\bibitem[Dallas \& Tobias(2016)]{dallas2016forcing}
{\sc \au{Dallas, Vassilios} \& \au{Tobias, Steven~M}} \yr{2016}
  \at{Forcing-dependent dynamics and emergence of helicity in rotating
  turbulence}.  \jt{Journal of Fluid Mechanics}  \bvol{798},  \pg{682--695}.

\bibitem[Deusebio {\em et~al.\/}(2014)Deusebio, Boffetta, Lindborg \&
  Musacchio]{deusebio2014dimensional}
{\sc \au{Deusebio, E.}, \au{Boffetta, G.}, \au{Lindborg, E.} \& \au{Musacchio,
  S.}} \yr{2014}  \at{Dimensional transition in rotating turbulence}.
  \jt{Phys. Rev. E}  \bvol{90}~(2),  \pg{023005}.

\bibitem[Dickinson \& Long(1983)]{dickinson1983oscillating}
{\sc \au{Dickinson, S.~C.} \& \au{Long, R.~R.}} \yr{1983}  \at{Oscillating-grid
  turbulence including effects of rotation}.  \jt{J. Fluid Mech.}  \bvol{126},
  \pg{315--333}.

\bibitem[Ekman(1905)]{ekman1905}
{\sc \au{Ekman, V.~W.}} \yr{1905}  \at{On the influence of the
  earth$\backslash$'s rotation on ocean currents}.  \jt{Ark. Mat. Astron. Fys.}
   \bvol{2},  \pg{1--53}.

\bibitem[Favier {\em et~al.\/}(2010)Favier, Godeferd \&
  Cambon]{favier2010space}
{\sc \au{Favier, B.}, \au{Godeferd, F.~S.} \& \au{Cambon, C.}} \yr{2010}
  \at{On space and time correlations of isotropic and rotating turbulence}.
  \jt{Phys. Fluids}  \bvol{22}~(1),  \pg{015101}.

\bibitem[Gallet(2015)]{Gallet2015exact}
{\sc \au{Gallet, B.}} \yr{2015}  \at{Exact two-dimensionalization of rapidly
  rotating large-reynolds-number flows}.  \jt{J. Fluid Mech.}  \bvol{783},
  \pg{412--447}.

\bibitem[Gallet {\em et~al.\/}(2014)Gallet, Campagne, Cortet \&
  Moisy]{Gallet2014Scale}
{\sc \au{Gallet, B.}, \au{Campagne, A.}, \au{Cortet, P.-P.} \& \au{Moisy, F.}}
  \yr{2014}  \at{Scale-dependent cyclone-anticyclone asymmetry in a forced
  rotating turbulence experiment}.  \jt{Physics of Fluids}  \bvol{26}~(3),
  \pg{035108}.

\bibitem[Gallet \& Young(2013)]{gallet2013two}
{\sc \au{Gallet, B.} \& \au{Young, W.~R.}} \yr{2013}  \at{A two-dimensional
  vortex condensate at high reynolds number}.  \jt{J. Fluid Mech.}  \bvol{715},
   \pg{359--388}.

\bibitem[Galtier(2003)]{galtier2003weak}
{\sc \au{Galtier, S.}} \yr{2003}  \at{Weak inertial-wave turbulence theory}.
  \jt{Phys. Rev. E}  \bvol{68}~(1),  \pg{015301}.

\bibitem[Godeferd \& Lollini(1999)]{godeferd1999direct}
{\sc \au{Godeferd, FS} \& \au{Lollini, L}} \yr{1999}  \at{Direct numerical
  simulations of turbulence with confinement and rotation}.  \jt{Journal of
  Fluid Mechanics}  \bvol{393},  \pg{257--308}.

\bibitem[Greenspan(1968)]{greenspan1968theory}
{\sc \au{Greenspan, HP}} \yr{1968}  \at{The theory of rotating fluids cambridge
  university press}.  \jt{Cambridge, England} .

\bibitem[Hopfinger \& Heijst(1993)]{hopfinger1993vortices}
{\sc \au{Hopfinger, EJ} \& \au{Heijst, GJFV}} \yr{1993}  \at{Vortices in
  rotating fluids}.  \jt{Annual review of fluid mechanics}  \bvol{25}~(1),
  \pg{241--289}.

\bibitem[Hopfinger {\em et~al.\/}(1982)Hopfinger, Browand \&
  Gagne]{hopfinger1982turbulence}
{\sc \au{Hopfinger, E.~J.}, \au{Browand, F.~K.} \& \au{Gagne, Y.}} \yr{1982}
  \at{Turbulence and waves in a rotating tank}.  \jt{J. Fluid Mech.}
  \bvol{125},  \pg{505--534}.

\bibitem[{Hough}(1897)]{Hough1897}
{\sc \au{{Hough}, S.~S.}} \yr{1897}  \at{{On the Application of Harmonic
  Analysis to the Dynamical Theory of the Tides. Part I. On Laplace's
  ''Oscillations of the First Species,'' and on the Dynamics of Ocean
  Currents}}.  \jt{Philos. Trans. R. Soc. London A}  \bvol{189},
  \pg{201--257}.

\bibitem[Howroyd \& Slawson(1975)]{howroyd1975characteristics}
{\sc \au{Howroyd, G.C.} \& \au{Slawson, P.R.}} \yr{1975}  \at{The
  characteristics of a laboratory produced turbulent ekman layer}.
  \jt{Boundary-Layer Meteorology}  \bvol{8}~(2),  \pg{201--219}.

\bibitem[Ibbetson \& Tritton(1975)]{ibbetson1975experiments}
{\sc \au{Ibbetson, A} \& \au{Tritton, DJ}} \yr{1975}  \at{Experiments on
  turbulence in a rotating fluid}.  \jt{Journal of Fluid Mechanics}
  \bvol{68}~(4),  \pg{639--672}.

\bibitem[Ishihara {\em et~al.\/}(2016)Ishihara, Morishita, Yokokawa, Uno \&
  Kaneda]{ishihara2016energy}
{\sc \au{Ishihara, Takashi}, \au{Morishita, Koji}, \au{Yokokawa, Mitsuo},
  \au{Uno, Atsuya} \& \au{Kaneda, Yukio}} \yr{2016}  \at{Energy spectrum in
  high-resolution direct numerical simulations of turbulence}.  \jt{Physical
  Review Fluids}  \bvol{1}~(8),  \pg{082403}.

\bibitem[Kaneda {\em et~al.\/}(2003)Kaneda, Ishihara, Yokokawa, Itakura \&
  Uno]{kaneda2003energy}
{\sc \au{Kaneda, Yukio}, \au{Ishihara, Takashi}, \au{Yokokawa, Mitsuo},
  \au{Itakura, Ken’ichi} \& \au{Uno, Atsuya}} \yr{2003}  \at{Energy
  dissipation rate and energy spectrum in high resolution direct numerical
  simulations of turbulence in a periodic box}.  \jt{Physics of Fluids}
  \bvol{15}~(2),  \pg{L21--L24}.

\bibitem[Kolmogorov(1941)]{K41}
{\sc \au{Kolmogorov, A.~N.}} \yr{1941}  \at{The local structure of turbulence
  in incompressible viscous fluid for very large reynolds number}.
  \jt{Proceedings of the USSR Academy of Sciences}  \bvol{30},  \pg{299303}.

\bibitem[Kraichnan(1967)]{kraichnan1967inertial}
{\sc \au{Kraichnan, R.~H.}} \yr{1967}  \bt{Inertial ranges in two-dimensional
  turbulence}. {\em Tech. Rep.\/}.  \org{DTIC Document}.

\bibitem[Machicoane {\em et~al.\/}(2016)Machicoane, Moisy \&
  Cortet]{Machicoane2016Two}
{\sc \au{Machicoane, Nathana\"el}, \au{Moisy, Fr\'ed\'eric} \& \au{Cortet,
  Pierre-Philippe}} \yr{2016}  \at{Two-dimensionalization of the flow driven by
  a slowly rotating impeller in a rapidly rotating fluid}.  \jt{Phys. Rev.
  Fluids}  \bvol{1},  \pg{073701}.

\bibitem[Marino {\em et~al.\/}(2013)Marino, Mininni, Rosenberg \&
  Pouquet]{marino2013inverse}
{\sc \au{Marino, Raffaele}, \au{Mininni, Pablo~Daniel}, \au{Rosenberg, Duane}
  \& \au{Pouquet, Annick}} \yr{2013}  \at{Inverse cascades in rotating
  stratified turbulence: Fast growth of large scales}.  \jt{EPL (Europhysics
  Letters)}  \bvol{102}~(4),  \pg{44006}.

\bibitem[{Marino} {\em et~al.\/}(2015){Marino}, {Pouquet} \&
  {Rosenberg}]{Marino2015resolving}
{\sc \au{{Marino}, R.}, \au{{Pouquet}, A.} \& \au{{Rosenberg}, D.}} \yr{2015}
  \at{{Resolving the Paradox of Oceanic Large-Scale Balance and Small-Scale
  Mixing}}.  \jt{Physical Review Letters}  \bvol{114}~(11),  \pg{114504}.

\bibitem[{Mininni} {\em et~al.\/}(2009){Mininni}, {Alexakis} \&
  {Pouquet}]{Mininni2009}
{\sc \au{{Mininni}, P.~D.}, \au{{Alexakis}, A.} \& \au{{Pouquet}, A.}}
  \yr{2009}  \at{{Scale interactions and scaling laws in rotating flows at
  moderate Rossby numbers and large Reynolds numbers}}.  \jt{Phys. Fluids}
  \bvol{21}~(1),  \pg{015108}.

\bibitem[{Mininni} \& {Pouquet}(2010)]{Mininni2010}
{\sc \au{{Mininni}, P.~D.} \& \au{{Pouquet}, A.}} \yr{2010}  \at{{Rotating
  helical turbulence. I. Global evolution and spectral behavior}}.  \jt{Phys.
  Fluids}  \bvol{22}~(3),  \pg{035105}.

\bibitem[Mininni {\em et~al.\/}(2011)Mininni, Rosenberg, Reddy \&
  Pouquet]{mininni2011hybrid}
{\sc \au{Mininni, P.~D.}, \au{Rosenberg, D.}, \au{Reddy, R.} \& \au{Pouquet,
  A.}} \yr{2011}  \at{A hybrid mpi--openmp scheme for scalable parallel
  pseudospectral computations for fluid turbulence}.  \jt{Parallel Computing}
  \bvol{37}~(6),  \pg{316--326}.

\bibitem[Mishra {\em et~al.\/}(2015)Mishra, Herault, Fauve \&
  Verma]{mishra2015dynamics}
{\sc \au{Mishra, Pankaj~Kumar}, \au{Herault, Johann}, \au{Fauve, Stephan} \&
  \au{Verma, Mahendra~K}} \yr{2015}  \at{Dynamics of reversals and condensates
  in two-dimensional kolmogorov flows}.  \jt{Physical Review E}  \bvol{91}~(5),
   \pg{053005}.

\bibitem[Moisy {\em et~al.\/}(2011)Moisy, Morize, Rabaud \&
  Sommeria]{moisy2011decay}
{\sc \au{Moisy, F.}, \au{Morize, C.}, \au{Rabaud, M.} \& \au{Sommeria, J.}}
  \yr{2011}  \at{Decay laws, anisotropy and cyclone--anticyclone asymmetry in
  decaying rotating turbulence}.  \jt{J. Fluid Mech.}  \bvol{666},  \pg{5--35}.

\bibitem[Morize \& Moisy(2006)]{morize2006energy}
{\sc \au{Morize, C.} \& \au{Moisy, F.}} \yr{2006}  \at{Energy decay of rotating
  turbulence with confinement effects}.  \jt{Phys. Fluids}  \bvol{18}~(6),
  \pg{065107}.

\bibitem[Nazarenko(2011)]{nazarenko2011wave}
{\sc \au{Nazarenko, S.}} \yr{2011} {\em Wave turbulence\/}, ,  \vol{vol. 825}.
  \publ{Springer Science \& Business Media}.

\bibitem[Pedlosky(1987)]{pedlosky2013geophysical}
{\sc \au{Pedlosky, J.}} \yr{1987} {\em Geophysical fluid dynamics\/}.
  \publ{New York, Springer}.

\bibitem[Pouquet \& Marino(2013)]{pouquet2013geophysical}
{\sc \au{Pouquet, A.} \& \au{Marino, R.}} \yr{2013}  \at{Geophysical turbulence
  and the duality of the energy flow across scales}.  \jt{Phys. Rev. Lett.}
  \bvol{111}~(23),  \pg{234501}.

\bibitem[Pouquet {\em et~al.\/}(2013)Pouquet, Sen, Rosenberg, Mininni \&
  Baerenzung]{pouquet2013inverse}
{\sc \au{Pouquet, Annick}, \au{Sen, A}, \au{Rosenberg, D}, \au{Mininni,
  Pablo~Daniel} \& \au{Baerenzung, J}} \yr{2013}  \at{Inverse cascades in
  turbulence and the case of rotating flows}.  \jt{Physica Scripta}
  \bvol{2013}~(T155),  \pg{014032}.

\bibitem[{Proudman}(1916)]{Proudman1916}
{\sc \au{{Proudman}, J.}} \yr{1916}  \at{{On the Motion of Solids in a Liquid
  Possessing Vorticity}}.  \jt{Philos. Trans. R. Soc. London A}  \bvol{92},
  \pg{408--424}.

\bibitem[Richardson(1926)]{richardson1926atmospheric}
{\sc \au{Richardson, L.~F.}} \yr{1926}  \at{Atmospheric diffusion shown on a
  distance-neighbour graph}.  \jt{Proc. Roy. Soc. London. A}  \bvol{110}~(756),
   \pg{709--737}.

\bibitem[Roberts(1972)]{roberts1972dynamo}
{\sc \au{Roberts, Gareth~O}} \yr{1972}  \at{Dynamo action of fluid motions with
  two-dimensional periodicity}.  \jt{Philosophical Transactions of the Royal
  Society of London A: Mathematical, Physical and Engineering Sciences}
  \bvol{271}~(1216),  \pg{411--454}.

\bibitem[Ruppert-Felsot {\em et~al.\/}(2005)Ruppert-Felsot, Praud, Sharon \&
  Swinney]{ruppert2005extraction}
{\sc \au{Ruppert-Felsot, Jori~E}, \au{Praud, Olivier}, \au{Sharon, Eran} \&
  \au{Swinney, Harry~L}} \yr{2005}  \at{Extraction of coherent structures in a
  rotating turbulent flow experiment}.  \jt{Physical Review E}  \bvol{72}~(1),
  \pg{016311}.

\bibitem[Scott(2014)]{scott2014wave}
{\sc \au{Scott, Julian~F}} \yr{2014}  \at{Wave turbulence in a rotating
  channel}.  \jt{Journal of Fluid Mechanics}  \bvol{741},  \pg{316--349}.

\bibitem[{Sen} {\em et~al.\/}(2012){Sen}, {Mininni}, {Rosenberg} \&
  {Pouquet}]{Mininni2012}
{\sc \au{{Sen}, A.}, \au{{Mininni}, P.~D.}, \au{{Rosenberg}, D.} \&
  \au{{Pouquet}, A.}} \yr{2012}  \at{{Anisotropy and nonuniversality in scaling
  laws of the large-scale energy spectrum in rotating turbulence}}.  \jt{Phys.
  Rev. E}  \bvol{86}~(3),  \pg{036319}.

\bibitem[Seshasayanan \& Alexakis(2016)]{seshasayanan2016critical}
{\sc \au{Seshasayanan, K.} \& \au{Alexakis, A.}} \yr{2016}  \at{Critical
  behavior in the inverse to forward energy transition in two-dimensional
  magnetohydrodynamic flow}.  \jt{Phys. Rev. E}  \bvol{93}~(1),  \pg{013104}.

\bibitem[Seshasayanan {\em et~al.\/}(2014)Seshasayanan, Benavides \&
  Alexakis]{seshasayanan2014edge}
{\sc \au{Seshasayanan, K.}, \au{Benavides, S.~J.} \& \au{Alexakis, A.}}
  \yr{2014}  \at{On the edge of an inverse cascade}.  \jt{Phys. Rev. E}
  \bvol{90}~(5),  \pg{051003}.

\bibitem[Shats {\em et~al.\/}(2007)Shats, Xia, Punzmann \&
  Falkovich]{shats2007suppression}
{\sc \au{Shats, MG}, \au{Xia, Hua}, \au{Punzmann, Horst} \& \au{Falkovich,
  Gregory}} \yr{2007}  \at{Suppression of turbulence by self-generated and
  imposed mean flows}.  \jt{Phys. Rev. Lett.}  \bvol{99}~(16),  \pg{164502}.

\bibitem[Smith {\em et~al.\/}(1996)Smith, Chasnov \&
  Waleffe]{smith1996crossover}
{\sc \au{Smith, L.~M.}, \au{Chasnov, J.~R.} \& \au{Waleffe, F.}} \yr{1996}
  \at{Crossover from two-to three-dimensional turbulence}.  \jt{Phys. Rev.
  Lett.}  \bvol{77}~(12),  \pg{2467}.

\bibitem[{Smith} \& {Waleffe}(1999)]{Smith1999}
{\sc \au{{Smith}, L.~M.} \& \au{{Waleffe}, F.}} \yr{1999}  \at{{Transfer of
  energy to two-dimensional large scales in forced, rotating three-dimensional
  turbulence}}.  \jt{Phys. Fluids}  \bvol{11},  \pg{1608--1622}.

\bibitem[Smith~R. \& Yakhot(1994)]{smithr1994finite}
{\sc \au{Smith~R., Leslie~M} \& \au{Yakhot, Victor}} \yr{1994}  \at{Finite-size
  effects in forced two-dimensional turbulence}.  \jt{Journal of Fluid
  Mechanics}  \bvol{274},  \pg{115--138}.

\bibitem[Sous {\em et~al.\/}(2013)Sous, Sommeria \& Boyer]{sous2013}
{\sc \au{Sous, D.}, \au{Sommeria, J.} \& \au{Boyer, D.~L.}} \yr{2013}
  \at{{Friction law and turbulent properties in a laboratory Ekman boundary
  layer}}.  \jt{{Phys. Fluids}}  \bvol{25}~(4),  \pg{xx}.

\bibitem[Sozza {\em et~al.\/}(2015)Sozza, Boffetta, Muratore-Ginanneschi \&
  Musacchio]{Sozza2015dimensional}
{\sc \au{Sozza, A}, \au{Boffetta, G}, \au{Muratore-Ginanneschi, P} \&
  \au{Musacchio, Stefano}} \yr{2015}  \at{Dimensional transition of energy
  cascades in stably stratified forced thin fluid layers}.  \jt{Physics of
  Fluids}  \bvol{27}~(3),  \pg{035112}.

\bibitem[Sreenivasan \& Davidson(2008)]{sreenivasan2008formation}
{\sc \au{Sreenivasan, Binod} \& \au{Davidson, PA}} \yr{2008}  \at{On the
  formation of cyclones and anticyclones in a rotating fluid}.  \jt{Physics of
  Fluids}  \bvol{20}~(8),  \pg{085104}.

\bibitem[Sreenivasan(1984)]{sreenivasan1984scaling}
{\sc \au{Sreenivasan, Katepalli~R}} \yr{1984}  \at{On the scaling of the
  turbulence energy dissipation rate}.  \jt{The Physics of fluids}
  \bvol{27}~(5),  \pg{1048--1051}.

\bibitem[{Staplehurst} {\em et~al.\/}(2008){Staplehurst}, {Davidson} \&
  {Dalziel}]{Staplehurst2008}
{\sc \au{{Staplehurst}, P.~J.}, \au{{Davidson}, P.~A.} \& \au{{Dalziel},
  S.~B.}} \yr{2008}  \at{{Structure formation in homogeneous freely decaying
  rotating turbulence}}.  \jt{J. Fluid Mech.}  \bvol{598},  \pg{81--105}.

\bibitem[{Sugihara} {\em et~al.\/}(2005){Sugihara}, {Migita} \&
  {Honji}]{Sugihara2005}
{\sc \au{{Sugihara}, Y.}, \au{{Migita}, M.} \& \au{{Honji}, H.}} \yr{2005}
  \at{{Orderly flow structures in grid-generated turbulence with background
  rotation}}.  \jt{Fluid Dyn. Res.}  \bvol{36},  \pg{23--34}.

\bibitem[{Taylor}(1917)]{Taylor1917}
{\sc \au{{Taylor}, G.~I.}} \yr{1917}  \at{{Motion of Solids in Fluids When the
  Flow is Not Irrotational}}.  \jt{Proc. R. Soc. London A}  \bvol{93},
  \pg{99--113}.

\bibitem[{Thiele} \& {M{\"u}ller}(2009)]{Thiele2009}
{\sc \au{{Thiele}, M.} \& \au{{M{\"u}ller}, W.-C.}} \yr{2009}  \at{{Structure
  and decay of rotating homogeneous turbulence}}.  \jt{J. Fluid Mech.}
  \bvol{637},  \pg{425}.

\bibitem[Tsang \& Young(2009)]{tsang2009forced}
{\sc \au{Tsang, Y.~K.} \& \au{Young, W.~R.}} \yr{2009}  \at{Forced-dissipative
  two-dimensional turbulence: A scaling regime controlled by drag}.  \jt{Phys.
  Rev. E}  \bvol{79}~(4),  \pg{045308}.

\bibitem[Valente \& Dallas(2017)]{valente2017spectral}
{\sc \au{Valente, Pedro~C} \& \au{Dallas, Vassilios}} \yr{2017}  \at{Spectral
  imbalance in the inertial range dynamics of decaying rotating turbulence}.
  \jt{Physical Review E}  \bvol{95}~(2),  \pg{023114}.

\bibitem[Van~Bokhoven {\em et~al.\/}(2008)Van~Bokhoven, Cambon, Liechtenstein,
  Godeferd \& Clercx]{van2008refined}
{\sc \au{Van~Bokhoven, LJA}, \au{Cambon, Claude}, \au{Liechtenstein, Lukas},
  \au{Godeferd, Fabien~S} \& \au{Clercx, HJH}} \yr{2008}  \at{Refined vorticity
  statistics of decaying rotating three-dimensional turbulence}.  \jt{Journal
  of Turbulence}  \bvol{9}~(N6).

\bibitem[{van Bokhoven} {\em et~al.\/}(2009){van Bokhoven}, {Clercx}, {van
  Heijst} \& {Trieling}]{Bokhoven2009experiments}
{\sc \au{{van Bokhoven}, L.~J.~A.}, \au{{Clercx}, H.~J.~H.}, \au{{van Heijst},
  G.~J.~F.} \& \au{{Trieling}, R.~R.}} \yr{2009}  \at{{Experiments on rapidly
  rotating turbulent flows}}.  \jt{Physics of Fluids}  \bvol{21}~(9),
  \pg{096601}.

\bibitem[Weeks {\em et~al.\/}(1997)Weeks, Tian, Urbach, Ide, Swinney \&
  Ghil]{weeks1997transitions}
{\sc \au{Weeks, Eric~R}, \au{Tian, Yudong}, \au{Urbach, JS}, \au{Ide, Kayo},
  \au{Swinney, Harry~L} \& \au{Ghil, Michael}} \yr{1997}  \at{Transitions
  between blocked and zonal flows in a rotating annulus with topography}.
  \jt{Science}  \bvol{278}~(5343),  \pg{1598--1601}.

\bibitem[Xia {\em et~al.\/}(2008)Xia, Punzmann, Falkovich \&
  Shats]{xia2008turbulence}
{\sc \au{Xia, H.}, \au{Punzmann, H.}, \au{Falkovich, G.} \& \au{Shats, M.~G.}}
  \yr{2008}  \at{Turbulence-condensate interaction in two dimensions}.
  \jt{Phys. Rev. Lett.}  \bvol{101}~(19),  \pg{194504}.

\bibitem[Yarom \& Sharon(2014)]{yarom2014experimental}
{\sc \au{Yarom, Ehud} \& \au{Sharon, Eran}} \yr{2014}  \at{Experimental
  observation of steady inertial wave turbulence in deep rotating flows}.
  \jt{Nature Physics}  \bvol{10}~(7),  \pg{510--514}.

\bibitem[Yarom {\em et~al.\/}(2013)Yarom, Vardi \&
  Sharon]{Yarom2013experimental}
{\sc \au{Yarom, E.}, \au{Vardi, Y.} \& \au{Sharon, E.}} \yr{2013}
  \at{Experimental quantification of inverse energy cascade in deep rotating
  turbulence}.  \jt{Phys. Fluids}  \bvol{25}~(8),  \pg{085105}.

\bibitem[Yeung \& Zhou(1998)]{yeung1998numerical}
{\sc \au{Yeung, P.~K.} \& \au{Zhou, Y.}} \yr{1998}  \at{Numerical study of
  rotating turbulence with external forcing}.  \jt{Phys. Fluids}
  \bvol{10}~(11),  \pg{2895--2909}.

\bibitem[{Yokoyama} \& {Takaoka}(2017)]{2017arXiv170108497Y}
{\sc \au{{Yokoyama}, N.} \& \au{{Takaoka}, M.}} \yr{2017}  \at{{Bistability
  between quasi-two-dimensional flow and three-dimensional flow in forced
  rotating turbulence}}.  \jt{ArXiv e-prints} ,  \arxiv{arXiv: 1701.08497}.

\bibitem[{Yoshimatsu} {\em et~al.\/}(2011){Yoshimatsu}, {Midorikawa} \&
  {Kaneda}]{Yoshimatsu2011}
{\sc \au{{Yoshimatsu}, K.}, \au{{Midorikawa}, M.} \& \au{{Kaneda}, Y.}}
  \yr{2011}  \at{{Columnar eddy formation in freely decaying homogeneous
  rotating turbulence}}.  \jt{J. Fluid Mech.}  \bvol{677},  \pg{154--178}.

\bibitem[Zavala~Sans{\'o}n {\em et~al.\/}(2001)Zavala~Sans{\'o}n, van Heijst \&
  Backx]{zavala2001ekman}
{\sc \au{Zavala~Sans{\'o}n, L}, \au{van Heijst, GJF} \& \au{Backx, NA}}
  \yr{2001}  \at{Ekman decay of a dipolar vortex in a rotating fluid}.
  \jt{Phys. Fluids}  \bvol{13}~(2),  \pg{440--451}.

\bibitem[Zeman(1994)]{zeman1994note}
{\sc \au{Zeman, O.}} \yr{1994}  \at{A note on the spectra and decay of rotating
  homogeneous turbulence}.  \jt{Phys. Fluids}  \bvol{6}~(10),  \pg{3221--3223}.

\end{thebibliography}

\end{document}